\documentclass[12pt]{article}
\usepackage{jheppub}
\newcommand\Tr{\mathrm{Tr}}
\newcommand\bC{\mathbb{C}}
\newcommand\bR{\mathbb{R}}
\newcommand\Pexp{\mathrm{Pexp}}

\usepackage{float}
\def\ie{\begin{equation}\begin{aligned}}
\def\fe{\end{aligned}\end{equation}}

\newcommand{\E}{{\epsilon}}

\title{Aspects of $\Omega$-deformed M-theory}
\abstract{We explore the properties of $\Omega$-deformed M-theory, 
with particular focus on the $\bC_{\epsilon_1} \times \bC_{\epsilon_2}\times \bC_{\epsilon_3}$
background and coupling to $\Omega$-deformed M2 and M5 brane world-volume theories.}
\author[1]{Davide Gaiotto,}
\author[1,2]{Jihwan Oh}
\affiliation[1]{Perimeter Institute for Theoretical Physics, Waterloo, Ontario, Canada N2L 2Y5}
\affiliation[2]{University of California, Berkeley, CA, MC 7300}
\begin{document}
\maketitle
\section{Introduction}

The notion of $\Omega$ deformation has been very useful in the study of quantum field theories endowed with extended supersymmetry \cite{Nekrasov:2002qd,Nekrasov:2003rj,Alday:2009aq,Nekrasov:2009rc,Nekrasov:2010ka,Yagi:2014toa}. 
This includes two-dimensional $(2,2)$ SQFTs \cite{Shadchin:2006yz,Fujimori:2015zaa}, 
three-dimensional ${\cal N}=4$ SQFTs  \cite{Yagi:2014toa,Beem:2018fng}, four- \cite{Nekrasov:2002qd,Nekrasov:2003rj} 
and five- \cite{2012arXiv1202.2756S} dimensional ${\cal N}=2$ SQFTs, six-dimensional $(2,0)$ SQFTs \cite{Yagi:2012xa} 
and a variety of composite configurations \cite{Alday:2010vg}. 
\footnote{In this paper we denote as $\Omega$ deformations the analogues of the four-dimensional setup in \cite{Nekrasov:2002qd}
defined as a dimensional reduction of a supersymmetric twisted compactification. We do {\it not} denote as $\Omega$ backgrounds
the twisted circle compactifications themselves. These tend to give $q$-deformed or ``trigonometric'' versions of the 
mathematical structures associated to $\Omega$ deformations. It would be very interesting to identify such a $q$-deformation of the 
structures analyzed in this paper.} Typically, the $\Omega$-deformed theory captures a variety 
of supersymmetry-protected couplings and OPE coefficients of the parent theory, organized in the form of a lower-dimensional 
(relative) QFT.

Two examples are particularly relevant for this paper. An $\Omega$ deformation of three-dimensional ${\cal N}=4$ SQFTs reduces 
them to topological quantum mechanics models, with a non-commutative algebra of local operators which deforms the commutative ring of 
half-BPS ``Higgs branch'' operators \cite{Yagi:2014toa,Beem:2018fng} 
\footnote{A mirror $\Omega$ background gives an analogous algebra of Coulomb branch operators}. 
An $\Omega$ deformation of six-dimensional $(2,0)$ SQFTs reduces them to two-dimensional chiral ${\mathcal W}$
algebras, which also deform a commutative ring of half-BPS operators \cite{Yagi:2012xa}. 

The notion of {\it twisted supergravity}  allows one to define the analogue of $\Omega$ deformation for extended supergravity theories and, 
by extension, perturbative string theories and M-theory \cite{Costello:2016mgj}. A beautiful example is the claim that the $\Omega$-deformation 
\begin{equation}
\bR \times \bC^2 \times \bC_{\epsilon_1} \times \bC_{\epsilon_2}\times \bC_{\epsilon_3\equiv - \epsilon_1-\epsilon_2}
\end{equation}
of flat space reduces M-theory to a five-dimensional (somewhat non-local) QFT which is topological in one direction and holomorphic in the remaining four. At least for some range of parameters the 5d theory admits a perturbative 
description as a non-commutative $U(1)$ Chern-Simons theory \cite{Costello:2016nkh,Costello:2017fbo}. Roughly speaking, the 5d gauge theory action should encode a protected part of the 
low energy effective Lagrangian for M-theory, with the $U(1)$ gauge field descending from the three-form field of supergravity. 


The low energy effective world-volume theories of M2 and M5 branes must admit a supersymmetric coupling to 11d supergravity.
The bulk $\Omega$ deformation induces a corresponding $\Omega$ deformation of the 3d and 6d world-volume theories. 
As a consequence, the respective topological quantum mechanics or chiral algebras should 
admit a consistent coupling to the non-commutative 5d Chern-Simons theory \cite{Costello:2016nkh,Costello:2017fbo}. 

\begin{table}[H]\centering
\begin{tabular}{|c|c|cccc|cc|cc|cc|} 
\hline
   &  0 &  1 &  2 &  3 &  4 &  5 &  6 &  7 &  8 &  9 & 10 \\
\hline\hline
 Geometry& $\bR_t$& $\bC^2_{NC}$ &   & & & $\bR^2_{\E_1}$ &  & $\bR^2_{\E_2}$ & &$\bR^2_{\E_3}$ &  \\
\hline
5d~CS&$\times$&  $\times$ & $\times$   & $\times$ & $\times$ & & &  &  & &\\
\hline
$M5_1$&   & $\times$ & $\times$ & & & &  & $\times$ & $\times$ &$\times$ &$\times$\\
\hline
$M5_2$&   & $\times$ & $\times$ & & & $\times$ & $\times$ & & &$\times$ &$\times$\\
\hline
$M5_3$&   & $\times$ & $\times$ & & & $\times$ & $\times$ &$\times$ &$\times$ & & \\
\hline
$M2_1$& $\times$  & & & & & $\times$ & $\times$ & & & &\\
\hline
$M2_2$& $\times$  & & & & & & & $\times$ & $\times$ & &\\
\hline
$M2_3$& $\times$  & & & & & & & & & $\times$ &$\times$  \\
\hline
\end{tabular}
\caption{The main actors in this paper. }
\label{table:M5Directions}
\end{table}

This note has three objectives
\begin{enumerate}
\item Review and elaborate some of the ideas in \cite{Costello:2016nkh,Costello:2017fbo}, drawing relations to other constructions such as that of ``corner vertex algebras'' \cite{Gaiotto:2017euk}. 
\item Discuss the geometric triality symmetry permuting the three $\Omega$-deformed planes. 
It implies non-perturbative dualities of the $U(1)$ 5d Chern-Simons theory.
\item Discuss the most general topological line defects built from M2 branes wrapping $\bR \times \bC_{\epsilon_i}$.
\item Study the intersections of these M2 branes and M5 branes wrapping $\bC \times \bC_{\epsilon_i}\times \bC_{\epsilon_j}$.
\end{enumerate} 

Our main tool is an algebra ${\mathcal A}_{\epsilon_1, \epsilon_2}$ built in \cite{Costello:2017fbo} as the Koszul dual of the algebra of observables 
of the 5d gauge theory which are localized near the origin of the holomorphic directions. 
It is a nonlinear quantum generalization of the algebra of holomorphic gauge symmetries. 
In particular, ${\mathcal A}_{\epsilon_1, \epsilon_2}$ must act on the world-line degrees of freedom of 
topological line defect in the 5d gauge theory. 

Our first result is to prove the triality symmetry of ${\mathcal A}_{\epsilon_1, \epsilon_2}$, by an explicit identification 
to a form of the $\mathfrak{gl}(1)$ affine Yangian \cite{2014arXiv1404.5240T}. Next, we 
identify algebraically the line defects which arise from M2 branes, generalizing the prescription of \cite{Costello:2017fbo}
for M2 branes wrapping $\bR \times \bC_{\epsilon_1}$. We accomplish these objective 
with the help of an alternative ``Coulomb branch'' presentation of ${\mathcal A}_{\epsilon_1, \epsilon_2}$ based on 3d mirror symmetry
\cite{Bullimore:2015lsa,Braverman:2016wma,Kodera:2016faj}. 

Our second result is an algebraic description of the intersection between M2 and M5 branes. 
In particular, we describe modules ${\mathcal M}^{N_1,N_2,N_3}_{\epsilon_1, \epsilon_2}$
which govern the endpoints of topological line defects onto surface defects built from M5 branes with various geometric orientation. 
We provide an initial comparison between the properties of the ${\mathcal M}^{N_1,N_2,N_3}_{\epsilon_1, \epsilon_2}$ modules and the  
``degenerate modules'' of the corner vertex algebras $Y^{N_1,N_2,N_3}_{\epsilon_1, \epsilon_2}$ associated to the same 
geometric configuration of M2 and M5 branes. 

We also propose a family of bi-modules ${\mathcal B}^{0;N_1,N_2,N_3}_{\epsilon_1, \epsilon_2}$ which govern the intersection of M2 line defects and surface defects built from M5 branes.

Finally, we speculate that some generalizations of the Coulomb branch presentations may provide 
a description of topological line defects in other backgrounds for the non-commutative $U(1)$ gauge theories,
such as $\bR \times \frac{\bC^2}{\mathbb{Z}_m}$ and $\bR \times \bC \times \bC^*$.

Our work leaves many interesting questions unanswered. We can give a non-exhaustive list here:
\begin{itemize}
\item Extract the full physical meaning of the algebraic structures of the $\Omega$-deformed M-theory.
Which part of the 11d M-theory effective action do they capture?
\item Define and study the ``GL twisted IIB string theory'' which appears in our analysis below.
\item Establish a full description of the world-line theory of generic M2 brane line defects.
\item Establish a full description of the modules and bimodules we propose. 
\item Prove a match between the module and bi-module structures and perturbation theory in the 5d gauge theory coupled to a surface defect. 
\item Identify the most general bimodules ${\mathcal B}^{N_1,N_2,N_3}_{\epsilon_1, \epsilon_2}$ associated to intersections points at which the line defect changes 
in an arbitrary manner. 
\item Prove in 5d perturbation theory that surface defects are controlled by the ${\mathcal W}^{\epsilon_1,\epsilon_2}_{1+\infty}$ algebra
which encompasses the $Y^{N_1,N_2,N_3}_{\epsilon_1, \epsilon_2}$ corner vertex algebras.  
\item Combine the algebraic structure of ${\mathcal A}_{\epsilon_1, \epsilon_2}$ and ${\mathcal W}^{\epsilon_1,\epsilon_2}_{1+\infty}$
to include both the OPE in the topological and holomorphic directions.
\item Include in the analysis M5 branes extending along both holomorphic directions in $\bR \times \bC^2$.
\item Formulate a full twisted holographic correspondence for M2 and M5 branes, analogous to the one in \cite{Costello:2018zrm}.
\item Generalize the algebraic description to other internal toric Calabi-Yau geometries.
\item Our tentative construction of line defect gauge algebras for $\bR \times \frac{\bC^2}{\mathbb{Z}_m}$ and $\bR \times \bC \times \bC^*$
from families of 3d Coulomb branch algebras should be tested in depth. The construction can also be extended to 4d (aka K-theoretic) Coulomb branch
algebras, perhaps leading to a $\bR \times \bC^* \times \bC^*$ background. Other generalizations may be possible. 
\item Develop a theory of ``5d gluing'' to study $\Omega$-deformed M-theory on general toric Calabi-Yau manifolds.
\end{itemize}

\subsection{Structure of the paper}
Section \ref{sec:omega} reviews the gauge theory description of $\Omega$-deformed M-theory and discusses in detail a duality to twisted IIB string theory 
which is instrumental in our analysis. Section 
\ref{sec:algebra} discusses the universal algebra ${\mathcal A}_{\epsilon_1, \epsilon_2}$ which governs the coupling of the $\Omega$-deformed 
M2 brane world-volume theory to the bulk theory. In particular, we demonstrate the triality properties of the algebra.  In Section \ref{sec:module} 
we propose modules for ${\mathcal A}_{\epsilon_1, \epsilon_2}$ which govern the properties of M2 branes ending on M5 branes. 
In Section \ref{sec:bimodule} we propose bi-modules for ${\mathcal A}_{\epsilon_1, \epsilon_2}$ which govern the properties of M2 branes crossing M5 branes. 
We formulate a natural conjecture for the universal bi-module which describes all such intersections. 

\section{5d gauge theory from $\Omega$ deformation} \label{sec:omega}
The ``twisted M-theory'' setup in \cite{Costello:2016nkh} can be specialized to a geometry of the form $\bR \times \bC^2 \times X_3$, where $X_3$ is a 
toric Calabi-Yau three-fold and we turn on a generic $\Omega$ deformation on $X_3$ preserving the holomorphic three-form. The result is a 5d QFT 
which is topological in one direction and holomorphic in the remaining four.
\footnote{The term QFT is used here in a somewhat loose sense, as the theory is somewhat non-local in the holomorphic directions.}
The 5d QFT depends on the choice of $X_3$ and on the (ratio of) the two $\Omega$-deformation parameters. 

When a theory has a Lagrangian description, one can analyze an $\Omega$ deformation directly. The answer is typically a path integral over 
some space of BPS configurations, with the $\Omega$ deformation parameters suppressing BPS configurations which are not fixed by the corresponding isometries. 
In principle, it should be possible to analyze 11d supergravity in that manner. An important caveat is that the BPS configurations fixed by 
the isometries are likely to be very singular, so that further, possibly unknown UV information about M-theory may be needed in the analysis. 
Furthermore, this type of analysis will be poorly suited to describe the interactions with M5 branes, whose world-volume theories are strongly 
coupled SCFTs with no known UV Lagrangian description. 

There is an alternative approach to study $\Omega$ deformed theories: compactification along the tori generated by the isometries involved in the definition of the 
$\Omega$ background \cite{Nekrasov:2010ka}. Away from the fixed loci of the isometries, the effect of the $\Omega$ deformation on the compactified theory is a standard topological twist, with a choice of supercharge 
which depends on the $\Omega$ deformation parameters. The fixed points of isometries give boundary conditions or lower-dimensional defects in the compactified theory. These 
defects are actually modified by the $\Omega$ deformation, in such a way that they are compatible with the bulk topological twist. The topological twist eliminates all degrees of freedom 
away from the image of the locus fixed by all isometries. This alternative approach is particularly useful if the compactified theory enjoys some useful dualities.  

\subsection{$\Omega$-deformation and compactification: the M5 brane case}
The classical example of this procedure are 6d $(2,0)$ SCFTs on an $\bC \times \bC_{\epsilon_1}\times \bC_{\epsilon_2}$ background \cite{Nekrasov:2010ka,Yagi:2012xa,Gaiotto:2017euk}.
These SCFTs do not admit a Lagrangian description, so a direct analysis of the $\Omega$-deformed theory is out of question. Before the $\Omega$ deformation is turned on, the 
twisted 6d theory is topological on $\bC_{\epsilon_1}\times \bC_{\epsilon_2}$ and holomorphic along the remaining $\bC$ factor. The BPS operators of the 6d SCFT descend to a tower of local operators in the twisted theory, with non-singular OPE. The $\Omega$ deformation should constrain these local operators to lie at the origin of $\bC_{\epsilon_1}\times \bC_{\epsilon_2}$
and form some holomorphic chiral algebra ${\mathcal W}^{\mathfrak{g}}_{\epsilon_1, \epsilon_2}$ on the $\bC$ factor of the geometry. The OPE of that chiral algebra 
can become non-trivial after the deformation. The main problem is to determine these OPE. 

Compactification along the two-torus generated by rotations around the origin in the $\bC_{\epsilon_1}\times \bC_{\epsilon_2}$ planes reduces the 6d $(2,0)$ SCFTs
to 4d ${\cal N}=4$ SYM. The 6d $\Omega$ deformation descends to a GL topological twist \cite{Kapustin:2006pk} of the 4d theory, with ``topological gauge coupling'' $\Psi = -\frac{\epsilon_2}{\epsilon_1}$.
The GL-twisted theory inherits the S-duality group of the physical theory, acting as fractional linear transformations of $\Psi$. 

The $\bC \times \bC_{\epsilon_1}\times \bC_{\epsilon_2}$ geometry maps to a ``corner'' $\bC \times \bR^+ \times \bR^+$: the two boundaries of the corner 
are the image of the fixed loci of one of the two isometries, the tip of the corner is the fixed locus of both isometries. 
At the boundaries of the corner one finds a topologically twisted {\it and} deformed version of certain half-BPS boundary conditions for the 4d gauge theory (``Neumann'' and ``regular Nahm'' respectively).
Neither the bulk nor the boundaries support any local operators. At the tip of the corner, instead, one finds a chiral algebra of  local operators which should realize 
${\mathcal W}^{\mathfrak{g}}_{\epsilon_1, \epsilon_2}$.

Crucially, the 4d gauge theory description is perturbative when either $\Psi \to \infty$ or $\Psi \to 0$. This allows two independent direct calculations of the
${\mathcal W}^{\mathfrak{g}}_{\epsilon_1, \epsilon_2}$ OPE from the Drinfeld-Sokolov reductions of an $\hat{\mathfrak{g}}$ Kac-Moody algebra at (critically shifted) levels $\Psi$ or $\Psi^{-1}$. 
Reassuringly, the two answers agree thanks to the Feigin-Frenkel duality of W-algebras \cite{feigin1991duality}. Furthermore, the 4d gauge theory is also perturbative when $\Psi \to 1$. This allows a third 
independent determination of ${\mathcal W}^{\mathfrak{g}}_{\epsilon_1, \epsilon_2}$, which gives the well-known coset construction of W-algebras, recently proven in \cite{Arakawa:2018iyk}.

\subsection{$\Omega$-deformation and compactification to IIB string theory}

We can apply the same strategy to the M-theory background. In the absence of $\Omega$ deformation, there is a well-known duality between M-theory compactified on a two-torus and 
IIB string theory compactified on a circle \cite{Schwarz:1995dk}. If one applies the duality to the two-torus generated by the toric isometries of $X_3$, the final result will be 
a web of IIB $(p,q)$-fivebranes with the same geometry as the toric diagram of $X_3$ \cite{Kol:1997fv,Aharony:1997ju}. 

The M-theory $\Omega$ deformation will 
descend to a specific topological twist (in the sense of \cite{Costello:2016mgj}) of IIB string theory. 
We will denote this twist as the ``GL twist'' of IIB string theory. We expect the bulk twisted theory to depend on a ``topological string coupling'' 
\begin{equation}
\Psi = -\frac{\epsilon_2}{\epsilon_1}
\end{equation}
and to inherit the S-duality group of the physical string theory, acting as fractional linear transformations of $\Psi$. 

This expectation is motivated by the observation that the world-volume theory on
M5 branes wrapping complex surfaces in $X_3$ fixed by the toric isometries is precisely the $\Omega$ deformation of a 6d $(2,0)$ SCFT. These M5 branes map to D3 branes 
wrapping faces of the $(p,q)$-fivebrane web, with world-volume theory given by the GL twist of 4d ${\cal N}=$ SYM with coupling $\Psi$.
From the point of view of the D3 branes world-volume theory, the $(p,q)$-five-brane web engineers a web of half-BPS interfaces which are topologically twisted and deformed 
to be compatible with the GL twist. The junctions of interfaces support interesting ``corner'' chiral algebras \cite{Gaiotto:2017euk,Creutzig:2017uxh,Prochazka:2017qum}.

In conclusion, we expect the $\Omega$-deformed M-theory background to be equivalent to a $(p,q)$-fivebrane web in the GL-twisted IIB string theory, 
where the fivebrane world-volume theories are both topologically twisted and deformed and the 5d holomorphic-topological theory we are after lives on the junctions of the web. 
In favourable circumstances, we may be able to find a description of the 5d theory using IIB perturbative string theory. 

There is an interesting lesson to be drawn from the analogy to the interface webs in GL-twisted 4d SYM. Only a few selected webs admit a direct perturbative description, and only in
specific S-duality frames for the 4d theory. More general webs can be studied by a careful gluing procedure which merges individual junctions which do not admit a simultaneous perturbative description.  
Similarly, only a few selected $X_3$'s will allow a direct perturbative determination of the corresponding 5d theory. It may be possible to develop a gluing procedure for 
5d theories to analyze more general webs, but that is a problem which goes beyond the scope of this note. 

\subsection{Perturbative junctions}
The main focus of this note is the case of flat space
\begin{equation}
X_3 = \bC_{\epsilon_1} \times \bC_{\epsilon_2} \times \bC_{\epsilon_3 \equiv -\epsilon_1-\epsilon_2} 
\end{equation}
which admits three distinct perturbative descriptions in IIB string theory, depending on which of the $\epsilon_i$ is parametrically smaller than the others. 
Notice the Calabi-Yau condition
\begin{equation}
\epsilon_1 + \epsilon_2 + \epsilon_3 = 0.
\end{equation}
The corresponding descriptions of the 5d theory will involve a non-commutative 5d $U(1)$ Chern-Simons theory with CS coupling controlled by the 
small $\epsilon_i$ parameter and non-commutativity controlled by one of the other two. 
We will denote the equivalence between such alternative descriptions as the ``triality''of the theory. 

There are other toric Calabi-Yau geometries which have perturbative IIB descriptions. We will briefly discuss them in footnotes $4$ and $5$.

\subsection{A Type IIA perspective}
The 5d $U(1)$ CS description was derived in \cite{Costello:2016nkh} with the help of a IIA duality frame, obtained by reducing on the 
$\epsilon_2$ isometry circle. The isometry fixed locus is a D6 brane, whose world-volume theory was determined to be a non-commutative 
deformation of 7d $U(1)$ SYM theory with non-commutativity parameter $\epsilon_2$. The 7d gauge theory on $\bC_{\epsilon_1}$ 
was then reduced further to the 5d $U(1)$ CS theory, with an action proportional to $\epsilon_1^{-1}$ and non-commutativity parameter $\epsilon_2$.
\footnote{The IIA analysis was also generalized to the case of $\bC_{\epsilon_1} \times \frac{\bC_{\epsilon_2} \times \bC_{\epsilon_3 \equiv -\epsilon_1-\epsilon_2}}{\mathbb{Z}_K}$. The first compactification step leads to a 7d $U(K)$ CS theory, which then is mapped to a 5d $U(K)$ CS theory.} 

The gauge theory involves a holomorphic-topological connection $A = A_t dt + A_{\bar z^1} d\bar z^1+ A_{\bar z^2} d\bar z^2$ and an action 
\begin{equation}
\frac{1}{\epsilon_1} \int dz_1 dz_2 \left[A *_{\epsilon_2} dA + \frac{2}{3} A *_{\epsilon_2} A *_{\epsilon_2} A \right]
\end{equation}
where $*_{\epsilon_2}$ is the ``holomorphic'' Moyal product associated to a non-commutativity $[z_1,z_2] = \epsilon_2$ of the holomorphic coordinates. 
As long as $\epsilon_2$ is non-zero, we can always rescale the holomorphic coordinates to set the non-commutativity parameter to $1$
and then the coupling becomes $\hbar^{-1} \equiv \frac{\epsilon_2}{\epsilon_1}$. Although everything could be expressed in terms of $\hbar$ only, it is convenient to keep both $\epsilon_i$ parameters. An important result from \cite{Costello:2016nkh} was that this theory can be systematically renormalized. 

We can quickly describe the IIA image of the M2's and M5's relevant to our work:
\begin{itemize}
\item The $M2$ branes wrapping $\bR \times \bR^2_{\epsilon_1}$ become D2 branes parallel to the D6 brane. The world-volume theory is the $\Omega$ deformation of a ${\cal N}=4$ 3d gauge theory based on the ADHM quiver with one flavour. The D6 brane (world-volume) gauge theory couples directly to the 
fields which arise from $2-6$ and $6-2$ open strings. This is the coupling described in \cite{Costello:2017fbo} and reviewed in Section \ref{sec:adhm}.
\item The $M2$ branes wrapping $\bR \times \bR^2_{\epsilon_{2,3}}$ become fundamental strings ending on the D6 brane 
from opposite directions. They couple to the D6 brane gauge theory as Wilson lines of positive or negative charge. We will describe 
their coupling to the 5d CS theory in Section \ref{sec:wilson}.
\item The M5 branes wrapping $\bC \times \bR^2_{\epsilon_2}\times \bR^2_{\epsilon_3}$ become D4 branes wrapping the three directions transverse to the 
D6 brane. The total number of transverse directions for the D4-D6 intersection is $8$ and thus the $4-6$ and $6-4$ strings give complex chiral fermions,
coupled minimally to the D6 gauge fields. The resulting surface defect in the 5d CS theory was discussed classically in \cite{Costello:2016nkh}.
We will discuss it further in Section \ref{sec:corner}.
\item The M5 branes wrapping $\bC \times \bR^2_{\epsilon_1}\times \bR^2_{\epsilon_{2,3}}$ become D4 branes wrapping one direction transverse to the 
D6 brane, ending on the D6 brane from either direction. They appear as 't Hooft surfaces in the D6 brane gauge theory. 
The resulting surface defect in the 5d CS theory was discussed classically in \cite{Costello:2016nkh}.
We will discuss it further in Section \ref{sec:corner}.
\end{itemize}
\subsection{A Type IIB perspective}

From the IIB perspective, the flat space setup descends to a trivalent junction based on the toric diagram of $\bC^3$. 
It involves a $(0,1)$ (aka D5) brane, a $(1,0)$ (aka NS5) brane and a $(1,1)$ brane. 

\begin{figure}[H]
\centering
\includegraphics[width=10cm]{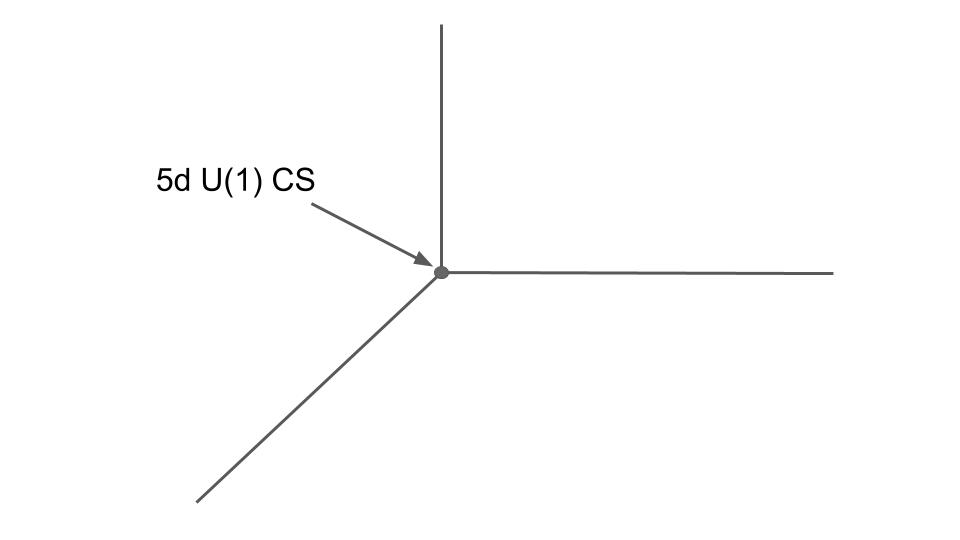}
  \vspace{-20pt}
\caption{The toric diagram of $\bC^3$ determines the fivebrane web in IIB. The 5d $U(1)$ CS theory lives at the junction of the basic 5-brane web.
Conventionally, the vertical fivebrane is of type $(1,0)$, i.e. NS5, the horizontal is of type $(0,1)$, i.e. D5, and the diagonal of type $(1,1)$. }
\label{fig00}
 \centering
\end{figure}

The setup has a triality symmetry which permutes the $\epsilon_i$ parameters. If
$\epsilon_1 \ll \epsilon_2, \epsilon_3$ the twisted IIB string theory is weakly coupled in the standard S-duality frame. If $\epsilon_2 \ll \epsilon_1, \epsilon_3$
or $\epsilon_3 \ll \epsilon_1, \epsilon_2$ then twisted IIB string theory is weakly-coupled in two alternative S-duality frames. These limits are completely analogous to the 
$\Psi \to \infty$, $\Psi \to 0$ and $\Psi \to 1$ limits we discussed for the M5 brane theory.

In the IIB setup, the 5d CS theory should emerge at the junction of the three fivebranes. In the physical theory, each fivebrane supports a supersymmetric 6d $U(1)$ gauge theory.
The physical junction imposes a supersymmetric boundary condition $\sum_i A_i =0$ which reduces the gauge group from $U(1)^3$ to $U(1)^2$. The reduction is related by supersymmetry to the geometric constraint that the three fivebranes should actually meet at the junction. We can see it in figure \ref{fig00}: the figure has two translational degrees of freedom int he plane, so the three scalar fields describing a transverse motion of the fivebranes must satisfy a 
single constraint at the junction. 

We can begin to understand the appearance of the 5d CS theory at such junction by looking at two simpler supersymmetric configurations in a 6d gauge theory: a Neumann boundary condition and an identity interface.
Upon bulk topological twist and appropriate deformation, a deformed Neumann boundary condition should already be able to support a 5d CS theory, analogously to the appearance of 4d CS theory at deformed Neumann b.c. for 5d gauge theory \cite{Costello:2018txb} or of (analytically continued) 3d CS theory at deformed Neumann b.c. for 4d gauge theory \cite{Witten:2010cx,Witten:2011zz}. On the other hand, an identity interface should support a trivial 5d theory, as it is indistinguishable from the bulk. 

The junction reducing $U(1)^3$ to $U(1)^2$ is somewhat analogous to the combination of an identity interface between two of the gauge theories and a Neumann boundary condition for the third. 
This statement becomes quite precise at weak coupling: the NS5 brane is much heavier than the D5 brane and the junction can be really interpreted as a D5 brane ending on the NS5 brane
with a Neumann boundary condition for the world-volume theory. In other duality frames, though, the role of the ``light'' fivebrane can be taken by either of the three. 

We thus expect that in each weakly-coupled frame the 5d CS theory gauge field can be thought of as the boundary value of the D5 brane world-volume gauge field, compatibly with the IIA description. 
The 5d CS theory is rigid \cite{Costello:2017fbo} and so this approximate description should be trustworthy, at least perturbatively in $\epsilon_1/\epsilon_2$.  \footnote{Analogously to lower-dimensional examples, it is natural to expect that a setup with $K$ D5 branes ending on one side of the NS5 brane and $K'$ on the other side will give a 5d $U(K|K')$ CS theory.}
At finite coupling, the 5d CS gauge field should be some appropriate linear combination of the boundary values of the three gauge fields, smoothly interpolating between the three triality frames. 

In Section \ref{sec:algebra} we will find a striking verification of this naive expectation: in an appropriate renormalization scheme the 5d CS gauge fields in different triality 
frames are simply proportional to each other, with proportionality coefficients given by certain ratios of $\epsilon_i$ parameters!
We leave to future work a direct triality invariant analysis of the IIB trivalent junction. 

\subsection{Gauge theory vs. supergravity}
At first blush, it may appear surprising that a theory of supergravity in 11 dimensions may be reduced to a gauge theory 
by the $\Omega$ background. This becomes a bit less surprising if we observe that the gauge transformations of a non-commutative $U(1)$ 
gauge theory are somewhat geometric in nature: at the leading order in the non-commutativity parameter $\epsilon_2$, 
a gauge transformation with parameter $\lambda(z_1,z_2)$ acts on functions of $\bC^2$ as a complex symplectomorphism
with Hamiltonian $\epsilon_2 \lambda$. 

The triality analysis in later sections suggests that it would be more natural to work with rescaled gauge parameters $\Lambda = \epsilon_2 \lambda$ 
and re-scaled connection $\Phi = \epsilon_2 A$, which should be triality-invariant in an appropriate renormalization scheme.  
In a limit $\epsilon_1\ll \epsilon_2 \ll 1$, we could write the classical action in a triality-invariant manner
\begin{equation}
\frac{1}{\epsilon_1 \epsilon_2 \epsilon_3} \int dz_1 dz_2 \left[\Phi d \Phi + \frac{2}{3} \Phi \{ \Phi, \Phi \}\right] + \cdots
\end{equation}
with a pre-factor $\frac{1}{\epsilon_1 \epsilon_2 \epsilon_3}$ which is would arise naturally from the $\Omega$ deformation of an 11d 
action. The $\{\cdot,\cdot\}$ denotes the holomorphic Poisson bracket. It would be interesting to elaborate this perspective further. We will not do so here.  

%
%

\subsection{M5 branes and surface defects} \label{sec:corner}
The M-theory $\Omega$-background is compatible with the presence of M5 branes which wrap toric complex surfaces in $X_3$ 
and an holomorphic plane $\bC$ in $\bR \times \bC^2$. The world-volume theory of the M5 branes is $\Omega$-deformed to some 
2d chiral algebra living on $\bC$. The coupling of the bulk M-theory fields to the M5 brane worldvolume theory should descend to a coupling of the 
5d CS gauge theory to the 2d chiral algebra to produce a surface defect. 

When $X_3$ is $\bC_{\epsilon_1} \times \bC_{\epsilon_2} \times \bC_{\epsilon_3 \equiv -\epsilon_1-\epsilon_2}$, we can have up to 
three stacks of M5 branes wrapping coordinate surfaces and intersecting along the same $\bC$. This will lead to some surface defect 
$S_{M_1,M_2,M_3}$ labelled by the numbers of M5 branes wrapping respectively $\bC_{\epsilon_2} \times \bC_{\epsilon_3}$,
$\bC_{\epsilon_1} \times \bC_{\epsilon_3 }$, $\bC_{\epsilon_1} \times \bC_{\epsilon_2}$.

In the IIB duality frame, the M5 branes descend to stacks of D3 branes wrapping the three faces of the toric diagram. More precisely, 
we get $M_3$ D3's in the sector between the $(1,0)$ and the $(0,1)$ branes, $M_2$ between the $(0,1)$ and the $(1,1)$ branes, 
$M_1$ between the $(1,0)$ and $(1,1)$ branes. The chiral algebra associated to the corresponding 4d gauge theory configuration 
was identified in \cite{Gaiotto:2017euk}. We can denote it as $Y^{M_1,M_2,M_3}_{\epsilon_1,\epsilon_2}$.

\begin{figure}[H]
\centering
\includegraphics[width=10cm]{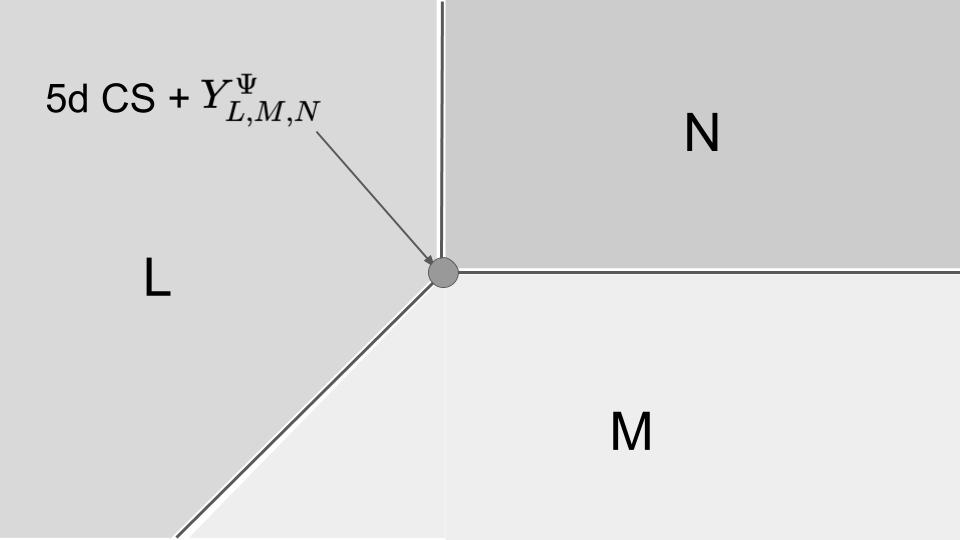}
  \vspace{-5pt}
\caption{L, M, N stacks of D3 branes on three faces of the web realize $Y^{L,M,N}_{\epsilon_1,\epsilon_2}$ chiral algebra at the corner, which couple to 5d $U(1)$ CS theory through its embedding into the quantum ${\mathcal W}^{\epsilon_1,\epsilon_2}_{1+\infty}$ algebra.}
\label{fig01}
 \centering
\end{figure}

The main result in \cite{Gaiotto:2017euk} was to identify and compare three independent definitions of $Y^{M_1,M_2,M_3}_{\epsilon_1,\epsilon_2}$, each derived from 
a distinct weakly-coupled duality frame for $\Psi =- \frac{\epsilon_2}{\epsilon_1}$. The triality simultaneously permutes $M_1,M_2,M_3$ and 
$\epsilon_1, \epsilon_2, - \epsilon_1-\epsilon_2$. It generalizes the Feigin-Frenkel duality of W-algebras and their relation to cosets.

The perturbative description of $Y^{M_1,M_2,M_3}_{\epsilon_1,\epsilon_2}$, say in the $\epsilon_1 \to 0$ frame, depends sensitively on $M_3$ and $M_2$ being equal or different. 

For $M_3=M_2$, the chiral algebra is a $\mathfrak{u}(M_3|M_1)$-BRST
reduction of a combination 
\begin{equation}
\mathrm{Sb}^{M_3|M_1} \times \widehat{\mathfrak{u}}(M_3|M_1)_{\Psi}\times \widehat{\mathfrak{u}}(M_3|M_1)_{-\Psi+1}
\end{equation}
with $\mathrm{Sb}^{M_3|M_1}$ denoting a combination of $M_3$ symplectic bosons (i.e. $\beta \gamma$ systems of dimension $\frac12$) and $M_1$ complex fermions and 
$\widehat{\mathfrak{u}}(M_3|M_1)_{\kappa}$ the appropriate Kac-Moody algebra at critically-shifted level $\kappa$. 

When $\epsilon_1 \to 0$, the $Y^{M_1,M_3,M_3}_{\epsilon_1,\epsilon_2}$ VOA is perturbatively close to the $\mathfrak{u}(M_3|M_1)$-invariant part of the 
$\mathrm{Sb}^{M_3|M_1}$ VOA of $M_3$ free symplectic bosons and $M_1$ fermions. 

This vertex algebra also has a simple IIA interpretation: The $M_1$ D4 branes fully transverse to the D6 brane give rise to the 
$M_1$ complex fermions, while the intersection of the D6 brane with the remaining $M_3$ D4 branes supports $M_3$ 4d hypermultiplets, 
which are $\Omega$-deformed to $M_3$ symplectic bosons \cite{Beem:2013sza,Oh:2019bgz}. The match would be complete if 
one can identify the $\Omega$ deformation of the D4 brane degrees of freedom with an (analytically continued) $U(M_3|M_1)$ 
3d Chern-Simons theory of level $\Psi$, with the odd gauge generators arising from $4-4'$ strings in a manner analogous to 
the setup of \cite{2015CMaPh.340..699M}. Coupling the 2d free fields to the 3d CS theory would automatically implement the above coset
\cite{Costello:2016nkh,Gaiotto:2017euk}. 

This is the natural starting point to couple the VOA perturbatively to the 5d non-commutative CS gauge theory:
we simply couple the 5d gauge fields to 2d free fields placed, say, at $t=0$, $z_2 = \bar z_2 =0$. 
All we need in order to define the classical coupling is an action of the non-commutative $U(1)$ gauge group on the 2d fields
and a corresponding covariant derivative. 

There are natural left- and right- actions of the Moyal algebra of functions 
on $\bC^2$ on the space of functions of a single complex variable $z_1$ (or better, spin $\frac12$ fields). 
We will review them in detail Section \ref{sec:module}. 
The corresponding gauge currents are the classical ${\mathcal W}_{1+\infty}$ bilinear currents of the form $\beta \partial_{z_2}^n \gamma$, where $\beta$ and $\gamma$ here denote wither the two components of a symplectic boson or of a complex fermion \cite{Costello:2016nkh}. 

These are coupled to the normal components 
$\partial_{z_2}^n A_{\bar z_1}$ of the 5d CS connection into an action of the schematic form
\begin{equation}
\int \beta(z_1) ( \partial_{\bar z_1}- A_{\bar z_1} *_{\epsilon_2})  \gamma(z_1) 
\end{equation}

Perturbatively, loop corrections can introduce anomalies which would make such a coupling inconsistent, unless the OPE algebra of the bilinear currents $\beta \partial_{z_2}^n \gamma$ is modified appropriately \cite{Costello:2016nkh}. Consistency of the M-theory setup predicts that these modifications must precisely match the OPE of currents in $Y^{M_1,M_3,M_3}_{\epsilon_1,\epsilon_2}$ which deform the $\beta \partial_{z_2}^n \gamma$
away from $\epsilon_1 \to 0$. We will come back to this point momentarily. 

When $M_3>M_2$, the $Y^{M_1,M_2,M_3}_{\epsilon_1,\epsilon_2}$ algebra is defined as the $\mathfrak{u}(M_2|M_1)$-BRST
reduction of a combination 
\begin{equation}
\mathrm{DS}_{M_3-M_2} \widehat{\mathfrak{u}}(M_3|M_1)_{\Psi}\times \widehat{\mathfrak{u}}(M_2|M_1)_{-\Psi-1}
\end{equation}
involving a Drinfeld-Sokolov reduction of $\widehat{\mathfrak{u}}(M_3|M_1)_{\Psi}$ which preserves an 
$\widehat{\mathfrak{u}}(M_2|M_1)$ subalgebra. A similar statement applies for $M_3<M_2$.

Because D3 branes ending on D5 branes carry magnetic charge under the D5 brane world-volume fields, 
the difference $M_3-M_2$ controls the magnetic charge of the corresponding surface defect in the 5d CS theory, 
which should be defined classically as a sort of 't Hooft surface defect dressed by a coupling to the $\epsilon_1 \to 0$ limit of 
$Y^{M_1,M_2,M_3}_{\epsilon_1,\epsilon_2}$. Consistency of the M-theory setup would again require the quantum-corrected 
5d CS theory to admit a consistent coupling to $Y^{M_1,M_2,M_3}_{\epsilon_1,\epsilon_2}$, 
with a systematic cancellation of perturbative gauge anomalies order-by-order in $\epsilon_1$. 

Remarkably, the $Y^{M_1,M_2,M_3}_{\epsilon_1,\epsilon_2}$ have been recognized as truncations of a universal ${\mathcal W}^{\epsilon_1, \epsilon_2}_{1+\infty}$ 
algebra \cite{Prochazka:2017qum} which deforms the classical ${\mathcal W}_{1+\infty}$ algebra. 
In particular, in the $\epsilon_1 \to 0$ limit it is equipped with the correct family of currents 
needed to formulate a classical coupling to the 5d $U(1)$ CS theory. It is natural to expect that as one turns on $\epsilon_1$, the 
corrections to the classical ${\mathcal W}_{1+\infty}$ OPE will precisely cancel the perturbative anomalies of the classical coupling,
i.e. that ${\mathcal W}^{\epsilon_1, \epsilon_2}_{1+\infty}$ is the correct quantum-corrected gauge algebra along surface defects. 
We leave a perturbative proof of this conjecture, analogous to the results in \cite{Costello:2017fbo}, to future work. 

We should mention also that the mode algebra of the ${\mathcal W}_{1+\infty}$ chiral algebra is closely related to the affine Yangian 
for $\mathfrak{gl}(1)$. See e.g. \cite{Prochazka:2015deb}. Later in the paper we will encounter a conjectural deformed gauge algebra 
${\mathcal C}_{0;\epsilon_1, \epsilon_2}$ controlling line defects in $\bR \times \bC \times \bC^*$, which is also related to the affine Yangian.
It is natural to suspect that the the relation between ${\mathcal W}_{1+\infty}$ and ${\mathcal C}_{0;\epsilon_1, \epsilon_2}$ 
should manifest two alternative ways to organize the 5d gauge algebra in terms of OPE in the holomorphic or topological directions. 
We leave a full exploration of this relation to future work.
 
\subsection{M2 branes and line defects}
The M-theory $\Omega$-background is compatible with the presence of M2 branes which wrap complex lines in $X_3$ invariant under the toric action  
and give rise to a topological line in 5d, wrapping $\bR \times (0,0)$ in $\bR \times \bC^2$. 
The world-volume theory of the M2 branes will be $\Omega$-deformed to a
topological quantum mechanics. The coupling of the M2 branes to the bulk M-theory fields will descend to a coupling of 
the topological quantum mechanics to the 5d theory, taking the form of a topological line defect. 

Upon reduction to IIB string theory, the M2 branes are mapped to $(p,q)$ strings which lie on top of the 
$(p,q)$ fivebranes in the toric diagram. In principle, one will produce an independent line defect for every combination of stacks of M2 branes in $X_3$.

When $X_3$ is $\bC_{\epsilon_1} \times \bC_{\epsilon_2} \times \bC_{\epsilon_3 \equiv -\epsilon_1-\epsilon_2}$, we can have three independent stacks 
of M2 branes, each wrapping a coordinate plane. We thus expect to be able to define line defects $L_{N_1,N_2,N_3}$ in the non-commutative 5d
$U(1)$ CS theory by placing $N_i$ M2 branes on $\bC_i$, with an obvious action of triality.  

\begin{figure}[H]
\centering
\includegraphics[width=10cm]{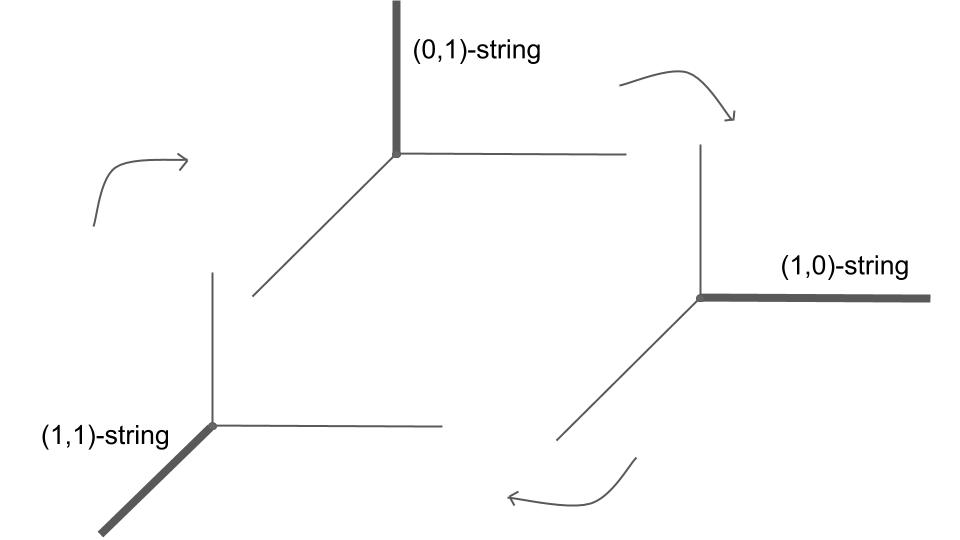}
  \vspace{-5pt}
\caption{One can add $(p,q)$ strings (thick lines) on top of a $(p,q)$ 5-brane web (thin lines) in order to build a line defect in the corner 5d $U(1)$ CS theory. Three types of the line defects in 5d $U(1)$ CS theory are related by triality.
We use here a convention where a $(q,p)$ string can lie super-symmetrically on a $(p,q)$ fivebrane: fundamental strings on NS5 branes, D1 branes on D5 branes, etc.}
\label{fig02}
 \centering
\end{figure}

As we will discuss in detail in the next section, these topological line defects should be envisioned as generalized Wilson lines
\begin{equation}
L_{N_1,N_2,N_3} \equiv \mathrm{Pexp} \int \sum_{n,m} \frac{1}{m! n!} t[m,n] \partial_{z_1}^m \partial_{z_2}^n A_t(t,0,0) dt
\end{equation}
where the $t[m,n]$ describe the action of the generalized gauge algebra ${\mathcal A}_{\epsilon_1, \epsilon_2}$ in the topological quantum mechanics on the 
line defect worldline. \footnote{We apologize to the reader for using ``t'' both as a name of these operators and as the name of the topological time direction.}

The M2 branes with $(0,1)$ orientations (i.e. wrapping $\bC_{\epsilon_1}$) appear as instanton surfaces in the D5 brane world-volume theory. 
In the IIA description, the $(0,1)$ M2 branes lead to a D2-D6 system and then, upon $\Omega$ deformation,
to an ADHM topological quantum mechanics with $N_1$ colours and $1$ flavour, with quantization parameter $\epsilon_1$ 
and quantum FI parameter $\epsilon_2$ \cite{Costello:2017fbo}. 

The D6 gauge field couples to the $2 N_f$ worldline fields $I$,$J$ associated to $2-6$ and $6-2$ strings, 
while the $2 N_f^2$ world-line fields $X$,$Y$ associated to $2-2$ strings control the position of the 
D2 branes in $\bC^2$. Intuitively, we would write a coupling of the form
\begin{equation}
L_{N_1,0,0} \equiv \mathrm{Pexp} \frac{1}{\epsilon_1} \int I(t) A_t(t,X(t),Y(t)) J(t) dt
\end{equation}
where we included an overall factor of $\epsilon_1^{-1}$ arising from the equivariant area of $\bR^2_{\epsilon_1}$. 
This coupling covariantizes the world-line kinetic term $\epsilon_1^{-1} I D_t J$. Matrix contractions are implied. 

This turns out to be precisely the leading term in the non-anomalous coupling between the ADHM topological quantum mechanics and the 
5d CS theory discussed in detail in \cite{Costello:2017fbo}. We will review it momentarily. Notice that the coupling 
does not have a good $\epsilon_1 \to 0$ limit.

The M2 branes with $(1,0)$ orientation are first discussed here. In the standard IIB duality frame they become fundamental strings.
From the perspective of the D5 brane world-volume theory they are boundary Wilson lines of charge $N_2$.
We will propose a definition of charge $N_2$ classical Wilson lines for the non-commutative 5d gauge theory. We will also 
show that they can be deformed to gauge-invariant line defects at finite $\epsilon_1$ which are good candidates for  
$L_{0,N_2,0}$. In particular, they are manifestly related by triality to the $L_{N_2,0,0}$ line defects.

The M2 branes with $(1,1)$ orientation can be analyzed in a similar manner. They map to charge $-N_3$ boundary Wilson lines 
for the D5 brane world-volume theory. We will construct the corresponding charge $-N_3$ classical Wilson lines for the non-commutative 5d gauge theory. 
We will also show that they can be deformed to gauge-invariant line defects at finite $\epsilon_1$ which are good candidates for  
$L_{0,0,N_3}$. In particular, they are manifestly related by triality to the $L_{N_3,0,0}$ and $L_{0,N_3,0}$ line defects. 

Finally, we can consider configurations involving both M5 and M2 branes meeting at the origin of $\bR \times \bC^2$. The M2 branes may either 
cross the M5 branes or end on them. This will require some specific couplings between the line defect and surface defect degrees of freedom, 
constrained again by anomaly cancellations. 

The initial motivation of this project was to study such couplings and compare them with 
constructions in the corner chiral algebras, which associate M2 brane configurations to specific 
``degenerate'' chiral algebra modules for $Y^{M_1,M_3,M_3}_{\epsilon_1,\epsilon_2}$ labelled by triples $(\lambda_{23}, \lambda_{13}, \lambda_{12})$,
where $\lambda_{ij}$ is a Young diagram labelling a representation of $\mathfrak{u}(M_i|M_j)$, i.e. fitting within a ``fat L'' with arms of thickness
$M_i$ and $M_j$. The total number of boxes in the three Young diagrams should coincide with the numbers of M2 branes of each kind.  

\begin{figure}[H]
\centering
\includegraphics[width=10cm]{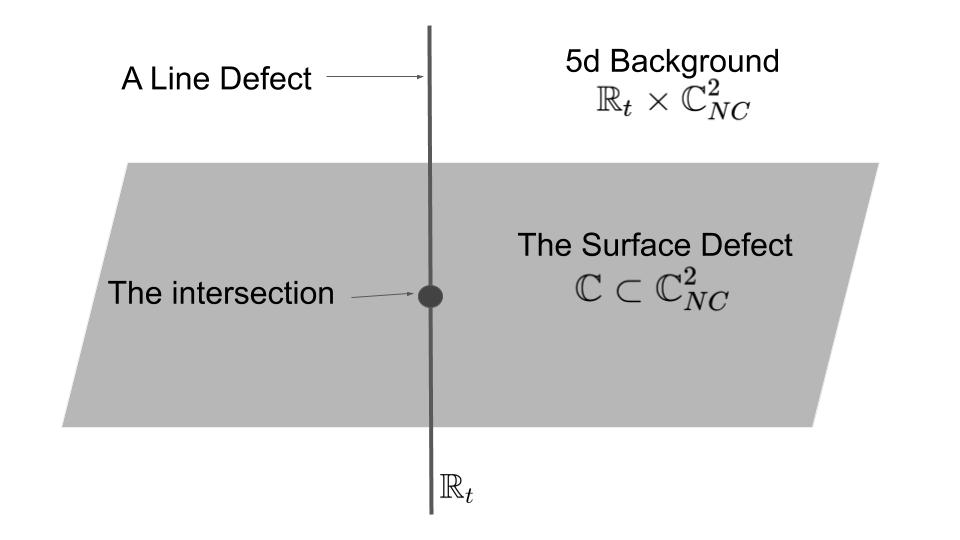}
  \vspace{-5pt}
\caption{M2-M5 systems as a line and a surface defect in 5d $U(1)$ Non-Commutative(NC) CS theory}
\label{fig03}
 \centering
\end{figure}

We will discuss some partial results in that direction. 

\section{Operator algebras and Koszul duality} \label{sec:algebra}

A local holomorphic-topological QFT on $\bR \times \bC^2$ will be equipped with some space of local operators, 
with a BRST differential and a variety of cohomological operations associated to OPEs along the topological or holomorphic directions.
This would be an holomorphic-topological generalization of the $E_d$ algebras which occur in the topological setup. 
See \cite{Beem:2018fng} for a nice review. 

The non-locality of the 5d non-commutative CS theory should modify in important ways both the definition of local operator and the 
appropriate operations available on local operators. We will not attempt to understand the full algebraic structure
required in that situation. 

Instead, we will focus on the $E_1$ algebraic operations associated to OPE the topological direction only of 
operators defined in the neighbourhood of the line $\bR \times (0,0)$ in $\bR \times \bC^2$. These structures are relevant to 
the definition of topological line defects in the 5d theory.  

\subsection{An universal algebra for line defects}

A topological line defect can be defined as a generalized Wilson line 
\begin{equation}
\mathrm{Pexp} \int \sum_{n,m} \frac{1}{m! n!} t[m,n](t) \partial_{z_1}^m \partial_{z_2}^n A_t(t,0,0)
\end{equation}
Here the $t[m,n]$ stand for a collection of local operators the world-line topological quantum mechanics
we use to engineer the defect. 

Classically, the $t[m,n]$ should form a representation of the Lie algebra of non-commutative $U(1)$ gauge transformations:
\begin{equation}\label{eq:classical}
\left[ t[u,v],t[p,q] \right] = \sum_{m,n} f^{m,n}_{u,v;p,q}(\epsilon_2) t[m,n]
\end{equation}
where the $f^{m,n}_{u,v;p,q}$ are the structure constants of the non-commutative $U(1)$ gauge algebra:
\begin{equation}
(z_1^u z_2^v) *_{\epsilon_2} (z_1^p z_2^q) -  (z_1^p z_2^q) *_{\epsilon_2} ( z_1^u z_2^v)  = \sum_{m,n} f^{m,n}_{u,v;p,q}(\epsilon_2) z_1^m z_2^n
\end{equation}

At the quantum level, the definition of the generalized Wilson line will require a careful renormalization. New gauge anomalies may appear 
at each order in perturbation theory. An important observation in \cite{Costello:2017fbo} is that these quantum effects will modify the relations
\ref{eq:classical} so that the $t[m,n]$ generate an algebra ${\mathcal A}_{\epsilon_1, \epsilon_2}$ which deforms the universal enveloping 
algebra of the classical gauge Lie algebra. 

Formally, ${\mathcal A}_{\epsilon_1, \epsilon_2}$ is defined as the {\it Koszul dual} of the $E_1$ algebra $\mathrm{Obs}_{\epsilon_1, \epsilon_2}$ of local observable of the 5d CS theory which are localized in a neighbourhood of the putative line defect. An $E_1$ algebra is a rich mathematical object which encodes the OPE of topological local operators and the descent relations which can be employed to build a topological action for a line defect. 

The topological action will couple the 5d CS theory to a topological world-line theory equipped with some algebra of observable $A$.  
Koszul duality establishes an one-to-one correspondence between renormalized topological actions built from $A \otimes \mathrm{Obs}_{\epsilon_1, \epsilon_2}$
and homomorphisms ${\mathcal A}_{\epsilon_1, \epsilon_2} \to A$ of associative algebras. Different renormalization schemes are related by 
re-definitions of the $t[m,n]$ generators in ${\mathcal A}_{\epsilon_1, \epsilon_2}$.

A detailed description of the renormalization of topological line defects and relation to Koszul duality seems to be missing from the physics  literature. 
We sketch it in Appendices \ref{app:topren} and \ref{app:koszul}. A basic step is to relate topological effective actions to solutions of 
the Maurer-Cartan equation for an auxiliary $A_\infty$ algebra of local operators, which in the current example deforms the classical differential graded algebra of 
polynomials in gauge theory ghost fields and their normal derivatives: 
\begin{equation}
c_{m,n}(t) \equiv \frac{1}{m! n!}\partial_{z_1}^m \partial_{z_2}^n c(t,0,0) 
\end{equation}
i.e. the coefficient of $z_1^m z_2^n$ in the Taylor expansion of $c(t,z_1,z_2)$. 

The $A_\infty$ algebra itself is a cumbersome and renormalization scheme-dependent object. A crucial simplification occurs because 
all the $c_{m,n}$ generators have positive ghost-number. This is enough to guarantee that the Koszul dual ${\mathcal A}_{\epsilon_1, \epsilon_2}$ 
of the auxiliary $A_\infty$ algebra is an associative algebra, defined up to algebra isomorphisms. 

In particular, the conjectural triality invariance of the 5d theory should manifest itself as an isomorphism of six distinct algebras
\begin{equation}
{\mathcal A}_{\epsilon_1, \epsilon_2} \simeq {\mathcal A}_{\epsilon_2, \epsilon_1} \simeq  {\mathcal A}_{\epsilon_1, -\epsilon_1-\epsilon_2} \simeq {\mathcal A}_{\epsilon_2, -\epsilon_1-\epsilon_2} \simeq {\mathcal A}_{ -\epsilon_1-\epsilon_2,\epsilon_1} \simeq {\mathcal A}_{ -\epsilon_1-\epsilon_2,\epsilon_2} 
\end{equation}

In principle one could compute the defining relations of ${\mathcal A}_{\epsilon_1, \epsilon_2}$ order-by-order in perturbation theory. 
Luckily, the deformation theory of the classical gauge algebra is very restrictive, so that some basic 1-loop calculations 
allow one to determine the full quantum algebra for the $U(K)$ 5d CS theory \cite{Costello:2017fbo}.  

Furthermore, the analysis of M2 brane couplings in \cite{Costello:2017fbo} gives a 
canonical definition of ${\mathcal A}_{\epsilon_1, \epsilon_2}$
at finite values for the $\epsilon_i$. This will allow us a very concrete test of triality. 

\subsection{Moyal and Weyl} \label{sec:wilson}
As an useful notational and computational tool, we can identify the algebra of polynomials in $z_1, z_2$ with Moyal product with the Weyl algebra $W_{\epsilon_2}$,
the universal enveloping algebra of the Heisenberg algebra with generators $x,y$ and commutator 
\begin{equation}
[x,y] = \epsilon_2
\end{equation}
The identification maps monomials in $z_1$, $z_2$ to the total symmetrization of the corresponding monomials in $x$, $y$. 
\begin{equation}
z_1^n z_2^m \to \frac{n! m!}{(n+m)!} \sum_{\sigma\in S_{n+m}} \sigma \circ (x^n y^m)
\end{equation}
We will also denote the Lie algebra defined by commutators in $W_{\epsilon_2}$ as $\mathrm{Diff}^{\epsilon_2}_\bC$.
Given any element $w$ in $W_{\epsilon_2}$, we denote the corresponding Lie algebra element as $t[w]$, with relations such as 
\begin{equation}
t[x y]= t[y x] + \epsilon_2 t[1]
\end{equation}
etc. 

With these notations, the gauge Lie algebra is $\mathrm{Diff}^{\epsilon_2}_\bC$ and ${\mathcal A}_{\epsilon_1, \epsilon_2}$ is a deformation of the universal enveloping algebra 
$U(\mathrm{Diff}^{\epsilon_2}_\bC)$. \footnote{Because of the unicity properties of this deformation, we could also refer to ${\mathcal A}_{\epsilon_1, \epsilon_2}$ as to $U_{\epsilon_1}(\mathrm{Diff}^{\epsilon_2}_\bC)$.} 

\subsection{Classical Wilson lines and their deformation}
Non-commutative gauge theory does not admit the most naive type of Wilson lines, such as 
\begin{equation}
\mathrm{Pexp} \left[ \int A_t(t,0,0) dt \right]
\end{equation}
That is because BRST transformations of $A_t(t,0,0)$ give an infinite sum of terms of the form 
$\epsilon_2^{n+m} \partial_{z_1}^m \partial_{z_2}^n c(t,0,0) \partial_{z_1}^m \partial_{z_2}^n A(t,0,0)$.
Equivalently, $t[m,n] = \delta_{m,0} \delta_{n,0}$ is not a representation of $U(\mathrm{Diff}^{\epsilon_2}_\bC)$.

The M-theory setup, though, suggests a straightforward way to fix that. An M2 brane with $(1,0)$ orientation 
descends to a fundamental string in IIB string theory, ending at the junction. Such a string does look like a boundary Wilson line 
of charge $1$ in the D5 brane worldvolume theory, but the string still has positional degrees of freedom which describe the motion 
in directions parallel to the junction. Usually, such degrees of freedom decouple from the gauge dynamics, but not here. 

Instead, the positional degrees of freedom should take the form of the generators $x,y$ of a Weyl algebra $W_{\epsilon_2}$
of local operators on the worldline. Happily, there is a nice algebra morphism $U(\mathrm{Diff}^{\epsilon_2}_\bC) \to W_{\epsilon_2}$,
which simply maps a monomial in $x$, $y$ to the same monomial:
\begin{equation} \label{eq:L1morph}
t[w] \to w
\end{equation} 
and indeed we have a perfectly good classical line defect: a ``dynamical'' Wilson line of charge $1$:
\begin{equation}
L_1 \equiv \mathrm{Pexp} \left[ \int A_t(t,x,y) dt \right]
\end{equation}
where $A_t(t,x,y)$ is a short notation for a Taylor series expansion with symmetrized monomials in $x$ and $y$.

The Weyl algebra is rigid. That means that if the line defect $L_1$ can survive quantum-mechanically, it will have to still be based on 
$W_{\epsilon_2}$ and there should be a morphism ${\mathcal A}_{\epsilon_1, \epsilon_2} \to W_{\epsilon_2}$ which deforms (\ref{eq:L1morph}).

Next, we can look an $N$ M2 branes with $(1,0)$ orientation, corresponding to $N$ fundamental strings in IIB string theory. 
Although each fundamental string should have positional degrees of freedom, we should treat the strings as indistinguishable. 
That suggests using the symmetrized product of $N$ Weyl algebras
\begin{equation}
W^{(N)}_{\epsilon_2} \equiv \frac{W_{\epsilon_2}^{\otimes N}}{S_N}
\end{equation}
as the worldline theory for a dynamical Wilson line of charge $N$. We have a nice morphism $U(\mathrm{Diff}^{\epsilon_2}_\bC) \to W^{(N)}_{\epsilon_2}$
\begin{equation} \label{eq:LNmorph}
t[w] \to \sum_{i=1}^N w_i
\end{equation} 
which maps, say, $t[1] \to N$, $t[x] \to \sum_i x_i$, etc. It gives a Wilson line of the form 
\begin{equation}
L_N \equiv \mathrm{Pexp} \left[ \int \sum_{i=1}^N A_t(t,x_i,y_i) dt \right]
\end{equation}

Using the symmetrized algebra $W^{(N)}_{\epsilon_2}$ as opposed to the most naive choice $W_{\epsilon_2}^{\otimes N}$ has an important
consequence at the quantum level. The algebras $W_{\epsilon_2}^{\otimes N}$ are all rigid. If we were to require the corresponding line defects to survive quantum-mechanically, 
the corresponding tower of ${\mathcal A}_{\epsilon_1, \epsilon_2} \to W_{\epsilon_2}^{\otimes N}$ morphisms would put extremely 
strong constraints on ${\mathcal A}_{\epsilon_1, \epsilon_2}$, probably preventing a non-trivial $\epsilon_1$ dependence. 

On the other hand, the symmetrized algebras $W^{(N)}_{\epsilon_2}$ are not rigid. They have a very nice, unique deformation given by the quantization of the algebra of functions 
on the Hilbert scheme of $N$ points in $\bC^2$. Interestingly, this is the same algebra which emerged in \cite{Costello:2017fbo} as the world-line theory of line defects associated to 
$(0,1)$ M2 branes, but with an important difference: the quantization parameter there was $\epsilon_1$ and the quantum FI parameter was $\epsilon_2$. For $(1,0)$ defects, the opposite 
identification is natural, as the quantum FI parameter governs the deformation away from $W^{(N)}_{\epsilon_2}$. 

We take this as a first hint of triality and propose that $L_N$ should be the classical limit of the line defects associated to $N_2$ M2 branes with $(1,0)$ orientation. 
If we denote the as quantization of the algebra of functions on the Hilbert scheme of $N$ points in $\bC^2$ as ${\mathcal A}^{(N)}_{\epsilon_2, \epsilon_1}$, 
we thus predict the existence of a morphism ${\mathcal A}_{\epsilon_1, \epsilon_2} \to {\mathcal A}^{(N)}_{\epsilon_2, \epsilon_1}$ deforming the classical morphism
$U(\mathrm{Diff}^{\epsilon_2}_\bC) \to W^{(N)}_{\epsilon_2}$. This morphism should be related by triality to the morphism ${\mathcal A}_{\epsilon_1, \epsilon_2} \to {\mathcal A}^{(N)}_{\epsilon_1, \epsilon_2}$
described in \cite{Costello:2017fbo}. We will verify that prediction momentarily. 

It is worth observing that for any algebra $A$ with associated Lie algebra $\mathfrak{a}$ one has a collection of morphisms 
\begin{equation}
U(\mathfrak{a}) \to \frac{A^{\otimes N}}{S_N} : t[a] \to \sum_{i=1}^N a_i
\end{equation}
Furthermore, $U(\mathfrak{a})$ is in an appropriate sense the ``uniform-in-$N$'' limit of the $\frac{A^{\otimes N}}{S_N}$ algebras, 
in the sense that it captures all the relations between products of $t[a]$'s which are not specific to a fixed value of $N$. From that point of view, the classical 
line defects $L_N$ fully capture the gauge algebra of the classical theory. 

One would similarly expect ${\mathcal A}_{\epsilon_1, \epsilon_2}$ to be fully captured
as the uniform-in-$N$ limit of the ${\mathcal A}^{(N)}_{\epsilon_2, \epsilon_1}$. This would be the triality image of the 
statement in \cite{Costello:2017fbo} that ${\mathcal A}_{\epsilon_1, \epsilon_2}$ can be described as
as uniform-in-$N$ limit of the ${\mathcal A}^{(N)}_{\epsilon_1, \epsilon_2}$ algebras. 

We can employ a similar strategy to describe classical Wilson lines of negative charge, tentatively associated to M2 branes with $(1,1)$ orientation. 
There is a morphism $U(\mathrm{Diff}^{\epsilon_2}_\bC) \to W_{-\epsilon_2}$,
\begin{equation} \label{eq:L1morph}
t[w] \to - w^t
\end{equation} 
where $w^t$ is obtained from $w$ by reversing the order of the symbols. In particular, this maps $t[1] \to -1$,$t[x] \to - x$, etc.
This gives us a Wilson line 
\begin{equation}
L_{-1} \equiv \mathrm{Pexp} \left[-\int A_t(t,x,y) dt \right]
\end{equation}
where we can omit the ${}^t$ symbol because of the symmetrization of the arguments. 

Similarly, we have a nice morphism $U(\mathrm{Diff}^{\epsilon_2}_\bC) \to W^{(N)}_{-\epsilon_2}$
\begin{equation} \label{eq:LNmorph}
t[w] \to -\sum_{i=1}^N w^t_i
\end{equation} 
which maps, $t[1] \to -N$, $t[x] \to -\sum_i x_i$, etc. It gives a Wilson line of the form 
\begin{equation}
L_{-N} \equiv \mathrm{Pexp} \left[-\int \sum_{i=1}^N A_t(t,x_i,y_i) dt \right]
\end{equation}

Based on the expected triality, we propose that the morphism $U(\mathrm{Diff}^{\epsilon_2}_\bC) \to W^{(N)}_{-\epsilon_2}$
should be deformed to a morphism ${\mathcal A}_{\epsilon_1, \epsilon_2} \to {\mathcal A}^{(N)}_{-\epsilon_1-\epsilon_2, \epsilon_1}$
related by triality to the morphism ${\mathcal A}_{\epsilon_1, \epsilon_2} \to {\mathcal A}^{(N)}_{\epsilon_1, \epsilon_2}$
described in \cite{Costello:2017fbo} and to the above-mentioned ${\mathcal A}_{\epsilon_1, \epsilon_2} \to {\mathcal A}^{(N)}_{\epsilon_2, \epsilon_1}$  morphism. 
We will verify that prediction momentarily. 

We are now ready to review the construction of $(0,1)$ line defects. 
\footnote{Although we can define a composite classical line defect 
\begin{equation}
\mathrm{Pexp} \left[\int \sum_{i=1}^N A_t(t,x^{(+)}_i,y^{(+)}_i)-\sum_{i=1}^{N'} A_t(t,x^{(-)}_i,y^{(-)}_i) dt \right]
\end{equation}
this is likely not the correct classical limit of the line defect associated to two stacks of M2 branes, which cannot be simultaneously 
described as fundamental strings in the same duality frame.}

\subsection{$(0,1)$ line defects and ADHM quantum mechanics} \label{sec:adhm}

The M2 branes with $(0,1)$ orientation descend to D2 branes in the IIA string theory duality frame, sitting on top of the D6 brane. 
The world-volume theory on the D2 branes is a 3d ${\cal N}=4$ gauge theory based on the ADHM (aka Jordan) quiver with one flavour:
the gauge group is $U(N)$ and the matter content consists of one adjoint hypermultiplet $(X,Y)$ and one (anti)fundamental hypermultiplet $(I,J)$. 

Upon $\Omega$ deformation, the 3d theory is reduced to a topological quantum mechanics, whose operator algebra ${\mathcal A}^{(N)}_{\epsilon_1, \epsilon_2}$
is the quantization of the algebra of functions on the Higgs branch of the 3d theory. Operationally, the algebra is obtained by a quantum symplectic reduction of 
the Weyl algebra built from the hypermultiplet fields \cite{Yagi:2014toa,Bullimore:2016nji}. The quantization depends on a parameter we identify with $\epsilon_1$. The quantum symplectic reduction depends on an FI parameter which we identify, in an appropriate normal ordering scheme, with $\epsilon_2$. 

A basic result of \cite{Costello:2017fbo} was to find an ``uniform-in-$N$'' definition of ${\mathcal A}^{(N)}_{\epsilon_1, \epsilon_2}$, 
giving some master algebra  ${\mathcal A}^{(*)}_{\epsilon_1, \epsilon_2}$ together with morphisms ${\mathcal A}^{(*)}_{\epsilon_1, \epsilon_2} \to {\mathcal A}^{(N)}_{\epsilon_1, \epsilon_2}$
which impose the trace relations valid at a specific value of $N$. A harder result was to show that ${\mathcal A}^{(*)}_{\epsilon_1, \epsilon_2}$ can be precisely identified with the 5d gauge theory algebra
${\mathcal A}_{\epsilon_1, \epsilon_2}$, as a unique deformation of $U(\mathrm{Diff}^{\epsilon_2}_\bC)$ compatible with a one-loop calculation in the 5d CS theory. 

In other words, the Koszul dual ${\mathcal A}_{\epsilon_1, \epsilon_2}$ to the $A_\infty$ algebra of observable of the 5d gauge theory 
is fully determined as an uniform-in-$N$ version ${\mathcal A}^{(*)}_{\epsilon_1, \epsilon_2}$ of the algebra ${\mathcal A}^{(N)}_{\epsilon_1, \epsilon_2}$ which quantizes the 
algebra of functions on the Hilbert scheme of $N$ points in $\bC^2$.

We will now review the basic elements of this relation. We will then invoke 3d mirror symmetry to give an alternative ``Coulomb branch'' presentation 
${\mathcal C}^{(N)}_{\epsilon_1, \epsilon_2}$ of ${\mathcal A}^{(N)}_{\epsilon_1, \epsilon_2}$. An uniform-in-$N$ description of ${\mathcal C}^{(N)}_{\epsilon_1, \epsilon_2}$
will give us a new presentation ${\mathcal C}^{(*)}_{\epsilon_1, \epsilon_2}$ of ${\mathcal A}_{\epsilon_1, \epsilon_2}$ which is manifestly triality-invariant, 
with the desired morphisms to various line defect world-line theories. It takes the form of a ``shifted'' affine Yangian \cite{Kodera:2016faj}.

\subsubsection{ADHM, classical}
The Higgs branch for the ADHM quiver gauge theory with gauge group $U(N)$, one flavour and non-zero complex FI parameter equal to $\epsilon_2$
is a smooth complex symplectic manifold of dimension $2N$. It is a deformation of $(\bC^{2})^N/S_N$. 

We can describe the Poisson algebra ${\mathcal A}^{(N)}_{0,\epsilon_2}$  of function on the Higgs branch in terms of two adjoint fields $(X,Y)$, transforming as a doublet of an $\mathfrak{sl}_2$ global symmetry, 
an anti-fundamental field $I$ and a fundamental field $J$, constrained by the F-term relation 
\begin{equation}
[X,Y] +J I= \epsilon_2 1_{N \times N}
\end{equation}
modulo gauge transformations. The Poisson bracket is $\{X^a_b, Y^c_d\} = \delta^a_d \delta^c_b$ and $\{J^b,I_a\} = \delta^b_a$.

With no loss of generality, invariant functions can be written as polynomials in closed words such as $\Tr XYXY^3X^2Y$ and open words such as 
$I X Y X J$. The F-term relation allows one to reorganize $X$ and $Y$ symbols in a word, at the price of producing shorter words or products of words. 
Furthermore, al long as $\epsilon_2 \neq 0$, symmetrized traces are in the same $\mathfrak{sl}_2$ orbit as 
\begin{equation}
\Tr X^n = \epsilon_2^{-1} \Tr X^n [X,Y] + \epsilon_2^{-1} I X^n J = \epsilon_2^{-1} I X^n J
\end{equation}
so with no loss of generality and without reference to a specific value of $N$ we can restrict ourselves to polynomials in open words of the form $I X^n Y^m J$.
At any given $N$, one will have further trace relations. 

Poisson brackets between the $I X^n Y^m J$ generators can be computed combinatorially 
with no reference to $N$, until we impose trace relations and $IJ = \epsilon_2 N$.
That means we can define an uniform-in-$N$ Poisson algebra ${\mathcal A}^{(*)}_{0,\epsilon_2}$ together with a map to ${\mathcal A}^{(N)}_{0,\epsilon_2}$  which imposes the trace relations. 

\subsubsection{ADHM, quantum}
Now we replace the Poisson brackets with commutators $[X^a_b, Y^c_d] = \epsilon_1 \delta^a_d \delta^c_b$ and $[J^b,I_a] =\epsilon_1 \delta^b_a$ 
and to a quantum symplectic reduction to get the quantized Higgs branch algebra ${\mathcal A}^{(N)}_{\epsilon_1,\epsilon_2}$. This is the expected worldvolume 
theory for the line defects associated to M2 branes with $(0,1)$ orientation.

There is some normal ordering ambiguity in the precise definition of the quantum FI parameter. We choose to 
keep the same operator order in the two terms of the commutator and adjust the order of $I$,$J$ appropriately:
the operator order of $X$,$Y$ 
\begin{equation}
X^i_k Y^k_j - X^k_j Y^i_k + I_j J^i = \epsilon_2 \delta^i_j 
\end{equation}
This choice has the immediate effect of preserving the relation 
\begin{equation}
I X^n J = \epsilon_2 \Tr X^n 
\end{equation}
and, by acting with global $\mathfrak{sl}_2$ rotations, the more general relation for symmetrized traces:
\begin{equation}
I S[X^n Y^n] J = \epsilon_2 \Tr S[X^n Y^m] 
\end{equation}
where the $S$ symbol denotes symmetrization of the $X$, $Y$ symbols, i.e. projection to the 
spin $\frac{n+m}{2}$ irrep of $\mathfrak{sl}_2$. 

The actual algebra only depends on the ratio 
\begin{equation}
\hbar \equiv -\Psi^{-1} = \frac{\epsilon_1}{\epsilon_2}
\end{equation}
because we can rescale the $\epsilon_i$ by a uniform rescaling of $X$, $Y$, $I$, $J$. It is convenient, though, 
not to do so. 

There is a simple isomorphism
\begin{equation}
{\mathcal A}^{(N)}_{\epsilon_1,\epsilon_2} \simeq {\mathcal A}^{(N)}_{\epsilon_1,-\epsilon_1 - \epsilon_2}.
\end{equation}
It is induced by the outer automorphism of $U(N)$, exchanging the fundamental and anti-fundamental representations:
\begin{equation}
X^i_j \to X^j_i \qquad Y^i_j \to Y^j_i \qquad I_a \to J^a \qquad J^a \to - I_a
\end{equation}
which maps the coefficient on the right hand side of the F-term relation to $-\epsilon_1 - \epsilon_2$. 

The analysis of \cite{Costello:2017fbo} implies that ${\mathcal A}^{(N)}_{\epsilon_1,\epsilon_2}$ can be given a
combinatorial presentation in terms of abstract open and closed words with uniform-in-$N$ relations, together with extra trace relations 
which occur at fixed values of $N$. The combinatorial presentation defines an algebra ${\mathcal A}^{(*)}_{\epsilon_1,\epsilon_2}$ 
together with an algebra morphism ${\mathcal A}^{(*)}_{\epsilon_1,\epsilon_2} \to {\mathcal A}^{(N)}_{\epsilon_1,\epsilon_2}$
which imposes the trace relations for some given $N$. \footnote{It is important to point out that this is not a large $N$ limit. It is an uniform in $N$ presentation. In particular, an object such as $IJ =N \epsilon_2$ would be infinite in a large $N$ limit, but it is a perfectly good central element in ${\mathcal A}^{(*)}_{\epsilon_1,\epsilon_2}$.}

Furthermore, ${\mathcal A}^{(*)}_{\epsilon_1,\epsilon_2}$ admits a linear basis of polynomials in open words $I X^n Y^m J$.
If we work perturbatively in $\epsilon_1$, we see that the map  
\begin{equation}
o[w(x,y)] \equiv \frac{1}{\epsilon_1} I p(X,Y) J
\end{equation} 
presents ${\mathcal A}^{(*)}_{\epsilon_1,\epsilon_2}$ as a deformation of $U(\mathrm{Diff}^{\epsilon_2}_{\bC})$. Indeed:
\begin{itemize}
\item The F-term relation applied to the right hand side maps to the $[x,y]=\epsilon_2$ relation on the left hand side, up to 
terms suppressed by powers of $\epsilon_1$.
\item The commutator $\left[ o[w(x,y)] , o[w'(x,y)]  \right]$ is dominated by the $[J,I]$ commutators, with all other terms subleading
in $\epsilon_1$. These give $o[w(x,y) w'(x,y) - w'(x,y) w(x,y)]$.
\end{itemize}

The unicity theorems in \cite{Costello:2017fbo} identify ${\mathcal A}^{(*)}_{\epsilon_1,\epsilon_2}$ with the 5d CS theory algebra 
${\mathcal A}_{\epsilon_1,\epsilon_2}$. More precisely, as ${\mathcal A}^{(*)}_{\epsilon_1,\epsilon_2}$ is defined for 
finite $\epsilon_1$, the identification should be thought of as a {\it non-perturbative definition} of ${\mathcal A}_{\epsilon_1,\epsilon_2}$.
A specific presentation of generators $t[m,n]$ in ${\mathcal A}^{(*)}_{\epsilon_1,\epsilon_2}$ which go to the 
generators of $U(\mathrm{Diff}^{\epsilon_2}_{\bC})$ as $\epsilon_1 \to 0$ should be thought of as a specific renormalization 
scheme in the 5d gauge theory. From this point on, we will accordingly drop the ${}^{(*)}$ superscript.

The first potential ambiguity lies in the normal ordering prescription we choose in the F-term relations. Our particular choice 
is motivated by two remarkable ``experimental'' observations, which we do not know how to prove in the ADHM presentation of the 
algebra, but which will become more manifest later on in a mirror ``Coulomb branch'' presentation of the algebra. 

As we mentioned before, ${\mathcal A}^{(N)}_{\epsilon_1,\epsilon_2}$ quantizes the Hilbert scheme of points in $\bC^2$ and 
should admit a presentation as a deformation of $\frac{W_{\epsilon_1}^{N}}{S_N}$, with the deformation parameter being the quantum FI parameter.
The first remarkable observation is that the symmetrized traces $\Tr S[X^n Y^m]$ provide precisely such a presentation:
The commutators of symmetrized traces precisely match these of $\frac{W_{\epsilon_1}^{N}}{S_N}$ both in the $\epsilon_2 \to 0$ limit and in the 
$\epsilon_3 \to 0$ limit, i.e. $\epsilon_2 \to - \epsilon_1$. 

Because of the simple relation between $\Tr S[X^n Y^m]$ and the open words $I S[X^n Y^n] J$, the symmetrized traces are a valid alternative basis for 
${\mathcal A}_{\epsilon_1,\epsilon_2}$. That means that the 
traces $\Tr S[X^n Y^m]$ provide a presentation of ${\mathcal A}_{\epsilon_1,\epsilon_2}$ as a deformation of $U(\mathrm{Diff}^{\epsilon_1}_{\bC})$
both in the $\epsilon_2 \to 0$ and in the $\epsilon_3 \to 0$ limits.

On the other hand, the rescaled traces 
\begin{equation}
t[m,n] = \frac{\epsilon_2}{\epsilon_1} \Tr S[X^n Y^m]
\end{equation}
also provide a presentation of ${\mathcal A}_{\epsilon_1,\epsilon_2}$ as a deformation of $U(\mathrm{Diff}^{\epsilon_2}_{\bC})$
in the $\epsilon_1 \to 0$ limit. 

We are beginning to see the first hints of triality: the non-perturbative definition of algebra ${\mathcal A}_{\epsilon_1,\epsilon_2}$
reduces to the perturbative 5d gauge algebra in the expected three regions of the $\epsilon_i$ parameter space. 

A second remarkable observation makes triality fully manifest: the rescaled symmetrized traces 
\begin{equation}
T[m,n] = \frac{1}{\epsilon_1} \Tr S[X^n Y^m]
\end{equation}
have commutation relations which are manifestly triality invariant: the coefficients can be expressed purely in terms of the combinations 
\begin{equation}
T[0,0] \equiv \frac{N}{\epsilon_1} \qquad \qquad \sigma_2 = \frac{1}{2} \sum_{i=1}^3 \epsilon_i^2 \qquad \qquad \sigma_3 = \prod_{i=1}^3 \epsilon_i
\end{equation}

Assuming the correctness of these two observations, we arrive at a very pleasing situation. The non-perturbative algebra has a
triality-invariant basis $T[m,n]$ which can be mapped to three rescaled bases 
\begin{equation}
t_i[m,n] = \epsilon_i T[m,n]
\end{equation}
Each of these bases has a nice limit when one of the other two $\epsilon_j$ is sent to $0$ and can be used 
as a dual basis to the $c_{n,m}$ operators of the corresponding weakly-coupled gauge theory. 

This definition of the $t[m,n]$ generators implicitly selects a renormalization scheme in the 5d CS theory 
such that no further field re-definitions are needed to compare the operator algebras in different triality frames: 
only an overall rescaling of the $c_{m,n}$ generators is needed.  

\subsection{The quantum Coulomb branch perspective}
The 3d ${\cal N}=4$ gauge theory based on the ADHM (aka Jordan) quiver with one flavour is self-mirror. This fact is somewhat orthogonal 
to our discussion until now. In particular, string theory proofs \cite{Hanany:1996ie} of this fact are based on an alternative embedding of the 3d gauge theory in 
string theory. Still, we can employ this fact to find an alternative presentation of ${\mathcal A}^{(N)}_{\epsilon_1,\epsilon_2}$,
as the quantum Coulomb branch algebra ${\mathcal C}^{(N)}_{\epsilon_1,\epsilon_2}$ of the same quiver \cite{Bullimore:2015lsa,Braverman:2016wma}. 

With very little extra effort, we can study the quantum Coulomb branch algebra ${\mathcal C}^{(N)}_{\ell;\epsilon_1,\epsilon_2}$ 
of the ADHM quiver gauge theory with $\ell$ flavours. For general $\ell>1$ this is {\it not} the same as quantum Higgs branch
of the same quiver. Rather, it is related by 3d mirror symmetry to the quantum Higgs branch of an affine ADE quiver of type $A_{\ell-1}$, a necklace with $\ell$ nodes. Both 3d theories and quantum algebras can be obtained as UV gauge theory descriptions of the world-volume theory of $M2$ branes at the tip of an $A_{\ell-1}$ singularity $\frac{\bC^2}{\mathbb{Z}_\ell}$,
by replacing either $\bC_{\epsilon_1} \times \bC_{\epsilon_2}$ or $\frac{\bC^2}{\mathbb{Z}_\ell}$ factors in the transverse geometry with an appropriate Taub-NUT geometry.

This algebra thus has potentially interesting 5d gauge theory applications for all $\ell$. We see that 
${\mathcal C}^{(N)}_{\ell;\epsilon_1,\epsilon_2}$ for $\ell>1$ should be associated to 5d $U(1)$ CS theory line defects in a spacetime of the form
$\bR \times \frac{\bC^2}{\mathbb{Z}_\ell}$. Similar considerations also suggest that ${\mathcal C}^{(N)}_{0;\epsilon_1,\epsilon_2}$ should be associated to line defects in a spacetime of the form
$\bR \times \bC \times \bC^*$. The latter algebra will also play an important role in our discussion of surface defects in Section \ref{sec:bimodule}.

In particular, we will be able to define uniform-in-$N$ master algebras ${\mathcal C}_{\ell;\epsilon_1,\epsilon_2}$ and to demonstrate
triality uniformly for all $\ell$.

\subsubsection{A triality-invariant presentation of ${\mathcal C}_{0;\epsilon_1,\epsilon_2}$}
The quantum Coulomb branch admits an ``Abelianized'' presentation \cite{Bullimore:2015lsa,Kodera:2016faj}
in terms of explicit difference operators acting on functions of $\bC^N$. Define coordinates $w_a$ and shift operators 
$v_a:w_a \to w_a + \epsilon_1$. The quantum Coulomb branch algebra ${\mathcal C}^{(N)}_{0;\epsilon_1,\epsilon_2}$ is generated by the 
Hamiltonians 
\begin{equation}
W_k = \sum_{a=1}^N w_a^k 
\end{equation} for positive integer $k$
and monopole operators of minimal charge 
\begin{equation}
E_k = \sum_{a=1}^N \prod_{b \neq a} \frac{w_a - w_b - \epsilon_2}{w_a - w_b} w_a^k v_a
\end{equation} 
and
\begin{equation}
F_k = \sum_{a=1}^N \prod_{b \neq a} \frac{w_a - w_b + \epsilon_2}{w_a - w_b} v^{-1}_a w_a^k
\end{equation} 
for non-negative integer $k$.

We should mention briefly the perturbative triality of this algebra. We simply need a re-definition 
\begin{equation}
v_a = \prod_{b \neq a} \frac{w_a - w_b + \epsilon_2+ \epsilon_1}{w_a - w_b - \epsilon_2} \hat v_a
\end{equation}
which gives also 
\begin{equation}
v^{-1}_a = \prod_{b \neq a} \frac{w_a - w_b - \epsilon_1- \epsilon_2}{w_a - w_b + \epsilon_2} \hat v_a^{-1}
\end{equation}
which shows that the algebra is manifestly invariant under $\epsilon_2 \to - \epsilon_1 - \epsilon_2$. 

The uniform-in-$N$ version of the algebra, which we can denote as ${\mathcal C}_{0;\epsilon_1,\epsilon_2}$, 
was already discussed in \cite{Kodera:2016faj}. It can be presented abstractly by giving explicit commutation relations $[W_*,E_*]$, $[W_*,F_*]$, 
$[E_*,F_*]$ together with quadratic relations between $E$'s, quadratic relations between $F$'s and 
some Serre-like relations. The presentation given in \cite{Kodera:2016faj} is not manifestly triality invariant, but it can be made so with a little extra effort. 

It is useful to define Hamiltonians 
\begin{equation}
W[p] = \sum_{a=1}^N p(w_a) 
\end{equation}
associated to any polynomial $p$ in one variable, as well as more general formal generating functions
\begin{equation}
W[f] = \sum_{a=1}^N f(w_a) 
\end{equation}
associated to any function $f(w)$ with a formal expansion in polynomials of $w$.

We should also rescale the raising and lowering operators 
\begin{equation}
e_k = \frac{1}{\epsilon_1} E_k
\end{equation} 
and
\begin{equation}
f_k = \frac{1}{\epsilon_1} F_k
\end{equation} 

We have 
\begin{equation}
[e_n,f_m] = h_{n+m}
\end{equation}
with the $h_n$ being polynomials in the basic Hamiltonians
\begin{equation}
H(z) \equiv 1- \epsilon_1 \epsilon_2 (\epsilon_1 + \epsilon_2) \sum_{n=0}^{\infty} \frac{h_n}{z^{n+1}} = \prod_{a=1}^N \frac{(z-w_a-\epsilon_2)(z-w_a+\epsilon_1+\epsilon_2)}{(z-w_a)(z-w_a+\epsilon_1)}
\end{equation}
Conversely, we can write 
\begin{equation}
\frac{\partial_z H(z)}{H(z)} = W\left[\frac{1}{z-w-\epsilon_2}\right]+W\left[\frac{1}{z-w+\epsilon_1 +\epsilon_2}\right]-W\left[\frac{1}{z-w}\right]-W\left[\frac{1}{z-w+\epsilon_1}\right]
\end{equation}

In order to make contact with the standard presentation of the algebra, we need some other linear combinations $d_n$ 
of the Hamiltonians, which satisfy 
\begin{equation}
[d_n,e_{m}]=-n e_{n+m-1} \qquad \qquad [d_n,f_{m}]=n f_{n+m-1} 
\end{equation}
These can be built as
\begin{equation}
d_n = W[p_n]
\end{equation}
for (essentially Bernoulli) polynomials $p_n$ such that 
\begin{equation}
 p_n(w+\epsilon_1) -p_n(w) = n w^{n-1}
\end{equation}

We can use the derivative $\psi'$ of the digamma function to define a generating function \footnote{This follows from the expansion \begin{equation}
 \sum_{i=1}^\infty \frac{p_n(w)}{z^{n+1}} = \frac{1}{\epsilon^2_1}\psi'\left(1+\frac{z - w}{\epsilon_1}\right)
\end{equation}}
\begin{equation}
D(z) \equiv W\left[\frac{1}{\epsilon^2_1}\psi'\left(1+\frac{z - w}{\epsilon_1}\right)\right] =  \sum_{i=0}^\infty \frac{d_n}{z^{n+1}}
\end{equation}

Because 
\begin{equation}
\frac{1}{\epsilon^2_1}\psi'\left(1+\frac{z - w}{\epsilon_1}\right) -\frac{1}{\epsilon^2_1}\psi'\left(\frac{z - w}{\epsilon_1}\right) =\frac{1}{(z-w)^2}
\end{equation}
we learn that 
\begin{equation}
D(z) - D(z-\epsilon_1) =-W\left[\frac{1}{(z-w)^2} \right]
\end{equation}
and thus 
\begin{equation}
\partial^2_z \log H(z) = D(z-\epsilon_2)-D(z+\epsilon_2)+ D(z-\epsilon_1)-D(z+\epsilon_1)+D(z+\epsilon_1 +\epsilon_2)-D(z-\epsilon_2-\epsilon_1)
\end{equation}
which gives a triality-invariant presentation of the relation between $[e_*,f_*]$ and the Hamiltonians $d_*$.

All the other quadratic and Serre relations are also trivially triality-invariant when expressed in terms of the $e_*$, $f_*$ generators. 
We conclude that ${\mathcal C}_{0;\epsilon_1,\epsilon_2}$ has a triality symmetry fixing the $e_*$, $f_*$, $d_*$ and $h_*$ generators.

For later use, we should also mention a useful collection of other triality-invariant elements $e^{(n)}$ obtained recursively from 
\begin{equation}
e^{(1)} \equiv e_0 \qquad \qquad e^{(n+1)} \equiv -\frac{1}{n} [e_1, e^{(n)}]
\end{equation}
and $f^{(n)}$ obtained recursively from 
\begin{equation}
f^{(1)} \equiv f_0 \qquad \qquad f^{(n+1)} \equiv \frac{1}{n} [f_1, f^{(n)}]
\end{equation}
These elements all commute with each other. 

The algebra takes the form of an affine Yangian for $\mathfrak{gl}(1)$. It seems to be slightly different from the 
affine Yangian algebra which appears in the context of the ${\mathcal W}^{\epsilon_1, \epsilon_2}_{1+\infty}$ algebra \cite{Prochazka:2015deb}: 
the element $h_0$ vanishes here, while it is non-zero in the latter. The significance of this fact is 
obscure to us. 

\subsubsection{${\mathcal C}_{0;\epsilon_1,\epsilon_2}$ as a deformation of $U(\mathrm{Diff}^{\epsilon_1}_{\bC^*})$}
We should set some further notation. We can define the shift algebra $S_{\epsilon_1}$ as
the algebra generated by $w, v, v^{-1}$, with commutation relations
\begin{equation}
v w = (w+\epsilon_1)v \qquad \qquad v^{-1} w = (w-\epsilon_1)v^{-1}
\end{equation}
We will also denote the Lie algebra defined by commutators in $S_{\epsilon_1}$as $\mathrm{Diff}^{\epsilon_1}_{\bC^*}$.
We want to show that ${\mathcal C}_{0;\epsilon_1,\epsilon_2}$ is a deformation of $U(\mathrm{Diff}^{\epsilon_1}_{\bC^*})$.

If we set $\epsilon_2 \to 0$ in the standard expressions for the generators of 
${\mathcal C}^{(N)}_{0;\epsilon_1,\epsilon_2}$ we get some drastic simplifications:
\begin{equation}
E_k \to \sum_{a=1}^N w_a^k v_a
\end{equation} 
and
\begin{equation}
F_k \to \sum_{a=1}^N v^{-1}_a w_a^k
\end{equation} 
All elements of ${\mathcal C}^{(N)}_{0;\epsilon_1,\epsilon_2}$ which can be defined from these generators and $W_k$ without dividing by $\epsilon_2$ 
have thus a natural $\epsilon_2 \to 0$ limit to elements of the symmetric product algebra $\frac{S_{\epsilon_1}^{\otimes N}}{S_N}$.

Furthermore, all elements of ${\mathcal C}^{(N)}_{0;\epsilon_1,\epsilon_2}$ should have an interpretation as monopole operators of the 
3d gauge theory. The expected general form of such monopole operators should always contain terms which will not vanish as $\epsilon_2 \to 0$.
In short, we expect ${\mathcal C}^{(N)}_{0;\epsilon_1,\epsilon_2}$ to be truly a deformation of  $\frac{S_{\epsilon_1}^{\otimes N}}{S_N}$
in the $\epsilon_2 \to 0$ limit.

As we mentioned before, the symmetric powers of an algebra $A$ have an uniform-in-$N$ description as the universal enveloping algebra of the Lie algebra obtained from $A$ by taking commutators. 
Hence we also expect ${\mathcal C}_{0;\epsilon_1,\epsilon_2}$ to be a deformation of $U(\mathrm{Diff}^{\epsilon_1}_{\bC^*})$, with limiting identifications
\begin{equation}
W_k \to t[w^k] \qquad \qquad E_k \to t[w^k v] \qquad \qquad F_k \to t[v^{-1} w^k]
\end{equation}

We would like to present generators $t[w^m v^n]$ in ${\mathcal C}_{0;\epsilon_1,\epsilon_2}$ whose commutation relations are manifestly a deformation away from $\epsilon_2 =0$ of 
the commutation relations of $\mathrm{Diff}^{\epsilon_1}_{\bC^*}$.

There is a certain degree of ambiguity in choosing such generators, but some choices are better-behaved under triality. We present here a possible convenient choice.

First, we should build a convenient basis in $S_{\epsilon_1}$. We define 
\begin{equation}
s_{m,0} = \epsilon_1 p_m(w)
\end{equation}
and then recursively 
\begin{equation}
s_{m,n+1} =\epsilon^{-1}_1 [v,s_{m+1,n}] \qquad \qquad s_{m,-n-1} =\epsilon^{-1}_1 [s_{m+1,-n},v^{-1}]
\end{equation}
for $n \geq 0$. The elements $s_{m,n}$ are some appropriately-ordered versions of $w^m v^n$. 

We set 
\begin{equation}
t[s_{m,0}]\equiv \epsilon_1 d_m \qquad t[v]\equiv E_0 = \epsilon_1 e_0 \qquad t[v^{-1}]\equiv F_0 = \epsilon_1 f_0
\end{equation}
and then recursively 
\begin{equation}
t[s_{m,n+1}] \equiv [e_0, t[s_{m+1,n}] ]  \qquad \qquad t[s_{m,-n-1}] \equiv [t[s_{m+1,-n}],f_0] 
\end{equation}
This is particularly convenient for two reasons. First of all, $e_0$ and $f_0$ commute. 
Second, Serre relations guarantee that repeated commutators with $e_0$ or $f_0$ 
will ultimately give $0$. The step before $0$ always produces one of the bare monopole operators 
\begin{equation}
t[v^n] \equiv \epsilon_1 e^{(n)} \qquad \qquad t[v^{-n}] \equiv \epsilon_1 f^{(n)}
\end{equation}
or $t[1]\equiv \epsilon_1 h_1$.

An interesting feature of this presentation is that it behave in a very simple way under triality: all generators take 
the form of $\epsilon_1$ times a triality-invariant expression. As a consequence, triality simply rescales all generators uniformly
by some ratio of $\epsilon_i$. This is analogous to the triality properties we conjectured for the symmetrized traces in the ADHM algebra. 

The analysis so far suggests strongly that ${\mathcal C}_{0;\epsilon_1,\epsilon_2}$ should be identified with the Koszul dual 
to the algebra of operators for the 5d non-commutative $U(1)$ CS theory on a $\bR \times \bC \times \bC^*$ background. 
This is compatible with triality and with the three $\epsilon_i \to 0$ limits 
to the universal enveloping algebra of the gauge Lie algebra for the 5d theory. 

Under such an identification, the above basis of generators for ${\mathcal C}_{0;\epsilon_1,\epsilon_2}$ maps to a particularly nice 
form for the $A_\infty$ algebra of operators of the 5d theory, dual to a basis of $c_{n,m}$ which are the coefficient of an expansion of 
$c(v,w)$ into the basis of $s_{m,n}$. In particular, triality will act on the $c_{m,n}$ 
by an overall rescaling by some ratio of $\epsilon_i$.

\subsubsection{More general truncations}
If the relation of ${\mathcal C}_{0;\epsilon_1,\epsilon_2}$ to the 5d non-commutative $U(1)$ CS theory on a $\bR \times \bC \times \bC^*$ is correct, 
we should be able to find more general morphisms ${\mathcal C}_{0;\epsilon_1,\epsilon_2} \to {\mathcal C}^{N_1,N_2,N_3}_{0;\epsilon_1,\epsilon_2}$
to the world-line algebras of line defects associated to multiple stacks of M2 branes with the three possible orientations. 

At the moment, we know very little about such truncations. The central element $d_0 = h_1$ is fixed to $\frac{N}{\epsilon_1}$ 
in ${\mathcal C}^{(N)}_{0;\epsilon_1,\epsilon_2}$ and it seems to be controlling the ``charge'' of the Wilson line. Using triality, and assuming that the charge is additive, 
we conjecture that the central element would be fixed to 
\begin{equation}
d_0 = h_1 = \frac{N_1}{\epsilon_1} + \frac{N_2}{\epsilon_2} - \frac{N_3}{\epsilon_1+\epsilon_2}
\end{equation} 
in ${\mathcal C}^{N_1,N_2,N_3}_{0;\epsilon_1,\epsilon_2}$. 

A stronger conjecture is that the whole $H(z)$ would be fixed to take a rational form which combines the three factors associated to individual lines
\begin{equation}
H(z) = \frac{Q_{N_1}(z-\epsilon_2)Q_{N_1}(z+\epsilon_1+\epsilon_2)}{Q_{N_1}(z)Q_{N_1}(z+\epsilon_1)} \frac{Q_{N_2}(z-\epsilon_1)Q_{N_2}(z+\epsilon_1+\epsilon_2)}{Q_{N_2}(z)Q_{N_2}(z+\epsilon_2)} \frac{Q_{N_3}(z-\epsilon_2)Q_{N_3}(z-\epsilon_1)}{Q_{N_3}(z)Q_{N_3}(z-\epsilon_1-\epsilon_2)}
\end{equation}

It would be nice to produce explicitly such a reduction, perhaps using some kind of coproduct for ${\mathcal C}_{0;\epsilon_1,\epsilon_2}$.


\subsubsection{A triality-invariant presentation of ${\mathcal C}_{\epsilon_1,\epsilon_2}$}

The algebra ${\mathcal C}^{(N)}_{\epsilon_1,\epsilon_2}$ we are most interested in is defined in a similar manner, except that 
the power of $w_a$ in $F_k$ is shifted by one: 
\begin{equation}
F_k = \sum_{a=1}^N \prod_{b \neq a} \frac{w_a - w_b + \epsilon_2}{w_a - w_b} v^{-1}_a w_a^{k+1}
\end{equation} 
In particular, ${\mathcal C}^{(N)}_{\epsilon_1,\epsilon_2}$ is a sub-algebra of ${\mathcal C}^{(N)}_{0;\epsilon_1,\epsilon_2}$.
We can define ${\mathcal C}_{\epsilon_1,\epsilon_2}$ in the same way as a sub-algebra of ${\mathcal C}_{0;\epsilon_1,\epsilon_2}$.

Notice that there is still a perfect symmetry between $E_k$ and $F_k$, the $F_k$ satisfy the same quadratic relations as the $E_k$ do, etc. 
Much of our analysis of ${\mathcal C}_{0;\epsilon_1,\epsilon_2}$ is immediately inherited. 

The algebra ${\mathcal C}^{(N)}_{\epsilon_1,\epsilon_2}$ is expected to coincide with ${\mathcal A}^{(N)}_{\epsilon_1,\epsilon_2}$. The precise identification, 
or ``quantum mirror map'' is not obvious. At the classical level, the $u_a$ are simply the eigenvalues of $XY$ or $YX$, 
with a parameterization
\begin{equation}
X^a_b = \frac{\prod_{c \neq b}(w_a-w_c-\epsilon_2)}{\prod_{c \neq a} (w_a - w_c)} v_b
\end{equation}
and
\begin{equation}
Y^a_b = \frac{\prod_{c \neq b}(w_a-w_c+\epsilon_2)}{\prod_{c \neq a} (w_a - w_c)} v_b
\end{equation}
and $J^a=1$, $I_a = m \prod_{b \neq a} \frac{w_a - w_b - \epsilon_2}{w_a - w_b}$.
The quantum version of these relations has likely something to do with ``spherical Cherednik algebras''
\cite{Kodera:2016faj,2016arXiv160905494W}. We will not attempt to describe or employ them here. 

The Coulomb branch presentation hides the global $\mathrm{su}(2)_{XY}$ symmetry 
rotating $X$ and $Y$ into each other. Only the Cartan generator is manifest, assigning charge $1$ to the $v_a$, $0$ to the $w_a$. 
The raising and lowering operators are expected to coincide with some generators we present below. 

We can define again rescaled generators 
\begin{equation}
e_k = \frac{1}{\epsilon_1} E_k
\end{equation} 
and
\begin{equation}
f_k = \frac{1}{\epsilon_1} F_k
\end{equation} 
and use the same $h_n$ and $d_n$ as before, except that now 
\begin{equation}
[e_n,f_m] = h_{n+m+1}
\end{equation}
but we still have
\begin{equation}
[d_n,e_{m}]=-n e_{n+m-1} \qquad \qquad [d_n,f_{m}]=n f_{n+m-1} 
\end{equation}
Triality invariance is inherited from ${\mathcal C}_{0;\epsilon_1,\epsilon_2}$.

We can also define operators $e^{(n)}$ and $f^{(n)}$ obtained recursively from 
\begin{equation}
e^{(1)} \equiv e_0 \qquad \qquad e^{(n+1)} \equiv -\frac{1}{n} [e_1, e^{(n)}]
\end{equation}
and 
\begin{equation}
f^{(1)} \equiv f_0 \qquad \qquad f^{(n+1)} \equiv \frac{1}{n} [f_1, f^{(n)}]
\end{equation}
The $e^{(n)}$'s still commute with each other, and so do the $f^{(n)}$'s, but the commutator of $e^{(m)}$ and $f^{(n)}$ is now non-trivial. 
For example, $[e^{(1)},f^{(1)}]$ equals the central element $d_0$. 

The $\mathrm{su}(2)_{XY}$ raising and lowering operators in ${\mathcal C}_{\epsilon_1,\epsilon_2}$ are $[e^{(2)}$ and $f^{(2)}]$, the Cartan generator is $\frac12 d_1$.
One can use these generators to organize the operators in ${\mathcal C}_{\epsilon_1,\epsilon_2}$ into  $\mathrm{su}(2)_{XY}$ irreps. 
In particular, the Serre relations imply that $[e^{(n)}$ and $f^{(n)}]$ are the highest and lowest weight elements of an irreducible $\mathrm{su}(2)_{XY}$
representation of dimension $n+1$. 

We should observe that we can write the difference operators in terms of $x_a = v_a$ and $y_a = v_a^{-1} w_a$, with $w_a = x_a y_a$.
In particular, in the $\epsilon_2 \to 0$ limit the algebra ${\mathcal C}^{(N)}_{\ell;\epsilon_1,\epsilon_2}$ is presented as a deformation of $\frac{W_{\epsilon_1}^{\otimes N}}{S_N}$.
Working uniformly in $N$, we expect ${\mathcal C}_{\epsilon_1,\epsilon_2}$ to be a deformation of $U(\mathrm{Diff}^{\epsilon_1}_{\bC})$.
We would like to present generators $t[x^m y^n]$ in ${\mathcal C}_{\epsilon_1,\epsilon_2}$ whose commutation relations are manifestly a deformation away from $\epsilon_2 =0$ of 
the commutation relations of $\mathrm{Diff}^{\epsilon_1}_{\bC}$.

The $\mathrm{su}(2)_{XY}$ quantum numbers, triality properties, $\epsilon_2 \to 0$ and $\epsilon_2 + \epsilon_1 \to 0$ limits of the operators in the same irrep as $[e^{(k)}$ and $f^{(k)}]$ 
are the same as these of the symmetrized traces $T[m,n]$ we introduced in ${\mathcal A}^{(N)}_{\epsilon_1,\epsilon_2}$. This is not quite enough to insure they 
coincide with them under the quantum mirror map, as one could have non-linear corrections proportional to $\sigma_3 = \epsilon_1 \epsilon_2 \epsilon_3$ 
which would not change these properties. 

Still, we can use these operators in the same manner to define a good presentation of ${\mathcal C}_{\epsilon_1,\epsilon_2}$. We can set 
\begin{equation}
\tilde t[x^n] \equiv \epsilon_1  e^{(n)} \qquad \qquad  \tilde t[y^n] \equiv \epsilon_1  f^{(n)}
\end{equation}
and $\tilde t[1] = \epsilon_1 d_0$.
and then define recursively 
\begin{equation}
\tilde t[ [y^2,p(x,y)] ] \equiv [\tilde t[y^2],\tilde t[p(x,y)]]
\end{equation}
for any symmetrized monomial $p(x,y)$. This gives  us a definition of generators $\tilde t[m,n] = \epsilon_1 \tilde T[m,n]$ 
which have properties analogous to these we conjectured for the symmetrized traces $t[m,n]$ in ${\mathcal A}^{(N)}_{\epsilon_1,\epsilon_2}$.

\subsection{More general truncations}
We have now established that the same universal algebra ${\mathcal C}_{\epsilon_1,\epsilon_2}$ admits three inequivalent truncations 
${\mathcal C}^{(N)}_{\epsilon_1,\epsilon_2}$, ${\mathcal C}^{(N)}_{\epsilon_2,\epsilon_1}$, ${\mathcal C}^{(N)}_{-\epsilon_1-\epsilon_2,\epsilon_1}$.
which correspond to line defects associated with $N$ M2 branes of three different orientations. 

The three truncations impose respectively $h_1 = \frac{N}{\epsilon_i}$. More generally, they require $H(z)$ to take the form of a rational function
with a very specific pattern of poles and zeroes.  

As we should be able to place three separate stacks of $N_1$, $N_2$, $N_3$ M2 branes wrapping the respective $\bC_{\epsilon_i}$ factor in 
the internal geometry. Correspondingly, there should exist a more general family of truncations ${\mathcal C}^{(N_1,N_2,N_3)}_{\epsilon_1,\epsilon_2}$
of ${\mathcal C}_{\epsilon_1,\epsilon_2}$, perhaps associated with a specialization 
\begin{equation}
h_1 = \sum_{i=1}^3 \frac{N_i}{\epsilon_i}
\end{equation}
and a rational $H(z)$ which combines the three factors associated to individual lines
\begin{equation}
H(z) = \frac{Q_{N_1}(z-\epsilon_2)Q_{N_1}(z+\epsilon_1+\epsilon_2)}{Q_{N_1}(z)Q_{N_1}(z+\epsilon_1)} \frac{Q_{N_2}(z-\epsilon_1)Q_{N_2}(z+\epsilon_1+\epsilon_2)}{Q_{N_2}(z)Q_{N_2}(z+\epsilon_2)} \frac{Q_{N_3}(z-\epsilon_2)Q_{N_3}(z-\epsilon_1)}{Q_{N_3}(z)Q_{N_3}(z-\epsilon_1-\epsilon_2)}
\end{equation}
The truncations should be inherited from the corresponding truncations of ${\mathcal C}_{0;\epsilon_1,\epsilon_2}$.

\subsection{A triality-invariant presentation of ${\mathcal C}_{\ell;\epsilon_1,\epsilon_2}$}

The algebra ${\mathcal C}^{(N)}_{\ell;\epsilon_1,\epsilon_2}$ is defined in a similar manner s the others, except that now 
\begin{equation}
F_k = \sum_{a=1}^N \prod_{b \neq a} \frac{w_a - w_b + \epsilon_2}{w_a - w_b} v^{-1}_a w_a^{k} \prod_{i=1}^\ell (w_a - m_i)
\end{equation} 
with some extra parameters $m_i$ with $\sum_i m_i=0$, which are mass parameters for an $SU(\ell)$ flavour symmetry of the 3d gauge theory. 

There is an obvious embedding of this algebra in ${\mathcal C}^{(N)}_{0;\epsilon_1,\epsilon_2}$. There are actually multiple embeddings, 
where we shift some of the $(w_a - m_i)$ factors from $F_k$ to $E_k$ by a re-definition of $v_a$, as well as compatible embeddings in 
${\mathcal C}^{(N)}_{\ell;\epsilon_1,\epsilon_2}$ with smaller $\ell$'s. They all lift to embeddings of the uniform-in $N$ algebras ${\mathcal C}_{\ell;\epsilon_1,\epsilon_2}$.
These embeddings are all manifestly triality invariant. 

If we write $\tilde v_a = v^{-1}_a \prod_{i=1}^\ell (w_a - m_i)$, the operators $v_a$, $\tilde v_a$ and $w_a$ describe the quantized algebra of functions 
$W_{\ell; \epsilon_1}$ of the deformed $A_{\ell-1}$ singularity, which is generated such operators with the relation
\begin{equation}
v_a \tilde v_a = \prod_{i=1}^\ell (w_a - m_i)
\end{equation}
As a result, as $\epsilon_2 \to 0$ we recognize ${\mathcal C}^{(N)}_{\ell;\epsilon_1,\epsilon_2}$ as a deformation of $\frac{W_{\ell,\epsilon_1}^{\otimes N}}{S_N}$.
It should be possible to similarly identify ${\mathcal C}^{(N)}_{\epsilon_1,\epsilon_2}$ as a deformation of the universal enveloping algebra of the Lie algebra associated to 
$W_{\ell,\epsilon_1}$, though some care may be needed for special valued of the $m_i$. 

Correspondingly, we would associate ${\mathcal C}_{\ell;\epsilon_1,\epsilon_2}$ to 5d gauge theory on the deformed $A_{\ell-1}$ singularity.
The algebra embeddings between different $\ell$'s should be dual to the opposite embeddings of operator algebras, associated to 
geometric relations between the deformed $A_{\ell-1}$ singularities.

\section{M2-M5 systems and modules} \label{sec:module}
Consider now a setup where both M2's and M5's are present. The M2's could be simply crossing the M5's, or may be ending on them. 
The 5d result is an intersection point between a line defect and a surface defect, across which the line defect may remain the same, change or disappear. 

We would like to study such intersections from the point of view of the line defects. Before coupling to the 5d gauge theory, 
the world-line theory of the $\Omega$-deformed M2 branes will have some linear space of local operators at the intersection, 
forming a (bi)-module for the M2 brane operator algebra(s) under composition along the topological direction. 
After coupling to the 5d gauge theory and to the M5 brane world-volume fields, one may or not have a gauge-invariant intersection point. 

As we will explain momentarily, the existence of a gauge-invariant intersection point with some prescribed surface defect $S$
can be expressed in algebraic terms with the help of a certain bi-module ${\mathcal B}^S_{\epsilon_1,\epsilon_2}$ for ${\mathcal A}_{\epsilon_1,\epsilon_2}$,
built from the collection of local operators on $S$ available at the intersection. 

Analogously to what we did in previous sections, we aim to test the conjectural description of $\Omega$-deformed M-theory 
by making sure that the expected gauge-invariant intersection points exist, are triality-covariant, agree with known 
boundary conditions and interfaces for the M2 brane world-volume theories and use the expected chiral local operators 
from the corner vertex algebras. 

We will accomplish these objectives to a reasonable degree, but only in two restrictive situations:
\begin{itemize}
\item When all M2 branes end on the M5's. In this section we will build ``Verma'' modules ${\mathcal M}^{n_1,n_2,n_3}_{\epsilon_1,\epsilon_2}$ 
which control gauge-invariant endpoints, with a structure which agrees with the expected field theory boundary conditions for M2's ending on M5's. 
\item When no M2 branes end on the M5's. In the next section we will use a gauge theory construction to build a simple, important example of 
bimodules ${\mathcal B}^{n_1,n_2,n_3}_{0;\epsilon_1,\epsilon_2}$ for M2's crossing M5's without ending on them and conjecture a more general construction.
\end{itemize}

We leave to future work the analysis of bimodules which describe intersections across which the numbers of M2 branes jump in an arbitrary manner.

\subsection{Universal (bi)modules}

In general, the surface defect will enrich the collection of local operators which exist at $t=0$. 
Schematically, the operators will consists of polynomials in the usual modes $c_{n,m}$ 
of the gauge theory ghosts multiplying the modes 
\begin{equation}
O_{i;n} \equiv \frac{1}{n!} \partial_{z_1}^n O_i(0)
\end{equation} 
of other operators from the surface defect's world-volume theory.

We assume that the world-volume theory of the defect is some standard chiral algebra, 
with operators in ghost number $0$ only. Then the world-volume theory modes will include the vacuum module of the chiral algebra, 
together with any other chiral algebra module which we may want to couple to the line defect at the origin. 

Classically, the local operators will be equipped with a BRST differential encoding gauge symmetry, which roughly acts as a gauge transformation
with parameter $c$. There will also be natural actions of the algebra of bulk local operators defined by bringing in bulk local operators from $t>0$ or $t<0$ 
and colliding them with the operators at $t=0$.

Quantum mechanically, the BRST differential and algebra actions will be modified, and higher operations may appear. 
The local operators at the defect will form an $A_\infty$ bi-module $\mathrm{Obs}^D_{\epsilon_1, \epsilon_2}$
for the bulk $A_\infty$ algebra $\mathrm{Obs}_{\epsilon_1, \epsilon_2}$. As we did before, we can employ Koszul duality 
to convert this intricate structure into something simpler: a bi-module ${\mathcal B}^D_{\epsilon_1, \epsilon_2}$
for the algebra ${\mathcal A}_{\epsilon_1, \epsilon_2}$.

Following Appendix \ref{app:koszul}, we can give a simple physical interpretation to ${\mathcal B}^D_{\epsilon_1, \epsilon_2}$.
Suppose that we are studying some line defects $L_\pm$ associated to world-line degrees of freedoms defined by algebras $A_\pm$.
As discussed before, gauge-invariant couplings of the 5d theory to these line defects are described by algebra morphisms ${\mathcal A}_{\epsilon_1, \epsilon_2} \to A_\pm$
which tell us which operators in $A_\pm$ are coupled to specific modes of the gauge field. 

We can try to build a junction where $L_+$ for $t>0$ and $L_-$ for $t<0$ meet the defect $D$. Classically, that means finding some operator on $D$ 
which transforms in the correct representation to live at the endpoints of the Wilson lines $L_\pm$. Formally, that means providing an  $A_+$-$A_-$-bimodule
$B$ and a gauge-invariant pairing to $\mathrm{Obs}^D_{\epsilon_1, \epsilon_2}$. After Koszul duality, the problem is mapped to the following algebraic problem.
The bi-module $B$ can be made into an ${\mathcal A}_{\epsilon_1, \epsilon_2}$-bimodule by the morphisms ${\mathcal A}_{\epsilon_1, \epsilon_2} \to A_\pm$.
Gauge-invariant junctions are labelled by morphism ${\mathcal B}^D_{\epsilon_1, \epsilon_2} \to B$ of ${\mathcal A}_{\epsilon_1, \epsilon_2}$-bimodules. 

Classically, we have an action of the gauge Lie algebra $\mathrm{Diff}^{\epsilon_2}_{\bC}$ on the space $O_D$ of local operators from the surface defect's world-volume theory,
giving a dual Lie algebra representation on $O_D^*$. In order to study the quantum deformation, we promote $\mathrm{Diff}^{\epsilon_2}_{\bC}$ to $U(\mathrm{Diff}^{\epsilon_2}_{\bC})$
and $O_D^*$ to a bi-module $O_D^* \otimes U(\mathrm{Diff}^{\epsilon_2}_{\bC})$, where we act from the right in the trivial way and we act from the left by 
using the Lie algebra action:
\begin{equation}
(o w) \circ a = o (w a) \qquad \qquad a \circ (o w') = o (a w') + (a \circ o) w'
\end{equation}
where $w$ is a word in $U(\mathrm{Diff}^{\epsilon_2}_{\bC})$, $a$ an element of $\mathrm{Diff}^{\epsilon_2}_{\bC}$, $o$ an element of $O_D^*$ and the rightmost $\circ$ represents the 
dual of the action of $\mathrm{Diff}^{\epsilon_2}_{\bC}$ on $O_D$. 

Parsing through the definition, we see that a bimodule map 
$O_D^* \otimes U(\mathrm{Diff}^{\epsilon_2}_{\bC}) \to B$ precisely gives a pairing between $B$ and $O_d$ such that 
the action of a gauge generator in the representation $L_+$ from the left equals the sum of the the action on $B$ and of the action on the right in the representation $L_-$,
i.e. the junction is gauge invariant. 

Quantum mechanically, ${\mathcal B}^D_{\epsilon_1, \epsilon_2}$ will be a deformation of $O_D^* \otimes U(\mathrm{Diff}^{\epsilon_2}_{\bC})$ compatible with the 
deformation of $U(\mathrm{Diff}^{\epsilon_2}_{\bC})$ to ${\mathcal A}_{\epsilon_1, \epsilon_2}$.

Much of our discussion until now was not quite specific to a surface defect localized in the $z_2$ direction. It would apply equally well to other sorts of defects. 
We will now explain that the characteristic property of a surface defect localized in the $z_2$ direction is that it gives a bimodule ${\mathcal B}^D_{\epsilon_1, \epsilon_2}$
with some kind of weight condition.

Classically, we need to look at the action of the gauge Lie algebra  $\mathrm{Diff}^{\epsilon_2}_{\bC}$ on the surface defect local operators. A  gauge transformation 
of parameter $z_1^m z_2^n$ acts on functions of $z_1$ roughly as a combination of $z_1^m$ and $\partial_{z_1}^n$. If $n>m$, it will involve at least some derivative. 
It will never produce, say, an operator with no derivatives. Dually, that means that the module $O_D^*$ should have a highest weight property: 
generators $t[m,n]$ in $\mathrm{Diff}^{\epsilon_2}_{\bC}$ with $n>m$ will act nilpotently on $O_D^*$ and there will be some highest weight vectors dual to operators with no derivative, 
which are annihilated by all $t[m,n]$ with $n>m$. 

Furthermore, operators of the form $t[n,n]$ act as powers of the dilatation operator in the $\bC_{z_1}$ plane. Something behaving as like a primary operator of dimension 
$\Delta$ should have eigenvalue $\Delta^n$ for $t[(xy)^n]$. A collection of primary fields brought close to the origin should have eigenvalue $\sum_i \Delta_i^n$ for $t[(xy)^n]$.

After passing to the universal enveloping algebra, the highest weight property becomes the requirement that the difference $\overrightarrow t[m,n]- \overleftarrow t[m,n]$ with $n>m$
should act nilpotently. We expect this condition to persist quantum-mechanically to ${\mathcal B}^D_{\epsilon_1, \epsilon_2}$. More precisely, we take it as our {\it definition}
of a surface defect localized in the $z_2$ direction.

As a special case we can look at situations where $L_-$ is trivial. That situation leads us to consider some universal module ${\mathcal M}^D_{\epsilon_1, \epsilon_2}$
for ${\mathcal A}_{\epsilon_1, \epsilon_2}$, such that endpoints of $L_+$ on the defect are controlled by $A_+$ module maps ${\mathcal M}^D_{\epsilon_1, \epsilon_2} \to M$.
This will be a deformation of $O_D^*$, which we expect to be a highest weight module for ${\mathcal A}_{\epsilon_1, \epsilon_2}$. 

In the remainder of this section, we will focus on such modules. In the next section we will go back to study more general bi-modules. 

\subsection{Classical modules}
In order to gain some intuition about the junctions in the classical theory, we can look at highest weight modules for 
the woldline algebras of classical Wilson lines. 

Recall that the world-line algebra for a charge $1$ $(1,0)$ M2 line defect is the Weyl algebra $W_{\epsilon_2}$.
The two generators $x$ and $y$ literally describe the transverse position of the worldline. 

The Weyl algebra $W_{\epsilon_2}$ has an unique highest weight module $M_1$, 
which is the usual highest weight module for the Heisenberg algebra: the span of vectors $|n\rangle = x^n |0\rangle$ built from a highest weight vector which satisfies $y |0\rangle = 0$. This has eigenvalue $-\frac12$ under the dilatation operator $\frac{1}{2 \epsilon_2} (x y + y x)$. 

If we identify the Weyl algebra with the worldline algebra for a charge $1$ Wilson loop, this the natural module which one could pair up with the modes $\frac{1}{n!}\partial^n \psi(0)$ of a chiral complex fermion, or $\frac{1}{n!} \partial^n \gamma(0)$ of a $\beta \gamma$ system, naturally coupled with charge $1$ to the 5d $U(1)$ gauge field at the surface defect. If we use the transposition map to identify the Weyl algebra with the worldline algebra for a charge $-1$ Wilson loop, this the natural module which one could pair up with the modes $\partial^n \psi^\dagger(0)$ of a chiral complex fermion, or $\partial^n \beta(0)$ of a $\beta \gamma$ system, naturally coupled with charge $-1$ to the 5d $U(1)$ gauge field at the surface defect.

\subsubsection{Verma modules and Young diagrams}
The symmetric algebra $\frac{W_{\epsilon_2}^{\otimes N}}{S_N}$ has a variety of highest weight modules $M_{Y;\epsilon_2}$, 
obtained by projecting $M_1^{\otimes N}$ to various irreducible representations $R_Y$ of $S_N$. They could represent the endpoints of a charge $N$ line defect 
on some appropriate surface defect. Furthermore, we may be able to find families of modules which have a uniform-in-$N$ definition 
and could be candidates for some ${\mathcal M}^D_{\epsilon_1, \epsilon_2}$.

The projection onto an irreducible representation $R_Y$ of $S_N$ can be accomplished with the help of the Young projector associated with a 
given Young diagram $Y$: we associate each copy of $W_{\epsilon_2}$ to a box in the diagram, antisymmetrize within each column and then symmetrize within each row. When we apply the projector onto a vector of the form $\prod_i x_i^{m_i}$, we will get zero if some $m_i$ within the same column 
coincide.  Otherwise, we will get some basis vector in  $M_{Y;\epsilon_2}$. Monomials related by a permutation within columns followed by a permutation within rows 
give the same basis vector, up to a sign. We can use this identity repeatedly to bring the $m_i$ to a canonical form, where the $m_i$ grow along each column and 
do not decrease along each row. A basis of $M_{Y;\epsilon_2}$ thus consists of such canonical decorations of the Young diagram. 

This presentation will be useful later on when we turn on $\epsilon_1$. In particular, generators of the form $t[(xy)^m]$, $t[(xy)^m x]$ and 
 $t[y(xy)^m]$ act in a particularly simple way. 

In order to get a practical understanding of the $M_{Y;\epsilon_2}$, though, 
we do not actually need to fully project onto irreducible representations. $\frac{W_{\epsilon_2}^{\otimes N}}{S_N}$
acts on the whole of $M_1^{\otimes N}$ and we just need to present appropriate highest weight vectors in $M_1^{\otimes N}$.

\subsubsection{Universal symmetric module}
Obviously $|0\rangle$ itself is highest weight. It generates the module labelled by the symmetric representation, i.e. a Young diagram with 
a single row of length $N$. The vector $|0\rangle$ is annihilated by all generators $t[\cdots y]$ which end in $y$. It is an eigenvector of eigenvalue $0$ 
for all $t[(x y)^n]$ and $N$ for $t[1]$. A generic descendant involves a symmetric polynomial in the $x_i$ and can be built from vectors of the form $\prod_{m_i} t[x^{m_i}] |0\rangle$.

Notice that writing the relations in terms of $t[\cdots]$'s is the same as promoting the module to a module for $U(\mathrm{Diff}^{\epsilon_2}_{\bC})$. 
This sequence of symmetric modules are obviously all truncations of the same uniform-in-$N$ universal module 
defined by the same relations, except that the vector $|0\rangle$ has generic eigenvalue $\delta_0$ for $t[0]$. 

This universal module looks like the Fock space of a free chiral boson. The corner chiral algebra associated to a single
$M5$ brane is a $U(1)$ current algebra, so this universal symmetric module may be potentially associated to a surface defect 
built from a single $M5$ brane. 

Furthermore, this universal module does {\it not} appear to admit a good truncation when we set $\delta_0 =-N$, 
as the $t[(x y)^n]$ highest weight eigenvalues do not agree with these for a highest weight module of $\frac{W_{-\epsilon_2}^{\otimes N}}{S_N}$
with the transposition action. In conclusion, this universal module admits a truncation associated to $N$ $(1,0)$ M2 branes, 
but not to $N$ $(1,1)$ M2 branes. This is compatible with M2 branes ending on a single M5 brane wrapping $\bC_{\epsilon_1} \times \bC_{\epsilon_2}$.

We thus denote the ``universal symmetric module'' built from $\frac{W_{\epsilon_2}^{\otimes N}}{S_N}$ symmetric highest weights as
${\mathcal M}^{0,0,1}_{0,\epsilon_2}$. We will discuss momentarily the generalization to non-zero $\epsilon_1$. 
This will allow us to check that the truncations associated to $(0,1)$ M2 branes really exist. 

Similarly we can denote the universal symmetric module built from $\frac{W_{-\epsilon_2}^{\otimes N}}{S_N}$ symmetric highest weights as
${\mathcal M}^{0,1,0}_{0,\epsilon_2}$. We will discuss momentarily the generalization to non-zero $\epsilon_1$. 
This will allow us to check that the truncations associated to $(0,1)$ M2 branes really exist. 

\subsubsection{Other universal Verma modules}
At level $1$, $|0\rangle$ has a unique descendant $\sum_i x_i |0\rangle$. That means $N-1$ of the level $1$ vectors
will be highest weight. Indeed, a vector such as $(x_1 - x_2) |0\rangle$ must be annihilated by any lowering operator: there are no 
level $0$ vectors which are antisymmetric under the $(12)$ permutation. It will also never mix with other level $1$ vectors, as none 
is antisymmetric under the $(12)$ permutation. It generates a module we can label by a Young diagram with a column of height 2 and all others of 
height 1. 

We can continue this process. At level $2$ we can look at $(x_1 - x_2)(x_3 - x_4) |0\rangle$ generating a module we can label by a Young diagram with two columns of height 2 and all others of height 1. On the other hand, something like $(x_1 - x_2)(x_1 -x_3) |0\rangle$ is {\it not} highest weight, 
as it does not have good symmetry properties and indeed it can be lowered to $x_1 |0\rangle$. At level $3$ we can consider 
$(x_1 - x_2)(x_1 -x_3)(x_2-x_3) |0\rangle$ or $(x_1 - x_2)(x_3 - x_4)(x_5-x_6) |0\rangle$, etcetera. 

A Young diagram with columns $c_i$ will give a highest weight vector $|c_i \rangle$ such that 
\begin{align}
t[1]|c_i \rangle &= \left( \sum_i c_i \right) |c_i \rangle \cr
t[x y]|c_i \rangle &= - \epsilon_2 \left( \sum_i \frac{c_i(c_i-1)}{2} \right) |c_i \rangle \cr
t[x^2 y^2]|c_i \rangle &= (- \epsilon_2)^2 \left( \sum_i \frac{c_i(c_i-1)(c_i-2)}{3} \right) |c_i \rangle \cr
t[x^3 y^3]|c_i \rangle &= (- \epsilon_2)^3 \left( \sum_i \frac{c_i(c_i-1)(c_i-2)(c_i-3)}{4} \right) |c_i \rangle \cr
\cdots
\end{align}

An interesting case is the sequence of antisymmetric modules. Notice that any totally antisymmetric polynomial is the product of 
$\prod_{i<j}(x_i-x_j)$ and a symmetric polynomial. This is an isomorphism between antisymmetrized monomials and Schur polynomials. 
The highest weight eigenvalues are 
\begin{align}
t[1]|c_i \rangle &= N |c_i \rangle \cr
t[x y]|c_i \rangle &= - \epsilon_2 \frac{N(N-1)}{2} |c_i \rangle \cr
\cdots
\end{align}

We can lift these modules to an uniform-in $N$ module for $U(\mathrm{Diff}^{\epsilon_2}_{\bC})$, with highest weight eigenvalues
\begin{align}
t[1]|c_i \rangle &= \delta_0 |c_i \rangle \cr
t[x y]|c_i \rangle &= - \epsilon_2 \frac{\delta_0(\delta_0-1)}{2} |c_i \rangle \cr
\cdots
\end{align}
The module still looks like the Fock space of a chiral boson. On the other hand, something remarkable happens: 
the antisymmetric modules for $\frac{W_{-\epsilon_2}^{\otimes N}}{S_N}$ with the transposed action have precisely the same eigenvalues, 
with $\delta = -N$. 

This strongly suggests that the ``universal symmetric module'' built from $\frac{W_{\epsilon_2}^{\otimes N}}{S_N}$ 
anti-symmetric highest weights will have both truncations suitable to describe endpoints of $N$ $(1,0)$ and $N$ $(1,1)$ M2 branes
onto an M5 brane wrapping $\bC_{\epsilon_2} \times \bC_{-\epsilon_1 - \epsilon_2}$. We will denote it as ${\mathcal M}^{1,0,0}_{0,\epsilon_2}$. We will discuss momentarily the generalization to non-zero $\epsilon_1$. This will allow us to check that the truncations associated to $(0,1)$ M2 branes do not exist,
as required by the M5 brane interpretation. 

The obvious next step is to look at L-shaped Young diagrams, as a way to combine two types of M5 branes. 
Now the highest weight eigenvalues can be parameterized by two numbers 
\begin{align}
t[1]|c_i \rangle &=(\delta_0  + \delta_1) |c_i \rangle \cr
t[x y]|c_i \rangle &= - \epsilon_2 \frac{\delta_0(\delta_0-1)}{2} |c_i \rangle \cr
\cdots
\end{align}

When we seek a truncation to a module for $\frac{W_{\epsilon_2}^{\otimes N}}{S_N}$ we can set $\delta_0$, $\delta_1$  
to any non-negative numbers adding to $N$. When we seek a truncation to a module for $\frac{W_{-\epsilon_2}^{\otimes N}}{S_N}$
we seem to have to set $\delta_1=0$ and $\delta_0 =-N$. These choices are compatible with the expected M2 brane modules 
for the corner vertex algebra $Y^{1,0,1}$ \cite{Gaiotto:2017euk}: the $(1,0)$ M2 brane endpoints are expected to be labelled by $u(1|1)$ irreps associated to
L-shaped Young diagrams, while the $(1,1)$ M2 brane endpoints are expected to be labelled by $u(1)$ irreps with charge $-N$. 

We will discuss momentarily the truncations corresponding to $(0,1)$ endpoints. For now we can confidently 
label the universal module induced from L-shaped diagrams as ${\mathcal M}^{1,0,1}_{0,\epsilon_2}$.

This has an obvious generalization to the universal module ${\mathcal M}^{N_1,0,N_3}_{0,\epsilon_2}$ induced from ``fat L''-shaped diagrams,
with up to $N_1$ columns of arbitrary length and up to $N_3$ rows of arbitrary length. This has highest weight eigenvalues 
\begin{align}
t[1]|\delta_a, \tilde \delta_b;N_1,0,N_3 \rangle &= \left( \sum_{a=1}^{N_1} \delta_a + \sum_{b=1}^{N_3} b \tilde \delta_b \right)|\delta_a, \tilde \delta_b;{N_1},0,N_3 \rangle \cr
t[x y]|\delta_a, \tilde \delta_b;N_1,0,N_3 \rangle  &= - \epsilon_2 \left( \sum_{a=1}^{N_1} \frac{\delta_a(\delta_a-1)}{2} + \sum_{b=1}^{N_3} \frac{b(b-1)}{2} \tilde \delta_b \right) |\delta_a, \tilde \delta_b;N_1,0,N_3 \rangle  \cr
t[x^2 y^2]|\delta_a, \tilde \delta_b;N_1,0,N_3 \rangle  &= (- \epsilon_2)^2 \left( \sum_{a=1}^{N_1}\frac{\delta_a(\delta_a-1)(\delta_a-2)}{3}+\sum_{b=1}^{N_3} \frac{b(b-1)(b-2)}{3} \tilde \delta_b \right)|\delta_a, \tilde \delta_b;N_1,0,N_3 \rangle  \cr
\cdots
\end{align}
The different Young diagrams with this structure and $N$ boxes label precisely the $u(N|L)$ irreps which control the $Y^{N_1,0,N_3}$ chiral algebra modules associated to $N$ M2 branes of the $(1,0)$ type.  

At this point we do not seem to have a good strategy to identify the most general universal modules 
${\mathcal M}^{N_1,N_2,N_3}_{0,\epsilon_2}$ just by looking at $\frac{W_{\epsilon_2}^{\otimes N}}{S_N}$ Verma modules. 
We will come back to that after turning on $\epsilon_1$. 

It would be nice to compare in greater detail the universal modules we proposed here and the corresponding chiral modules for $Y^{N_1,0,N_3}$.
We will leave that to future work. 

\subsection{Highest weight modules ${\mathcal M}^{(N,Y)}_{\epsilon_1, \epsilon_2}$ for ${\mathcal A}^{(N)}_{\epsilon_1, \epsilon_2}$}
In order to identify the correct deformations of the $M_{Y;\epsilon_1}$ modules, we need a strategy to build highest weight modules for 
${\mathcal A}^{(N)}_{\epsilon_1, \epsilon_2}$. For generic $\epsilon_1,\epsilon_2$, these algebras are known to have 
a nice collection of Verma modules labelled by Young diagrams. 

Verma modules for the quantum Higgs or Coulomb branches have a very direct physical interpretations in terms of 
spaces of BPS states for the 3d theory in the presence of an appropriate $\Omega$ deformation and of certain masses and FI parameters
which make the theory gapped. This was discussed in detail for the quantum Coulomb branch algebra in \cite{Bullimore:2016hdc}. 
\footnote{Alternatively, one can use the constructions in \cite{Bullimore:2016nji} for ``exceptional Dirichlet boundary conditions''}
The construction depends on a choice of ``vacuum'' for the 3d theory, which can be thought of as an isometry fixed point on the Higgs branch of the theory. 

In order to proceed, we will thus review the relationship between vacua for the 3d theory and Young diagrams and 
then we will apply that information to present the Verma modules ${\mathcal M}^{(N,Y)}_{\epsilon_1, \epsilon_2}$ for 
${\mathcal A}^{(N)}_{\epsilon_1, \epsilon_2}$. We will compute the highest weight charges for these modules and demonstrate 
that ${\mathcal M}^{(N,Y)}_{\epsilon_1, \epsilon_2}$ is a deformation of $M_{Y;\epsilon_1}$ as $\epsilon_2 \to 0$ 
and of $M_{Y^t;\epsilon_1}$ as $\epsilon_2+ \epsilon_2 \to 0$.

\subsubsection{The classical massive vacua for the 3d ADHM gauge theory and associated Verma modules}
The classical massive vacua of the 3d gauge theory ADHM gauge theory are the fixed points of the classical
ADHM Higgs branch under the $U(1)$ isometry which maps $X \to \lambda X$, $Y \to \lambda^{-1} Y$. 

A vacuum is thus described by some vevs for the $X,Y,I,J$ fields which preserve a combination of the $U(1)$ flavour symmetry and 
some gauge symmetry. An embedding of the $U(1)$ flavour symmetry into the Cartan of the gauge symmetry can be described by 
a decomposition $\bC^N = \oplus_k \bC^{N_k}$, with $\bC^{N_k}$ being the subspace where the symmetry acts as $\lambda^k$. 

In such a sector, the symmetry-preserving vevs can only include maps $X : \bC^{N_k} \to \bC^{N_{k+1}}$, 
$ Y: \bC^{N_k} \to \bC^{N_{k-1}}$ and $I$, $J$ living in $\bC^{N_0}$. These fields can be arranged in a linear quiver, 
with a single flavour at the $0$-th node and FI parameters $\epsilon_2$ at each node. The expected dimension of 
the Higgs branch of such a quiver is 
\begin{equation}
2 N_0 - \sum_i (N_i - N_j)^2
\end{equation}
This is non-positive, and equals to $0$ if and only if the $N_i$ decrease monotonically away from $0$, with $|N_i - N_{i+1}| \leq 1$. 
Such a sequence of numbers can be identified with the lengths of diagonals of a Young diagram $Y$ with $N$ boxes. 

As long as the $N_i$ satisfy this constraint, the F-term relations have a unique solution modulo gauge transformations. 
Hence the vacua are labelled by Young diagrams with $N$ boxes. 

We can give examples for low values of $N$:
\begin{enumerate}
\item For $N=1$, the adjoint fields decouple and have zero vev at the unique vacuum. We can only have $N_0=1$ and solve the
F-term equation by a vev for $I$ and $J$. The linear quiver has a single $U(1)$ node at position $0$ and one flavour. It has indeed a Higgs branch 
which is a point for any value of the FI parameter. It is labelled by a Young diagram with a single box. 
\item For $N=2$, the maximum value of $N_0$ is $1$. The flavour symmetry can either be embedded in the gauge group Cartan as $(\lambda,1)$ or $(1,\lambda)$. 
The single component of $I$, $J$ on which that acts trivially gets a vev. The charge-0 off-diagonal components of $X$ and $Y$ also get vevs.
The linear quivers have two $U(1)$ nodes, one at position $0$ and one at positions $\pm 1$. They have indeed Higgs branches
which are a point for any value of the FI parameter. They are labelled by Young diagrams with two boxes, either in a row or in a column. 
\item For $N=3$ the maximum value of $N_0$ is still $1$. The three possible linear quivers have three $U(1)$ nodes, with the single flavour 
attached to either of the three. They are labelled by the three Young diagrams with three boxes.
\item For $N=4$ we have four fixed points with $N_0=1$ and one with $N_0=2$, $N_{1}=N_{-1}=1$. The former are labelled by the four L-shaped 
Young diagrams with four boxes. The latter is labelled by the $2 \times 2$ square Young diagram. 
\end{enumerate}

Because of the self-mirror properties of the original theory, we can also use the same data to label the fixed points in a Coulomb branch
description, with $\epsilon_2$ playing the role of the adjoint mass. This suggests looking at the quantum Coulomb branch algebra of these linear quivers. 
The algebra is built as usual from variables $w_{k;a}$, $v_{k;a}$, with monopole operators such as 
\begin{equation}
E_{k,t} = \sum_{a=1}^{N_k}\frac{ \prod_{b} w_{k,a} - w_{k-1,b} - \epsilon_2}{ \prod_{b \neq a}  (w_{k,a} - w_{k,b})} w_{k,a}^t v_{k,a}
\end{equation} 
and
\begin{equation}
F_{k,t} = \sum_{a=1}^N \frac{ \prod_{b} w_{k,a} - w_{k+1,b} + \epsilon_2}{ \prod_{b \neq a}  (w_{k,a} - w_{k,b})}  v^{-1}_{k,a} w_{k,a}^{t + \delta_{0,k}}
\end{equation} 
As the classical theory has a single vacuum, these algebras are expected to have a single Verma module, generated from vectors annihilated 
by all $F_{k,t}$.

We should start with some simple examples. For $N=1$ we have a single node with $N_0=1$ and monopoles $v_{0,0}$ and $v^{-1}_{0,0} w_{0,0}$. 
The obvious unique Verma is built from a highest weight vector $|0\rangle$ annihilated by $w_{0,0}$. The other vectors have eigenvalues $- n_{0,0} \epsilon_1$ 
for $w_{0,0}$. 

Next, we can look at $N=2$, $N_0=N_{-1}=1$. The monopoles are  
\begin{align}
E_{0,t} &= (w_{0,0} - w_{-1,0} - \epsilon_2) w_{0,0}^t v_{0,0} \cr
F_{0,t} &=  v^{-1}_{0,0} w_{0,0}^{t + 1} \cr
E_{-1,t} &=  w_{-1,0}^t v_{-1,0} \cr
F_{-1,t} &= (w_{-1,0} - w_{0,0} + \epsilon_2) v^{-1}_{-1,0} w_{-1,0}^{t}
\end{align} 
The highest weight vector should be annihilated by $w_{0,0}$ and by $(w_{-1,0} - w_{0,0} + \epsilon_1+ \epsilon_2)$,
and thus by $(w_{-1,0} + \epsilon_1+ \epsilon_2)$. We can denote it as $|- \epsilon_1 - \epsilon_2;0\rangle$. 

Acting with $E_{-1,t}$ we can raise the $w_{-1,0}$ eigenvalue by any $- n_{-1,0} \epsilon_1$. On the other hand, 
$E_{0,t}$ can raise $w_{0,0}$ by $- n_{0,0} \epsilon_1$ but only up to a point where $(w_{0,0} - w_{-1,0} - \epsilon_2- \epsilon_1)$
vanishes. In particular, $n_{0,0} \leq n_{-1,0}$. That condition cannot be violated by $F_{-1,t}$ either. 
Hence the vectors in the module have eigenvalues $w_{0,0} = - \epsilon_1 n_{0,0}$ and $w_{-1,0} = -\epsilon_1- \epsilon_2- \epsilon_1 n_{-1,0}$
with $n_{0,0} \leq n_{-1,0}$

A similar analysis applies to $N_0 = N_{-1} = N_{-2} \cdots = N_{1-N} =1$. 
The highest vector has $w_{-k,0} = -k(\epsilon_1 + \epsilon_2) - n_{-k,0} \epsilon_1$ 
with $n_{-k,0} \leq n_{-k-1,0}$.

Next, we can look at $N=2$, $N_0=N_1=1$. The monopoles are  
\begin{align}
E_{0,t} &= w_{0,0}^t v_{0,0} \cr
F_{0,t} &= (w_{0,0} - w_{1,0} + \epsilon_2) v^{-1}_{0,0} w_{0,0}^{t + 1} \cr
E_{1,t} &= (w_{1,0} - w_{0,0} - \epsilon_2) w_{1,0}^t v_{1,0} \cr
F_{1,t} &= v^{-1}_{1,1} w_{1,1}^{t}
\end{align} 
but this presentation is inconvenient. We can re-define $(w_{0,0} - w_{1,0} + \epsilon_2) v^{-1}_{0,0} =  \hat v^{-1}_{0,0}$
and $ v^{-1}_{1,1} = (w_{1,0} - w_{0,0} -\epsilon_1 -  \epsilon_2) \hat v^{-1}_{1,1}$ to get an alternative presentation  
\begin{align}
E_{0,t} &= (w_{0,0} - w_{1,0} + \epsilon_1+ \epsilon_2) w_{0,0}^t \hat v_{0,0} \cr
F_{0,t} &= \hat v^{-1}_{0,0} w_{0,0}^{t + 1} \cr
E_{1,t} &= w_{1,0}^t \hat v_{1,0} \cr
F_{1,t} &= (w_{1,0} - w_{0,0} -\epsilon_1 -  \epsilon_2) \hat  v^{-1}_{1,1} w_{1,1}^{t}
\end{align} 
Now we can define a highest weight module from a vector annihilated by $w_{0,0}$ and $(w_{1,0} - w_{0,0} -  \epsilon_2)$.
Again the factors in the $E$'s and $F$'s conspire to keep $n_{0,0} \leq n_{1,0}$ in the descendants, which have 
eigenvalues $w_{0,0} = - \epsilon_1 n_{0,0}$ and $w_{1,0} = \epsilon_2- \epsilon_1 n_{1,0}$.

A similar analysis applies to $N_0 = N_{1} = N_{2} \cdots = N_{N-1} =1$. 
The highest vector has $w_{k,0} = k \epsilon_2 - n_{k,0} \epsilon_1$ 
with $n_{k,0} \leq n_{k+1,0}$.

Our final example is $N_0=2$, $N_1=1$, $N_{-1}=1$, which illustrates the subtleties associated with 
the denominator factors, which prevent us from using conditions such as $w_{0,0}=w_{0,1} =0$ to 
define a highest weight vector. The monopoles are
\begin{align}
E_{-1,t} &=  w_{-1,0}^t v_{-1,0} \cr
F_{-1,t} &= (w_{-1,0} - w_{0,0} + \epsilon_2)(w_{-1,0} - w_{0,1} + \epsilon_2) v^{-1}_{-1,0} w_{-1,0}^{t} \cr
E_{0,t} &= \frac{w_{0,0} - w_{-1,0} - \epsilon_2}{w_{0,0}-w_{0,1}} w_{0,0}^t v_{0,0}+ \frac{w_{0,1} - w_{-1,0} - \epsilon_2}{w_{0,1}-w_{0,0}} w_{0,1}^t v_{0,1}\cr
F_{0,t} &=  \frac{w_{0,0} - w_{1,0} + \epsilon_2}{w_{0,0}-w_{0,1}}v^{-1}_{0,0} w_{0,0}^{t + 1}+ \frac{w_{0,1} - w_{1,0} + \epsilon_2}{w_{0,1}-w_{0,0}}v^{-1}_{0,1} w_{0,1}^{t + 1} \cr
E_{1,t} &= (w_{1,0} - w_{0,0} - \epsilon_2) (w_{1,0} - w_{0,1} - \epsilon_2) w_{1,0}^t v_{1,0} \cr
F_{1,t} &= v^{-1}_{1,1} w_{1,1}^{t}
\end{align} 
We need to reorganize some numerator factors to 
\begin{align}
E_{-1,t} &=  w_{-1,0}^t v_{-1,0} \cr
F_{-1,t} &= (w_{-1,0} - w_{0,0} + \epsilon_2)(w_{-1,0} - w_{0,1} + \epsilon_2) v^{-1}_{-1,0} w_{-1,0}^{t}\cr
E_{0,t} &= \frac{(w_{0,0} - w_{1,0} + \epsilon_1+ \epsilon_2)(w_{0,0} - w_{-1,0} - \epsilon_2)}{w_{0,0}-w_{0,1}} w_{0,0}^t \hat v_{0,0}+ \frac{w_{0,1} - w_{-1,0} - \epsilon_2}{w_{0,1}-w_{0,0}} w_{0,1}^t v_{0,1}\cr
F_{0,t} &=  \frac{1}{w_{0,0}-w_{0,1}}\hat  v^{-1}_{0,0} w_{0,0}^{t + 1}+ \frac{w_{0,1} - w_{1,0} + \epsilon_2}{w_{0,1}-w_{0,0}}v^{-1}_{0,1} w_{0,1}^{t + 1} \cr
E_{1,t} &=  (w_{1,0} - w_{0,1} - \epsilon_2) w_{1,0}^t \hat v_{1,0} \cr
F_{1,t} &=  (w_{1,0} - w_{0,0} -\epsilon_1 -  \epsilon_2) \hat v^{-1}_{1,1} w_{1,1}^{t}
\end{align} 
and then to 
\begin{align}
E_{-1,t} &=  (w_{-1,0} - w_{0,1} + \epsilon_1+ \epsilon_2) w_{-1,0}^t \hat v_{-1,0} \cr
F_{-1,t} &= (w_{-1,0} - w_{0,0} + \epsilon_2) \hat v^{-1}_{-1,0} w_{-1,0}^{t}\cr
E_{0,t} &= \frac{(w_{0,0} - w_{1,0} + \epsilon_1+ \epsilon_2)(w_{0,0} - w_{-1,0} - \epsilon_2)}{w_{0,0}-w_{0,1}} w_{0,0}^t \hat v_{0,0}
+ \frac{1}{w_{0,1}-w_{0,0}} w_{0,1}^t \hat v_{0,1}\cr
F_{0,t} &=  \frac{1}{w_{0,0}-w_{0,1}}\hat  v^{-1}_{0,0} w_{0,0}^{t + 1}+ \frac{(w_{0,1} - w_{-1,0} - \epsilon_2- \epsilon_1)(w_{0,1} - w_{1,0} + \epsilon_2)}{w_{0,1}-w_{0,0}} \hat v^{-1}_{0,1} w_{0,1}^{t + 1} \cr
E_{1,t} &=  (w_{1,0} - w_{0,1} - \epsilon_2) w_{1,0}^t \hat v_{1,0} \cr
F_{1,t} &=  (w_{1,0} - w_{0,0} -\epsilon_1 -  \epsilon_2) \hat v^{-1}_{1,1} w_{1,1}^{t}
\end{align} 
Now we can try to define a highest weight vector annihilated by $w_{0,0}$, $(w_{1,0} - w_{0,0} -  \epsilon_2)$, 
$(w_{-1,0} - w_{0,0} +\epsilon_1 + \epsilon_2)$ and both factors in $(w_{0,1} - w_{-1,0} - \epsilon_2)(w_{0,1} - w_{1,0} +\epsilon_1+ \epsilon_2)$.
It will satisfy $w_{0,0}=0$, $w_{1,0}=\epsilon_2$, $w_{-1,0} = -\epsilon_1-\epsilon_2$, $w_{0,1}= -\epsilon_1$.
This is still a bit dangerous, as the second term in $F_{0,t}$ has two zeroes at numerator and one at denominator. 
It should be OK, though, if we interpret the formulae in some limiting fashion, by shifting the eigenvalues by small amounts 
proportional to some $\eta$ and then sending $\eta \to 0$. 

The eigenvalues for a generic module element are thus 
\begin{equation}
w_{0,0}=- \epsilon_1 n_{0,0} \qquad w_{1,0}=\epsilon_2- \epsilon_1 n_{1,0} \qquad w_{-1,0} = -\epsilon_1-\epsilon_2- \epsilon_1 n_{-1,0}
\qquad w_{0,1}= -\epsilon_1 - \epsilon_1 n_{0,1}
\end{equation}

For general $N_a$, we will use the same re-definition to shift around factors associated to pairs of adjacent boxes in the same row, 
define a highest weight module with $w_{k,a}$ which equals $\epsilon_2 c -(\epsilon_1 + \epsilon_2)r$, where $(c,r)$ indicate which column and which row does the 
corresponding box belong to. Other vectors in the module have eigenvalues 
\begin{equation}
w_{k,a}= - \epsilon_1 n_{k,a} + \epsilon_2 c -(\epsilon_1 + \epsilon_2)r
\end{equation} 
with $n_{k,a}$ non-decreasing both along rows and columns. 

We can now define a Verma module ${\mathcal M}^{(N,Y)}_{\epsilon_1, \epsilon_2}$ for ${\mathcal A}^{(N)}_{\epsilon_1, \epsilon_2}$.
We split the $w_a$ and $v_a$ into groups of $N_k$, take the same basis of $|n_{k,a} \rangle$ with the same restrictions
on the $n_{k,a}$ and eigenvalues as above, do the same re-definition $v_{k,a} = \frac{\cdots}{\cdots} \hat v_{k,a}$ to shift around factors associated to pairs of adjacent boxes in the same row.  

All the ``dangerous'' numerator and denominator factors which can become $0$ on this basis of $|n_{k,a} \rangle$ 
are these already contained in the $E_{k,t}$, $F_{k,t}$ and thus no further problems can occur. 

The lowest weight vector with the smallest possible values of $n_{k,a}$ will be annihilated by all $F_k$ generators 
and have eigenvalues 
\begin{equation}
w_{k,a}= \epsilon_2 c -(\epsilon_1 + \epsilon_2)r
\end{equation} 
In the limit $\epsilon_2 \to 0$ these are the same eigenvalues as we encountered for $M_{Y;\epsilon_1}$ and the module seems to have the same structure. 
We expect ${\mathcal M}^{(N,Y)}_{\epsilon_1, \epsilon_2}$ to be a deformation of $M_{Y;\epsilon_1}$ as $\epsilon_2 \to 0$. 
The construction is perfectly symmetric in $\epsilon_2$ and $- \epsilon_1 - \epsilon_2$, up to a transposition of the Young diagram. 
We thus expect ${\mathcal M}^{(N,Y)}_{\epsilon_1, \epsilon_2}$ to be a deformation of $M_{Y^t;\epsilon_1}$ as $\epsilon_1 +\epsilon_2 \to 0$. 

\subsubsection{Highest weight charges}
We can write an explicit expression for the $H(z)$ generating function acting on the highest weight vector: 
\begin{equation}
H_Y(z) =\prod_{(c,r) \in Y} \frac{(z-\epsilon_2 (c+1) +(\epsilon_1 + \epsilon_2)r)(z-\epsilon_2 c +(\epsilon_1 + \epsilon_2)(r+1)}{(z-\epsilon_2 c +(\epsilon_1 + \epsilon_2)r)(z-\epsilon_2 (c+1) +(\epsilon_1 + \epsilon_2)(r+1))}
\end{equation}
There is obviously a lot of scope for telescopic cancellations. 

We can collect contributions from within each column of a diagram with $N_2$ columns
\begin{equation}
H_Y(z) =\frac{z-N_2 \epsilon_2}{z}\prod_{i=0}^{N_2-1} \frac{z-\epsilon_2 i +(\epsilon_1 + \epsilon_2)c_i}{z-\epsilon_2 (i+1) +(\epsilon_1 + \epsilon_2)c_i}
\end{equation}
or within each row of a diagram with $N_3$ rows
\begin{equation}
H_Y(z) =\frac{z+N_3(\epsilon_1+ \epsilon_2)}{z}\prod_{j=0}^{N_3-1} \frac{z+(\epsilon_1+\epsilon_2) j - \epsilon_2 r_j}{z+(\epsilon_1+\epsilon_2) (j+1) - \epsilon_2 r_j}
\end{equation}
or even mix it up for a diagram fitting within a fat-L: 
\begin{equation}
H_Y(z) =\frac{z-\epsilon_2 N_2 +(\epsilon_1 + \epsilon_2)N_3}{z}\prod_{i=0}^{N_2-1} \frac{z-\epsilon_2 i +(\epsilon_1 + \epsilon_2)c_i}{z-\epsilon_2 (i+1) +(\epsilon_1 + \epsilon_2)c_i}\prod_{j=0}^{N_3-1} \frac{z+(\epsilon_1+\epsilon_2) j - \epsilon_2 r_j}{z+(\epsilon_1+\epsilon_2) (j+1) - \epsilon_2 r_j}
\end{equation}

We can now abstract the column and row lengths into the parameters of a conjectural uniform-in-$N$ module 
${\mathcal M}^{(0,N_2,N_3)}_{\epsilon_1, \epsilon_2}$ associated to two stacks of $M5$ branes, with highest weight charges
\begin{align}
H_{0,N_2,N_3}(z) &=\frac{z-\epsilon_2 N_2 +(\epsilon_1 + \epsilon_2)N_3}{z}\prod_{i=0}^{N_2-1} \frac{z-\epsilon_2 i +\epsilon_1 (\epsilon_1 + \epsilon_2)\delta_i}{z-\epsilon_2 (i+1) +\epsilon_1 (\epsilon_1 + \epsilon_2)\delta_i} \cr &\cdot\prod_{j=0}^{N_3-1} \frac{z+(\epsilon_1+\epsilon_2) j - \epsilon_1 \epsilon_2 \delta'_j}{z+(\epsilon_1+\epsilon_2) (j+1) - \epsilon_1 \epsilon_2 \delta'_j}
\end{align}
and even invoke triality to get conjectural  highest weight charges for a module ${\mathcal M}^{(N_1,N_2,N_3)}_{\epsilon_1, \epsilon_2}$ associated to three stacks of $M5$ branes:
\begin{align}
H_{N_1,N_2,N_3}(z) &=\frac{z-\epsilon_1 N_1 -\epsilon_2 N_2 +(\epsilon_1 + \epsilon_2)N_3}{z}\prod_{i=0}^{N_1-1} \frac{z-\epsilon_1 i +\epsilon_2 (\epsilon_1 + \epsilon_2)\delta_i}{z-\epsilon_1 (i+1) +\epsilon_2 (\epsilon_1 + \epsilon_2)\delta_i} \cr &\prod_{j=0}^{N_2-1} \frac{z-\epsilon_2 j +\epsilon_1 (\epsilon_1 + \epsilon_2)\delta'_j}{z-\epsilon_2 (j+1) +\epsilon_1 (\epsilon_1 + \epsilon_2)\delta'_j}\cdot\prod_{k=0}^{N_3-1} \frac{z+(\epsilon_1+\epsilon_2) k - \epsilon_1 \epsilon_2 \delta''_k}{z+(\epsilon_1+\epsilon_2) (k+1) - \epsilon_1 \epsilon_2 \delta''_k}
\end{align}

Based on the pattern we have seen until now, we propose that the truncation of such a module compatible with a truncation 
of the universal algebra to the case of $n_i$ stacks of $M2$ branes will correspond to the following specialization of parameters: 
\begin{equation}
\delta_i = \frac{m_{12,i}}{\epsilon_2} + \frac{m_{13,i}}{\epsilon_3} \qquad \qquad \delta'_j = \frac{m_{21,j}}{\epsilon_1} + \frac{m_{23,j}}{\epsilon_3}
\qquad \qquad \delta''_k = \frac{m_{31,k}}{\epsilon_1} + \frac{m_{32,i}}{\epsilon_2}
\end{equation}
Then we can build a $\mathfrak{u}(N_2|N_3)$ Young diagram with $N_2$ rows and $N_3$ columns of lenghts $m_{21,j}$ and $m_{31,k}$,
a $\mathfrak{u}(N_1|N_2)$ Young diagram with $N_1$ rows and $N_2$ columns of lenghts $m_{13,i}$ and $m_{23,j}$ and a
$\mathfrak{u}(N_1|N_3)$ Young diagram with $N_1$ rows and $N_3$ columns of lenghts $m_{12,i}$ and $m_{32,k}$.

That is precisely the data labelling the special ``degenerate'' modules of the corner vertex algebra $Y^{N_1,N_2,N_3}_{\epsilon_1,\epsilon_2,\epsilon_3}$
 which are expected to be associated to three stacks of $M2$ branes ending onto three stacks of $M5$ branes!
 
 We leave a precise definition of these universal modules to future work. 
 
 \subsection{Comparison with surface defect observables}
As discussed at the beginning of this section, the universal modules ${\mathcal M}^{(N_1,N_2,N_3)}_{\epsilon_1, \epsilon_2}$
should be dual, as vector spaces, to the space of observables for the surface defect world-volume theory near the origin. 

We thus expect it to take the form of some sort of direct sum of all the ``degenerate'' modules for the corner vertex algebra $Y^{N_1,N_2,N_3}_{\epsilon_1,\epsilon_2,\epsilon_3}$,
including the vacuum module. Although an analysis of this proposal would bring us too far from the scope of the current paper, we should at least stress some non-trivial aspects of it,
as a motivation for the rest of the paper. 

\subsubsection{Fermion VOA}
For example, consider ${\mathcal M}^{(1,0,0)}_{\epsilon_1, \epsilon_2}$. In a weak-coupling $\epsilon_1 \to 0$ limit, the 
corner VOA $Y^{1,0,0}_{\epsilon_1,\epsilon_2,\epsilon_3}$ reduces to the charge $0$ sector of a theory of a single complex chiral fermion. 
The degenerate modules add up to the full chiral complex fermion vacuum module. We should thus be able to find a 
basis in ${\mathcal M}^{(1,0,0)}_{\epsilon_1, \epsilon_2}$ deforming a free fermion basis 
\begin{equation}
\prod_i \psi_{- m_i-\frac12} \prod_j \psi^\dagger_{- n_j-\frac12} |0 \rangle \qquad \qquad m_i<m_{i+1} \qquad n_j<n_{j+1}
\end{equation}
so that the ${\mathcal A}_{\epsilon_1, \epsilon_2}$ action deforms the natural action of $U(\mathrm{Diff}^{\epsilon_2}_{\bC})$ 
on the fermionic Fock space given fermion bilinears
\begin{equation}
t[x^n y^m] \to C_n \delta_{n+m,0}+ (-\epsilon_2)^m \sum_{k} {k \choose m}\psi_{-n+m-k-\frac12} \psi^\dagger_{k+\frac12}
\end{equation}
with appropriate normal-ordering constants $C_n$ for $n=m$ which act on $\psi(z_1)$ and 
$\psi^\dagger(z_1)$ by the classical non-commutative $U(1)$ gauge transformations.

For comparison with the Coulomb branch description of the algebra, we can specialize to
\begin{align}
t[(xy)^n x ] & \to (-\epsilon_2)^n \sum_{k} k^n \psi_{-k-\frac12} \psi^\dagger_{k-\frac12} \cr
t[(xy)^n] & \to C'_n \delta_{n+m,0}+ (-\epsilon_2)^n \sum_{k} k^n \psi_{-k-\frac12} \psi^\dagger_{k+\frac12} \cr
t[y (xy)^n] & \to (-\epsilon_2)^{n+1} \sum_{k} k^{n+1} \psi_{-k+\frac12} \psi^\dagger_{k+\frac12} 
\end{align}

Unfortunately, at this point we can only study in detail the truncations of the algebra corresponding to $(1,0)$ or $(0,1)$ 
M2 branes, which we expect to force a truncation of ${\mathcal M}^{(1,0,0)}_{\epsilon_1, \epsilon_2}$ to 
the submodule built respectively from $\psi$'s only or to $\psi^\dagger$'s only, to be matched to the 
Verma modules associated to the Young diagram with a single column. 

It is still interesting to look at some details for the $(1,0)$ M2 brane truncation. The module consists of vectors $|n_a\rangle$ with eigenvalues 
\begin{equation}
w_{a} |n_a\rangle=( - \epsilon_2 n_{a} -(\epsilon_1 + \epsilon_2) a)|n_a\rangle
\end{equation} 
Notice that we are using the algebra with quantization parameter $\epsilon_2$ and deformation parameter $\epsilon_1$ here. 
It is natural to identify 
\begin{equation} \label{eq:idfer}
|n_a\rangle = \prod_{a=1}^N \psi_{-n_a -a-\frac12}|0\rangle
\end{equation}

The action of the $E_n$, $W_n$ and $F_n$ generators then approaches that of $t[(xy)^n x ]$,$t[(xy)^n]$,$t[y (xy)^n]$
as $\epsilon_1 \to 0$. The leading corrections depend on the value of pairs of $n_a$'a. They thus give leading corrections to 
$t[x^n y^m]$ which involve four fermions. It would be nice to compute them in detail and compare with 5d perturbation theory calculations. 

It may also be useful to modify systematically the identification \ref{eq:idfer}, for example by building from the vacuum the unique  
basis on which the deformed $t[x^n]$ we defined in Section \ref{sec:algebra} act without corrections. We leave this to future work.

\subsubsection{Beta-Gamma VOA}
Next, consider ${\mathcal M}^{(0,1,1)}_{\epsilon_1, \epsilon_2}$. In a weak-coupling $\epsilon_1 \to 0$ limit, the 
corner VOA $Y^{0,1,1}_{\epsilon_1,\epsilon_2,\epsilon_3}$ reduces to the charge $0$ sector of a theory of a single $\beta \gamma$ system. 
The degenerate modules include the full chiral $\beta \gamma$ vacuum module and all the spectral flowed images, 
where $\beta$ has a pole of fixed order and $\gamma$ a zero of the same order. 

We should thus be able to find a 
basis in ${\mathcal M}^{(0,1,1)}_{\epsilon_1, \epsilon_2}$ deforming the free $\beta \gamma$ basis 
\begin{equation}
\prod_i \gamma_{- m_i-\frac12} \prod_j \beta_{- n_j-\frac12} |t \rangle \qquad m_i<m_{i+1} \qquad n_j<n_{j+1} \qquad m_i\geq t \qquad n_j \geq -t
\end{equation}
so that the ${\mathcal A}_{\epsilon_1, \epsilon_2}$ action deforms the natural action of $U(\mathrm{Diff}^{\epsilon_2}_{\bC})$ 
on the fermionic Fock space given fermion bilinears
\begin{equation}
t[x^n y^m] \to \tilde C_n \delta_{n+m,0}+ (-\epsilon_2)^m \sum_{k} {k \choose m}\gamma_{-n+m-k-\frac12} \beta_{k+\frac12}
\end{equation}
with appropriate normal-ordering constants $\tilde C_n$ for $n=m$, which act on $\gamma(z_1)$ and 
$\beta(z_1)$ by the classical non-commutative $U(1)$ gauge transformations.

We should be able to study the module particularly well in the truncations of the algebra corresponding to $N$ $(0,1)$
M2 branes. Ideally, we should be able to describe the relevant modules as modules for the ADHM algebra. We will not do so in this paper. 
Instead, we will describe candidates for the bi-modules ${\mathcal B}^{(0,1,1)}_{\epsilon_1, \epsilon_2}$

%
%

\section{Bimodules}\label{sec:bimodule}

It should also be possible to consider situations where some or none of the M2 branes actually end on the M5 branes. As we discussed in the beginning of the previous Section \ref{sec:module},
this should correspond to some universal bi-module ${\mathcal B}_{\epsilon_1, \epsilon_2}$ for ${\mathcal A}_{\epsilon_1, \epsilon_2}$, possibly admitting specializations to bi-modules 
for pairs of algebras ${\mathcal A}^{(N_1,N_2,N_3)}_{\epsilon_1, \epsilon_2}$, ${\mathcal A}^{(N'_1,N'_2,N'_3)}_{\epsilon_1, \epsilon_2}$.

In this section, we will mostly focus on the bi-module ${\mathcal B}^{(0,1,1)}_{\epsilon_1, \epsilon_2}$, using a IIA brane construction to get interesting candidate
bi-modules for the ADHM algebra and then working uniformly in $N$ to get a corresponding bimodule for ${\mathcal A}_{\epsilon_1, \epsilon_2}$.

The basic idea is that in type IIA we are studying a D2-D4-D6 system, with $N$ D2 branes lying on a D6 brane and a single D4 brane intersecting the $D2$'s along 
the $\Omega$-deformed complex plane and the $D6$ along the $\Omega$-deformed complex plane and one of the two complex directions in $\bR \times \bC^2$. 
All the branes and on a single NS5 brane. 

The $4-6$ strings give a single 4d hypermultiplet, which upon $\Omega$ deformation maps to the $\beta \gamma$ system on the surface defect world-volume. 
The $2-4$ strings give $N$ 2d hypermultiplets, which upon $\Omega$ deformation map to $0d$ degrees of freedom $\varphi$, $\tilde \varphi$ 
transforming as (anti)fundamentals of the $U(N)$ gauge group of the ADHM quantum mechanics. Comparison with other examples of 
``D-branes ending on NS5-branes'' suggests the presence of a variety of super-potential couplings, giving a 0d path integral with action  
\begin{equation}
e^{\frac{1}{\epsilon_1} \left[\varphi X \tilde \varphi  + I \beta(Y) \tilde \varphi  + \varphi  \gamma(Y) J \right]}
\end{equation}
where the ADHM fields are restricted to $t=0$ and we tentatively adjusted the couplings so that the $\beta \gamma$ fields are restricted to the location of the D2 branes
in the transverse directions parameterized by the eigenvalues of $Y$. 

As we focus on the ADHM world-volume theory, the three superpotential terms play distinct roles. We will see momentarily that the 
first term defines a bimodule for the ADHM algebra with highest weight properties. We will interpret it as a truncation of 
${\mathcal B}^{(0,1,1)}_{\epsilon_1, \epsilon_2}$. The two remaining terms indicate how to couple the modes of the $\beta \gamma$ 
system to appropriate elements in the bi-module into a gauge-invariant junction. 

\subsection{An interesting Moyal bi-module}
In order to understand bi-modules associated to a superpotential, we can begin with a single copy of $W_{\epsilon_1}$. 

There is an interesting class of modules for $\mathrm{Diff}^{\epsilon_1}_{\bC}$ associated to a choice of polynomial ``superpotential'' $W(y,\phi)$, 
generators of the form $p(\phi) q(x,y)$ and relations 
\begin{align}
y p(\phi) q(x,y) &= p(\phi) y q(x,y) \cr
x p(\phi) q(x,y) &= p(\phi) x q(x,y)+  p(\phi) \partial_y W(y,\phi)q(x,y) \cr 
\partial_\phi W(y,\phi) p(\phi) q(x,y) &= - \epsilon_1 \partial_\phi p(\phi) q(x,y)
\end{align}
Such a bi-module behaves much as the space of integral expressions of the form 
\begin{equation}
\int p(\phi) e^{\frac{W(y,\phi)}{\epsilon_1}} d\phi \, q(x,y).
\end{equation}

If we pick $W = \varphi y \tilde \varphi$ the relations become 
\begin{align}
y p(\varphi,\tilde \varphi) q(x,y) &= p(\varphi, \tilde \varphi) y q(x,y) \cr
x p(\varphi,\tilde \varphi)  q(x,y) &= p(\varphi,\tilde \varphi)  x q(x,y)+  \varphi \tilde \varphi p(\varphi,\tilde \varphi)  q(x,y) \cr 
y \varphi  p(\varphi,\tilde \varphi)  q(x,y) &= - \epsilon_1 \partial_{\tilde \varphi}  p(\varphi,\tilde \varphi) q(x,y) \cr 
y \tilde \varphi  p(\varphi,\tilde \varphi)  q(x,y) &= - \epsilon_1 \partial_\varphi  p(\varphi,\tilde \varphi) q(x,y) 
\end{align}

Because 
\begin{equation}
\varphi y \tilde \varphi p(\varphi,\tilde \varphi) =  - \epsilon_1 \tilde \varphi\partial_{\tilde \varphi}  p(\varphi,\tilde \varphi) =  - \epsilon_1 \varphi\partial_{\varphi}  p(\varphi,\tilde \varphi)
\end{equation}
it follows that we can restrict ourselves to polynomials in $\varphi \tilde \varphi$. A basis for the module consists of vectors 
$x^n y^m$ with $m>0$ and $(\varphi \tilde \varphi)^{n+1} x^m$. 

The element ``1'' in this bimodule behaves much as the function 
\begin{equation}
\int d\varphi d \tilde \varphi e^{\frac{1}{\epsilon_1} \varphi y \tilde \varphi} = 1/y
\end{equation}.

In order to avoid confusion, denote as $B$ the basis element ``1'' in the bimodule.
Then we have an alternative basis of the form $y B x^n y^m$ and $x^m B x^n$. Notice that 
\begin{equation}
x B y = B x y + \varphi \tilde \varphi B y =B(x y -\epsilon_1) = B y x 
\end{equation}
so $yB$ commutes with $x$ and $y$ and $y B x^n y^m$ give a copy of the algebra itself as a module. On the other hand, 
\begin{align}
y x^m B x^n &= -m \epsilon_1 x^{m-1} B x^n + y B x^{n+m} \cr
x^m B x^n y &= n \epsilon_1 x^{m} B x^{n-1} + y B x^{n+m} 
\end{align}
which shows clearly that the bimodule is an extension built from the algebra itself and the product of two Verma modules. 

Finally, notice that the bimodule has another useful interpretation. If we embed the Heisenberg algebra in the shift algebra by 
$y = v$, $x = v^{-1} w$, then the bi-module is simply the shift algebra itself. We can identify $( \varphi \tilde \varphi )^n$ with a multiple of $v^{-n-1}$. 
This makes precise the intuitive relation between $B$ and $\frac{1}{y}$.

We could have attempted to define a more general bi-module by including $n_f$ sets of auxiliary fields, $\varphi^i$ and $\tilde \varphi_j$.
Acting with $y \varphi^i \tilde \varphi_j$, thought, shows immediately that only powers of the rotation-invariant contraction $\varphi^i \tilde \varphi_i$
can be non-zero in the bimodule. Furthermore, the element $B_{n_f} = ``1''$ will behave just like $(\overrightarrow x- \overleftarrow x)^{n_f-1} \circ B$ in the module defined with a single set of auxiliary variables, 
as we can obtain $y^{- n_f}$ from derivatives of $y^{-1}$. As a result, we get the same bi-module, with a different choice of ``origin'' $B$. 

Another interesting variation is to replace $\varphi$ and $\tilde \varphi$ with fermionic variables $\chi$ and $\tilde \chi$.
This leads to naively minor sign changes, but now $``1''$ behaves as the function $y$ rather than $y^{-1}$. The element $\chi \tilde \chi$ 
commutes with both $x$ and $y$ and there are no $(\chi \tilde \chi)^n$ elements which could behave as negative powers of $y$. 
The whole bimodule is isomorphic to the Weyl algebra itself.

\subsection{An interesting ADHM$_N$ bi-module}

We can readily generalize this bi-module to all $N$, by picking $\varphi$ and $\tilde \varphi$ to be (anti)fundamentals under the gauge group and 
superpotential $W = \varphi Y \tilde \varphi$. After quantum symplectic reduction, 
this defines a bi-module for the ADHM operator algebra, generated by gauge-invariant polynomials of the form 
$p(\varphi,\tilde \varphi) q(X,Y,I,J)$ with the same number of $\varphi$ and $\tilde \varphi$ in each monomial. 
Intuitively, the element $B \equiv ``1''$ in the bimodule represents the function $\frac{1}{\det Y}$. 

It is easy to show that the bi-module relations and F-term relations (as long as $\epsilon_2 \neq 0$) can be used
to simplify any basis element to a polynomial in open words $I X^n Y^m J$, $I X^n \tilde \varphi$ and $\varphi X^n J$.
These manipulations can be done in an uniform-in-$N$ manner, to be followed by imposing explicit trace relations. 

As a consequence, this defines a bimodule ${\mathcal B}^{0,1,1}_{0;\epsilon_1, \epsilon_2}$ for ${\mathcal A}_{\epsilon_1, \epsilon_2}$.
This is not fully universal, in the sense that its elements all commute with the central element $t[1]$ of  ${\mathcal A}_{\epsilon_1, \epsilon_2}$,
but we expect it to be as close to being universal as possible given that constraint. It should be the bi-module controlling all possible ways a set of M2 branes can cross 
the surface defect without ending on it. 

At the very least, it has the correct number of generators, as we can identify roughly $I X^n \tilde \varphi$ and $\varphi X^n J$ with generators $a[n]$ and $b[n]$
dual to the Laurent modes of the $\beta$ and $\gamma$ fields. 
Indeed, if we normalize
\begin{align}
t[p(x,y)] &\sim \frac{1}{\epsilon_1} I p(X,Y) J \cr
b[z^n] &\sim \frac{1}{\epsilon_1} \varphi X^n J \cr
a[z^n] &\sim \frac{1}{\epsilon_1} I X^n \tilde \varphi 
\end{align} 
then the commutators in the leading order in $\epsilon_1 \to 0$ will only receive contributions from $I,J$ commutators and 
reproduce the action of $\mathrm{Diff}^{\epsilon_1}_{\bC}$ which is precisely appropriate to get a gauge-invariant coupling to the 
modes of a $\beta \gamma$ system!

This is compatible with ${\mathcal B}^{0,1,1}_{0;\epsilon_1, \epsilon_2}$ being the charge $0$ piece of ${\mathcal B}^{0,1,1}_{\epsilon_1, \epsilon_2}$.

\subsection{Multiple flavours and fermionic flavours in the uniform-in-$N$ limit}
We can generalize the construction to bimodules defined by $n_b$ sets of auxiliary fields $\varphi^i$ and $\tilde \varphi_j$.
Something quite remarkable happens. Although for each $N$ we are likely to obtain the same bi-module, the uniform-in-$N$ 
presentation definitely depends on $n_b$. Intuitively, the relations between different powers of $\frac{1}{\det Y}$ operator
do not have an uniform-in-$N$ description.

Concretely, the bi-module will contain $U(n_b)$-invariant combinations of words $I X^n Y^m J$, $I X^n \tilde \varphi_i$ and $\varphi^i X^n J$. It is suitable to 
describe an intersection with a surface defect supporting $n_f$ copies of a $\beta \gamma$ system. The bi-module should be the charge $0$ piece of ${\mathcal B}^{0,n_b,n_b}_{\epsilon_1, \epsilon_2}$.

The tension between determinants and traces in the  uniform-in-$N$ limit becomes even sharper if we consider the variant with one set of fermionic auxiliary fields 
$\chi$ and $\tilde \chi$. Although the bimodule at finite $N$ is isomorphic to the algebra, with $``1'' \to \det Y$, the isomorphism is not compatible with the 
uniform-in-$N$ presentation. Instead, we will get $U(1)$-invariant combinations of words $I X^n Y^m J$, $I X^n \tilde \chi_i$ and $\chi^i X^n J$, suitable to 
describe an intersection with a surface defect supporting a chiral complex fermion. 

This is a perfectly sensible candidate for the charge-0 truncation of ${\mathcal B}^{1,0,0}_{\epsilon_1, \epsilon_2}$. Indeed, the corresponding M5 brane maps to IIA 
to a D4 brane which is transverse to the D2 brane in eight directions. The $2-4$ strings are naturally fermionic \cite{Costello:2016nkh}.
More generally, $n_b$ auxiliary bosons and $n_f$ auxiliary fermions should give a charge-0 truncation ${\mathcal B}^{n_f,n_b,n_b}_{0;\epsilon_1, \epsilon_2}$ of ${\mathcal B}^{n_f,n_b,n_b}_{\epsilon_1, \epsilon_2}$.
 
\subsection{Universal Bimodule form Coulomb branch algebras}

Although we do not have an explicit quantum mirror map, the operators $\det X$ and $\det Y$ are very natural candidates for monopole operators
built from the central $U(1)$ in the $U(N)$ gauge group of the ADHM theory. As a consequence, they should take the form 
$\prod_a x_a$ and $\prod_a y_a$ in the quantum Coulomb branch algebra. 

For convenience, lets embed ${\mathcal C}^{(N)}_{\epsilon_1,\epsilon_2}$ in ${\mathcal C}^{(N)}_{0;\epsilon_1,\epsilon_2}$
in a slightly different way than before, setting $x_a = w_a \hat v^{-1}_a$ and $y_a = \hat v_a$. Then we expect the 
generator $B$ of the bi-module for $n_b=1$, $n_f=0$ to behave as $\prod_a \hat v^{-1}_a$, which is an element in 
 ${\mathcal C}^{(N)}_{0;\epsilon_1,\epsilon_2}$!
 
As far as we can see, the action of ${\mathcal C}^{(N)}_{\epsilon_1,\epsilon_2}$ extends this identification to 
a full identification between the  $n_b=1$, $n_f=0$ bi-module and ${\mathcal C}^{(N)}_{0;\epsilon_1,\epsilon_2}$,
seen as a ${\mathcal C}^{(N)}_{\epsilon_1,\epsilon_2}$ bimodule by the above embedding ${\mathcal C}^{(N)}_{\epsilon_1,\epsilon_2} \subset {\mathcal C}^{(N)}_{0;\epsilon_1,\epsilon_2}$.
 
As we try to lift this description to a  ${\mathcal C}_{\epsilon_1,\epsilon_2}$ bimodule, we cannot quite talk of $\prod_a \hat v^{-1}_a$ ad an element of 
${\mathcal C}^{(N)}_{0;\epsilon_1,\epsilon_2}$. Instead, we can observe that commuting across $\prod_a \hat v^{-1}_a$ simply has the effect of shifting 
al the $w_a$ by $\epsilon_1$. As a consequence, we can introduce a twisted ${\mathcal C}^{(N)}_{0;\epsilon_1,\epsilon_2}$ bimodule action on 
${\mathcal C}^{(N)}_{0;\epsilon_1,\epsilon_2}$ where the left action is twisted by the automorphism $w_a \to w_a + \epsilon_1$ 
of ${\mathcal C}^{(N)}_{0;\epsilon_1,\epsilon_2}$.

This description can be lifted to the definition of a twisted ${\mathcal C}_{0;\epsilon_1,\epsilon_2}$ bimodule action on 
${\mathcal C}_{0;\epsilon_1,\epsilon_2}$ where the left action is twisted by the automorphism 
\begin{equation}
E[p(w)] \to E[p(w+\epsilon_1)] \qquad \qquad W[p(w)] \to W[p(w+\epsilon_1)] \qquad \qquad F[p(w)] \to F[p(w+\epsilon_1)]
\end{equation} 
of ${\mathcal C}_{0;\epsilon_1,\epsilon_2}$, where the $E[p(w)]$ are built linearly from $E[w^k]\equiv E_k$, etcetera. 

This can be combined with the embedding 
\begin{equation}
E_k \to E_{k+1}  
\end{equation}
of ${\mathcal C}_{\epsilon_1,\epsilon_2}$ in ${\mathcal C}_{0;\epsilon_1,\epsilon_2}$ to build our candidate Coulomb branch 
description of  ${\mathcal B}^{0,1,1}_{0;\epsilon_1, \epsilon_2}$.

If we repeat the analysis for the $n_b=1$, $n_f=1$ case, the candidate Coulomb branch 
description of  ${\mathcal B}^{1,1,1}_{0;\epsilon_1, \epsilon_2}$ is ${\mathcal C}_{0;\epsilon_1,\epsilon_2}$ itself as an ${\mathcal C}_{\epsilon_1,\epsilon_2}$
bi-module, without any twisting. This is manifestly triality invariant, just as expected from ${\mathcal B}^{1,1,1}_{0;\epsilon_1, \epsilon_2}$!

More generally, we could tentatively propose that ${\mathcal B}^{N_1,N_2,N_3}_{0;\epsilon_1, \epsilon_2}$ could be obtained by twisting the ${\mathcal C}_{0;\epsilon_1,\epsilon_2}$ bimodule action on 
${\mathcal C}_{0;\epsilon_1,\epsilon_2}$ where the left action is twisted by the automorphism 
\begin{align}
E[p(w)] \to E[p(w+N_1 \epsilon_1+N_2 \epsilon_2+N_3 \epsilon_3)] \cr
 W[p(w)] \to W[p(w+N_1 \epsilon_1+N_2 \epsilon_2+N_3 \epsilon_3)] \cr
 F[p(w)] \to F[p(w+N_1 \epsilon_1+N_2 \epsilon_2+N_3 \epsilon_3)]
\end{align} 
of ${\mathcal C}_{0;\epsilon_1,\epsilon_2}$. 

We should mention a possible caveat to these statements. A shift $N_i \to N_i + 1$ does not appear to change the structure of ${\mathcal C}_{0;\epsilon_1,\epsilon_2}$ as a bi-module. 
It is possible that ${\mathcal C}_{0;\epsilon_1,\epsilon_2}$ with the action shifted by $N_1 \epsilon_1+N_2 \epsilon_2+N_3 \epsilon_3$ should be instead interpreted as some kind of super-universal bimodule
which describe intersections of M2 branes with a collection of M5 branes with given magnetic charges, but otherwise generic value of the $N_i$. 
In that scenario, one would select specific values for $N_i$ by some appropriate truncation of the bimodule to a smaller ${\mathcal B}^{N_1,N_2,N_3}_{0;\epsilon_1, \epsilon_2}$.

\section*{Acknowledgements}
D.G. would like to thank K.Costello for innumerable explanations about his work and 
many important comments and A.Okounkov for useful clarifications on affine Yangians. 
J.O. would like to thank Kevin Costello, Yehao Zhou, Tadashi Okazaki, Junya Yagi, Miroslav Rapcak, 
Jingxiang Wu, Nafiz Ishtiaque. 
D.G. is supported by the NSERC
Discovery Grant program, by the Perimeter Institute for Theoretical
Physics and by the Krembil foundation. 
The research of JO is supported in part by Kwanjeong Educational Foundation and by the Visiting Graduate Fellowship Program at the Perimeter Institute for Theoretical Physics.
Research at Perimeter Institute is supported by the
Government of Canada through Industry Canada and by the Province of
Ontario through the Ministry of Research and Innovation.
\appendix

\section{Topological renormalization} \label{app:topren}

\subsection{Naive analysis}
If we are given a theory topological in the $t$ direction, with non-singular OPE along that direction, 
it is straightforward to define topological line defects by coupling the theory to some auxiliary 
topological quantum mechanics with operator algebra $A$. 

A topological theory, by definition, is equipped with an operator $\delta$ which makes translations exact
\begin{equation}
\{Q,\delta\} + \partial_t =0
\end{equation}
and gives a descent operation $O \to [\delta,O] dt$ on the operators $\mathrm{Obs}$ of the theory. 
Here and elsewhere the commutator $[\cdot, \cdot]$ denotes either the commutator or the anti-commutator 
depending on the ghost number of the arguments. 

A topological action is naturally produced by descent:
\begin{equation}
\int  [\delta,x(t)] dt
\end{equation}
with
\begin{equation}
x=\sum_i a_i O^i
\end{equation}  
and $a_i \in A$ and $O^i \in \mathrm{Obs}$ of total ghost number $1$.

Because both the operator algebra of the theory and the auxiliary $A$ can be non-commutative, the action of the line defect should really be thought of 
as the argument of a path-ordered exponential
\begin{equation}
\mathrm{Pexp}\int_{-\infty}^\infty [\delta,x(t)] dt
\end{equation}

If we take the BRST variation of such an expression, we produce two terms: 
\begin{equation}
\int dt' \left[\mathrm{Pexp}\int_{-\infty}^{t'}  [\delta,x(t)] dt \right] \left[ -\partial_{t'} x(t') - [\delta,[Q x(t')]\right] \left[\mathrm{Pexp}\int_{t'}^{\infty}  [\delta,x(t'')] dt \right]
\end{equation}
Integration by part catches boundary terms in the path-ordered exponential, which resum to 
\begin{equation}
-\int dt' \left[\mathrm{Pexp}\int_{-\infty}^{t'}  [\delta,x(t)] dt \right] \left[  -x(t')[\delta,x(t')]+ [\delta,x(t')]x(t') + [\delta,Q x(t')]\right] \left[\mathrm{Pexp}\int_{t'}^{\infty}  [\delta,x(t'')] dt \right]
\end{equation}
which vanishes if we satisfy a Maurer-Cartan equation
\begin{equation}
[Q,x] + x^2 =0
\end{equation}

The local operators on the topological lines are general linear combinations $\sum_i a'_i O^i$, and the boundary terms in the action of $Q$ modify the action of the 
BRST charge on such operators to $[Q+x, \cdot]$, which is nilpotent iff the MC equation is satisfied. 

\subsection{Improved analysis}
Generically, the OPE of two local operators in a topological theory will {\it not} be non-singular. At best, we can say that the 
OPE of BRST-closed operators will have singularities which are BRST exact. We will typically not be willing to pass to the BRST cohomology, as that 
can hamper our ability to write general line defect actions. In any case, explicit calculations are done in the underlying theory, so the 
divergences are problematic even when BRST exact. 

In particular, we cannot simply define an associative product of local operators by bringing them to the same point. Furthermore, when we define the line defect by path-ordered exponential, we need to renormalize the defect in order to avoid collisions of local operators and consequent divergences. 

It is useful to introduce some ``buffer'' between local operators, by defining for each open segment $(a,b)$ in $\bR$ a space $\mathrm{Obs}_{(a,b)}$
of observables which are supported strictly within $(a,b)$. Typical examples will be combinations of local operators inserted at distinct points within $(a,b)$.
In any QFT, if we are given a segment $(a,b)$ and smaller non-overlapping segments $(a_i,b_i)$ within $(a,b)$, we have a non-singular multilinear insertion map 
\begin{equation}
E_{\{(a_i,b_i)\}}^{(a,b)} : \prod_i \mathrm{Obs}_{(a_i,b_i)} \to \mathrm{Obs}_{(a,b)}
\end{equation}
which simply places the operators within $(a_i,b_i)$ at the corresponding locations within $(a,b)$. We also have a space $\mathrm{Obs}_{(-\infty,\infty)}$
built from local operators inserted anywhere on the line, and corresponding insertion maps. 

The special feature of TFTs is that translations are BRST exact. Because of that, we expect to have quasi-isomorphisms between 
$\mathrm{Obs}_{(a,b)}$ and $\mathrm{Obs}_{(a',b')}$ for any pairs of segments, with all sort of homotopies expressing the equivalence between 
any compositions of these identifications. To be specific, we can map some collection of local operators in $\mathrm{Obs}_{(a,b)}$ 
to the same collection in $\mathrm{Obs}_{(a',b')}$ by an uniform rescaling and translation of their positions. 

In the following we will fix once and for all a reference segment, say $(0,1)$. For every ordered collection $\gamma_0 = \{(a_i,b_i)\}$ of $n$ segments we can compose the 
above quasi-isomorphisms with the insertion map to get composition maps
\begin{equation}
E_{\gamma_0} : \prod_i \mathrm{Obs}_{(0,1)} \to \mathrm{Obs}_{(0,1)}
\end{equation}
If we have two ordered collections of $n$ segments, we can compare them by some homotopy, to be visualize as a 1-chain $\gamma_1$ in the 
configuration space $\mathrm{Conf}_n$ of $n$ ordered open segments within $(0,1)$. The homotopy will be implemented by some 
\begin{equation}
E_{\gamma_1} : \prod_i \mathrm{Obs}_{(0,1)} \to \mathrm{Obs}_{(0,1)}
\end{equation}
of ghost number $-1$. Iterating this, we get a collection of maps $E_{\gamma}$ labelled by chains $\gamma$ in $\mathrm{Conf}_n$,
such that 
\begin{equation}
[Q, E_{\gamma}] = E_{\partial \gamma}
\end{equation}
We take this as a basic piece of the definition of $E_1$ algebra of local observables. We will also use similar maps
\begin{equation}
\hat E_{\hat \gamma} : \prod_i \mathrm{Obs}_{(0,1)} \to \mathrm{Obs}_{(-\infty, \infty)}
\end{equation}
involving the configuration space $\hat{\mathrm{Conf}}_n$ of $n$ ordered open segments within $(-\infty, \infty)$.

Finally, we define the composition $\gamma \circ \gamma'$ of two cycles in $\mathrm{Conf}_n$ as a sum (with appropriate Koszul signs) 
over all segments in $\gamma$ of configurations where we replace the segment $(a_i, b_i)$ with $\gamma'$, uniformly rescaled so that 
$(0,1)$ is mapped to $(a_i, b_i)$. We define $\hat \gamma \circ \gamma'$ in a similar manner. 

We will assume for simplicity that $E_{\gamma \circ \gamma'} = E_{\gamma} \circ E_{\gamma'}$ where the composition again involves a sum 
with appropriate Koszul signs over all possible insertions of $E_{\gamma'}$ into a slot of $E_{\gamma'}$. This should hold if we used uniform rescalings to identify 
observables in different segments. If the equality only holds up to homotopies, we can adjust 
the construction below accordingly. 

We are now ready to set up our topological renormalization scheme. First, we define recursively the following collection of 
$(n-2)-$chains $M_n$ in $\mathrm{Conf}_n$. We take $M_2$ to be any configuration of two segments within $(0,1)$. 
We take $M_3$ to be any chain with boundary $-M_2 \circ M_2$. Next, observe that 
\begin{equation}
\partial (M_2 \circ M_3 + M_3 \circ M_2) = -M_2 \circ (M_2 \circ M_2) + (M_2 \circ M_2) \circ M_2
\end{equation}
but composition is associative and thus we can find $M_4$ such that
\begin{equation}
\partial M_4 + M_2 \circ M_3 + M_3 \circ M_2 =0
\end{equation}
Next, one defines $M_5$ such that 
\begin{equation}
\partial M_5 + M_2 \circ M_4 + M_3 \circ M_3+ M_4 \circ M_2 =0
\end{equation}
etcetera. 

These cycles define for us some useful $A_\infty$ algebra with operations 
\begin{equation}
\mu_1 = [Q,\cdot] \qquad \qquad \mu_n = E_{M_n}
\end{equation}
which will play the role of the naive dg-algebra of local operators in the case of non-singular OPE. 

The $A_\infty$ axioms follow from the observation that $[Q,E_{M_2}]=0$, 
\begin{equation}
[Q,E_{M_3}] = -E_{M_2 \circ M_2} = -\mu_2 \circ \mu_2
\end{equation}
etc. 

Next, we can build recursively $n-$cycles $L_n$ in $\hat{\mathrm{Conf}}_n$ representing the 
regularization of the integration cycles in the path-ordered exponential. We can take $L_1$ to be 
a uniform translation of $(0,1)$ from $-\infty$ to $\infty$. Next, we define $L_2$ 
so that 
\begin{equation}
\partial L_2 = L_1 \circ M_2
\end{equation}
Essentially, the second segment in $L_2$ sweeps from $-\infty$ to $\infty$, while the first
sweeps from $- \infty$ to some minimum cutoff distance from the second. When the segments are far from each other they can have 
length $1$, but should shrink as they approach each other so that they fit together as the segments in $M_2$ 
when they reach the minimum approach. Recursively, we require $\partial L_3 = L_2 \circ M_2 + L_1 \circ M_3$, etc.  

We now define our renormalized path-ordered exponential as
\begin{equation}
\mathrm{Pexp}\int_{-\infty}^\infty [\delta,x(t)] dt \equiv \sum_n \hat E_{L_n}(x^{\otimes n})
\end{equation}
with $O^i \in \mathrm{Obs}_{(0,1)}$.
Then the action of the BRST charge gives 
\begin{equation}
\sum_{n,n'} \hat E_{L_{n+n'}}(x^{\otimes n},[Q,x],x^{\otimes n'}) + \hat E_{L_n \circ M_{n'}}(x^{\otimes (n+n'-1)})
\end{equation}
which vanishes if we satisfy the MC equation 
\begin{equation}
\sum_n \mu_n(x^{\otimes n}) =0
\end{equation}
Similarly, the BQST operator on the line is deformed to 
\begin{equation}
\sum_{n,m} \mu_{n+n'+1}(x^{\otimes n},\cdot,x^{\otimes n'}) =0
\end{equation}

\section{Topological line defects, $A_\infty$ algebras and Koszul duality} \label{app:koszul}

We can present the requires facts about Koszul duality either starting from the $A_\infty$ algebra side or from the associative algebra side.
Before doing that, we should recall why $A_\infty$ algebras are relevant to the present context. 

\subsection{$A_\infty$ algebras in physics}
Essentially by definition, a QFT is equipped with some notion of OPE which allows one to combine a collection 
of operators with disjoint support into a composite operator. Mathematically, this structure is denoted as a
``factorization algebra'' \cite{Costello:2016vjw}. The factorization algebras for topological, cohomological QFTs
in $d$ dimensions  are denoted as $E_d$ algebras. In particular, $E_1$ algebras encode the OPE of local operators on a topological line defect. 

By definition, a line defect is topological if we have an odd symmetry $\delta_t$ which makes translations $Q$-exact:
\begin{equation}
\partial_t = \{Q, \delta \}
\end{equation}
In particular, this gives descent relations: we have 
\begin{equation}
Q (\delta O) + \delta (Q O) = \partial_t O
\end{equation}
The $E_1$ algebra structure ``integrates'' this infinitesimal BRST homotopy to homotopies which can 
move in an arbitrary way the locations of a collection of local operators without changing their relative order,
together with compatible fusion operations which can replace any sequence of local operators with a single local operator.

An important part of this large structure can be distilled to the operations of an $A_\infty$ algebra $\mathfrak{A}$, as described in detail 
in the previous appendix \ref{app:topren}. 
The first operation is simply the the BRST differential. The second operation $(\cdot, \cdot)_\mathfrak{A}$ 
is a regularized OPE of two operators. This operation is in general not associative, but the associator is controlled a the third operation $(\cdot, \cdot,\cdot)_\mathfrak{A}$, etcetera.
The various renormalization choices have a degree of arbitrariness, which mean that $\mathfrak{A}$ is only defined up to $A_\infty$ quasi-isomorphism. 

An important role of the $A_\infty$ operations is to control the deformations of the line defect, at least within perturbation theory. 
A simple deformation of the defect takes the form 
\begin{equation}
\mathrm{Pexp} \left[\int_{t=- \infty}^\infty [\delta, x] dt\right]
\end{equation}
for a ghost-number $1$ element $x$. 

At the leading order, the BRST variation of the deformed line defect is
\begin{equation}
\int Q\delta x dt = \int_{t=- \infty}^\infty \partial_t x dt - \int_{t=- \infty}^\infty \delta (Q x) dt 
\end{equation}
and the defect preserves BRST invariance if $Q x =0$. Perturbatively, this condition is deformed to the Maurer-Cartan equation
\begin{equation}
Q x + (x,x)_\mathfrak{A}+ (x,x,x)_\mathfrak{A}+ \cdots =0
\end{equation}

The deformed line defect has deformed $A_\infty$ algebra operations $\mathfrak{A}_x$ 
of the form 
\begin{equation}
(O_1,\cdots,O_n)_{\mathfrak{A}_x} = \sum_{k_0, \cdots, k_n} (x^{\otimes k_0}, O_1, x^{\otimes k_1},\cdots, O_n, x^{\otimes k_n})
\end{equation}
One also gets similar operations for local operators interpolating different deformations of the defect, etc. 

An important generalization of this construction introduce Chan-Paton degrees of freedom, in the form of a topological quantum mechanics 
with some differential graded operator algebra $A$. The Maurer-Cartan equation makes sense for $x \in A \otimes \mathfrak{A}$, with $Q$ 
defined by Leibniz rule and the higher operations being combined with algebra composition. Then $\mathfrak{A}_x$  
is the corresponding deformation by $x$ of $A \otimes \mathfrak{A}$.

\subsection{Koszul duality from the $A_\infty$ side.}

Consider an $A_\infty$ algebra $\mathfrak{A}$ with the property that the ghost number $0$ elements consist of the identity only
and all other elements lie in positive ghost number. Consider deformations of the corresponding line defect such that 
the Chan-Paton algebra $A$ is an actual algebra, living in ghost number $0$. 

If we pick a basis $c^i$ of the space $\mathfrak{A}_1$ of ghost-number $1$ operators and $e_I$ of the space $\mathfrak{A}_2$ of ghost-number $2$ operators in $\mathfrak{A}$,
the MC equation for $x = \sum_i a_i c^i$ takes the form (summing over repeated indices)
\begin{equation} \label{eq:mc}
a_i (Q c^i)^I + a_i a_j (c^i,c^j)^I_\mathfrak{A}+ a_i a_j a_k (c^i,c^j,c^k)^I_\mathfrak{A}+ \cdots =0
\end{equation}
We can interpret the $a_i$ as an algebra morphism ${}^! \mathfrak{A} \to A$, where the latter is the
{\it Koszul dual} (unital) algebra defined by generators $a_i$ and relations \ref{eq:mc}.

That means there is a one-to-one correspondence between solutions of MC equations, i.e. topological line defects built from the 
original one, and algebra morphisms ${}^! \mathfrak{A} \to A$.

Notice that the Koszul dual algebra ${}^! \mathfrak{A}$ is equipped with an ``augmentation'', i.e. an algebra morphism 
${}^! \mathfrak{A} \to \bC$ sending $a_i \to 0$. This represents the original, undeformed line defect. 

Finally, consider two $A_\infty$ algebras $\mathfrak{A}$ and $\tilde{\mathfrak{A}}$ as above, related by an $A_\infty$ morphism, 
i.e. a collection of maps $\phi(\cdot)$, $\phi(\cdot, \cdot)$, etc. which satisfy appropriate generalized associativity axioms. 
An important property of $A_\infty$ morphisms is that they map MC elements to MC elements:
\begin{equation}
\tilde x = \phi(x) + \phi(x,x) + \cdots
\end{equation}
Correspondingly, we can use the morphism maps to define a map of algebras ${}^! \phi$ in the opposite direction
\begin{equation}
\tilde a_{i'} =\phi(c^i)_{i'} a_i + \phi(c^i,c^j)_{i'} a_i a_j + \cdots
\end{equation}

\subsection{Starting from the associative algebra}
Consider some associative algebra $A$ equipped with an {\it augmentation}, 
i.e. a morphism $A \to \bC$ whose kernel we denote as $K$.
This is the same as saying that $A$ admits a one-dimensional module $\bC$, on which $K$ acts trivially. 

Consider the graded tensor algebra 
\begin{equation}
TK^*[1]=\oplus_{n=0}^\infty K^*[1]^{\otimes n}
\end{equation} 
This can be equipped with a differential $Q$ induced from the map 
\begin{equation}
K^* \to K^* \otimes K^* 
\end{equation}
dual to the algebra operation. The resulting differential graded algebra, supported in positive degree, is the {\it Koszul dual} ${}^! A$ of $A$.
This is the algebra of derived endomorphisms of the one-dimensional module $\bC$. 

Concretely, if $k_i$ are a basis of $K$ with structure constants $c^i_{jk}$ and $t^i$ are a basis of $K^*[1]$, we define $Q$ using the Leibniz rule and 
\begin{equation}
Q t^i = -c^i_{jk} t^j \otimes t^k
\end{equation}
Then 
\begin{equation}
Q^2 t^i = c^i_{jk} c^j_{uv} t^u \otimes t^v \otimes t^k - c^i_{jk} c^k_{uv} t^j \otimes t^u \otimes t^v
\end{equation}
which vanishes by associativity. 

The element $x = \sum_i k_i \otimes t^i$ in the composite differential graded algebra $A \times {}^! A$ has a nice property: it satisfies a Maurer-Cartan equation
\begin{equation}
Q x + x^2 = 0
\end{equation}
Furthermore, if we have any other augmented algebra $B$ and a Maurer-Cartan element $x_B = \sum_i b_i \otimes t^i$ in $B \times {}^! A$,
we must have $b_i b_j = c^k_{ij} b_k$, i.e. the $b_i$ define a map $A \to B$. 

A differential graded algebra is a special example of an $A_\infty$ algebra. $A_\infty$ algebras emerge naturally whenever we try to 
``simplify'' a differential graded algebra by eliminating some elements paired up by the differential. If we apply such simplification to 
${}^! A$ we will obtain an $A_\infty$ algebra $\mathfrak{A}$ quasi-isomorphic to it, whose Koszul dual is isomorphic to the original $A$.

We can discuss a couple of instructive examples. The first is $A$ being a symmetric algebra $A = \mathrm{Sym}^* V$. 
Then the dual algebra  ${}^! A$ has generators $t^i$, $t^{i,j}$, etc. dual to $v_i$, $v_i v_j$, etc. The differential maps
\begin{align}
t^{i,j} \to t^i \otimes t^j + t^j \otimes t^i \cr
t^{i,j,k} \to t^{i,j} \otimes t^k + \cdots \cr
\cdots
\end{align}
and the cohomology consists of antisymmetric powers $t^i \wedge t^j \cdots$. We can find an $A_\infty$ quasi-morphism 
between ${}^! A$ and the exterior algebra built from $V^*$, with no differential nor higher operations. This is the standard Koszul dual of $\mathrm{Sym}^* V$.
It has a simple Maurer-Cartan element $\sum_i v_i t^i$. 

The second example is $A$ being the universal enveloping algebra $U(\mathfrak{a})$ of some Lie algebra $\mathfrak{a}$. 
Then ${}^! A$ can be simplified to the exterior algebra built from $\mathfrak{a}^*$, with differential 
\begin{equation}
Q c^i = -f^i_{jk} c^j c^k 
\end{equation}

\subsection{Modules}
Suppose that we have a module $M$ for an associative algebra $A$, and that $A$ has an augmentation. 
The space
\begin{equation}
TK^*[1] \otimes M^*
\end{equation} 
with differential $Q$ induced from the dual of the algebra and module maps 
\begin{equation}
K^* \to K^* \otimes K^*  \qquad \qquad M^* \to K^* \otimes M^*
\end{equation}
is a differential graded module ${}^! M$ for ${}^! A$. 

Concretely, if $m_a$ are a basis of $M$ with structure constants $c^a_{jb}$ and $n^a$ are a basis of $M^*$, we define $Q$ using the Leibniz rule and 
\begin{equation}
Q n^a = -c^a_{jb} t^j \otimes n^b
\end{equation}

The element $y = \sum_i m_a \otimes n^a$ has a nice property: it is closed under the $Q_x$ differential associated to the Maurer-Cartan element $x$ above:
\begin{equation}
Q y + x y= 0
\end{equation}
Furthermore, if we have any other module $\tilde M$ and a $Q_x$-closed element in $\tilde M \otimes {}^! M$, 
we must have an $A$-module map $M \to \tilde M$, and viceversa. 

If we simplify  ${}^! A$ to some $A_\infty$ algebra $\mathfrak{A}$, then ${}^! M$ will map to an $\mathfrak{A}$ module $\mathfrak{M}$.  

Conversely, suppose that we have an $A_\infty$ module $\mathfrak{M}$ for $\mathfrak{A}$ concentrated in non-negative ghost number.

If we use some algebra $A$ to define a MC element associated to a morphism ${}^! \mathfrak{A} \to A$,
we can pick a module $M$ of $A$ and look for BRST-closed elements in $M \otimes \mathfrak{M}$.

Expanding out the equations, they will take the form of a morphism of ${}^! \mathfrak{A}$ modules from a Koszul dual module 
${}^! \mathfrak{M}$ to $M$, where the latter is seen as an ${}^! \mathfrak{A}$ module. The Koszul dual module ${}^! \mathfrak{M}$ is the 
dual of the ghost-number 0 part $\mathfrak{M}_0$. The module relations are dual to $\mathfrak{M}_1$, etc. 

Analogous statements hold for bi-modules. 

\bibliographystyle{JHEP}

\bibliography{mono}

\end{document}